\documentclass[aip,amsmath,amssymb,reprint]{revtex4-2}
\usepackage{float}
\usepackage{graphicx}
\usepackage{txfonts}
\usepackage{multirow}
\usepackage{hyperref}
\usepackage{braket}
\usepackage[usenames,dvipsnames]{color}
\usepackage{listings}
\usepackage{longtable}
\usepackage{xr}

\usepackage[normalem]{ulem}

\makeatletter
\newcommand*{\addFileDependency}[1]{
\typeout{(#1)}
\@addtofilelist{#1}
\IfFileExists{#1}{}{\typeout{No file #1.}}
}\makeatother

\newcommand*{\myexternaldocument}[1]{%
\externaldocument[SI-]{#1}%
\addFileDependency{#1.tex}%
\addFileDependency{#1.aux}%
}
\myexternaldocument{supplemental_}

\newcommand{\tvb}{\textsc{TurboRVB}} 
\newcommand{\tbw}{\textsc{TurboWorkflows}} 
\newcommand{\tbg}{\textsc{TurboGenius}} 
\newcommand{\pyturbo}{\textsc{PyTurbo}}
\newcommand{\tbf}{\textsc{TurboFilemanager}}

\newcommand{\pyscf}{\textsc{PySCF}}
\newcommand{\qpack}{\textsc{Quantum Package}}
\newcommand{\qe}{\textsc{Quantum Espresso}}
\newcommand{\trexio}{\textsc{TREXIO}}
\newcommand{\program}{\texttt}

\definecolor{dkgreen}{rgb}{0,0.6,0}
\definecolor{gray}{rgb}{0.5,0.5,0.5}
\definecolor{mauve}{rgb}{0.58,0,0.82}

\lstdefinestyle{PythonStyle}{frame=tb,
  language=Python,
  aboveskip=3mm,
  belowskip=3mm,
  showstringspaces=false,
  columns=flexible,
  basicstyle={\small\ttfamily},
  numbers=none,
  numberstyle=\tiny\color{gray},
  keywordstyle=\color{blue},
  commentstyle=\color{dkgreen},
  stringstyle=\color{mauve},
  breaklines=true,
  breakatwhitespace=true,
  tabsize=3
}

\lstdefinestyle{ShellStyle}{frame=tb,
  language=Bash,
  aboveskip=3mm,
  belowskip=3mm,
  showstringspaces=false,
  columns=flexible,
  basicstyle={\small\ttfamily},
  numbers=none,
  numberstyle=\tiny\color{gray},
  keywordstyle=\color{blue},
  commentstyle=\color{dkgreen},
  stringstyle=\color{mauve},
  breaklines=true,
  breakatwhitespace=true,
  tabsize=3
}

\begin{document}
\title{\tbg: Python suite for high-throughput calculations of \emph{ab initio} quantum Monte Carlo methods}
\author{Kousuke Nakano}
\email{kousuke\_1123@icloud.com}
\affiliation{International School for Advanced Studies (SISSA), Via Bonomea 265, 34136, Trieste, Italy}
\affiliation{Center for Basic Research on Materials, National Institute for Materials Science (NIMS), Tsukuba, Ibaraki 305-0047, Japan}
\author{Oto Kohul\'{a}k}
\affiliation{International School for Advanced Studies (SISSA), Via Bonomea 265, 34136, Trieste, Italy}
\affiliation{Laboratoire de Chimie et Physique Quantiques (LCPQ), Universit{\'e} de Toulouse (UPS) and CNRS, Toulouse, France}
\author{Abhishek Raghav}
\affiliation{International School for Advanced Studies (SISSA), Via Bonomea 265, 34136, Trieste, Italy}
\affiliation{Institut de Min{\'e}ralogie, de Physique des Mat{\'e}riaux et de Cosmochimie (IMPMC), Sorbonne Universit{\'e}, CNRS UMR 7590, IRD UMR 206, MNHN, 4 Place Jussieu, 75252 Paris, France}
\author{Michele Casula}
\affiliation{Institut de Min{\'e}ralogie, de Physique des Mat{\'e}riaux et de Cosmochimie (IMPMC), Sorbonne Universit{\'e}, CNRS UMR 7590, IRD UMR 206, MNHN, 4 Place Jussieu, 75252 Paris, France}
\author{Sandro Sorella}
\affiliation{International School for Advanced Studies (SISSA), Via Bonomea 265, 34136, Trieste, Italy}
%\author{Michele Casula}
%\affiliation{Institut de Min{\'e}ralogie, de Physique des Mat{\'e}riaux et de Cosmochimie (IMPMC), Sorbonne Universit{\'e}, CNRS UMR 7590, IRD UMR 206, MNHN, 4 Place Jussieu, 75252 Paris, France}

\date{\today}

\begin{abstract}
\tbg\ is an open-source Python package designed to fully control \emph{ab initio} quantum Monte Carlo (QMC) jobs using a Python script, which allows one to perform high-throughput calculations combined with \tvb\ [K. Nakano et al. \emph{J. Phys. Chem.} \underline{152}, 204121 (2020)]. This paper provides an overview of the \tbg\ package and showcases several results obtained in a high-throughput mode. For the purpose of performing high-throughput calculations with \tbg, we implemented another open-source Python package, \tbw, that enables one to construct simple workflows using \tbg. We demonstrate its effectiveness by performing (1) validations of density functional theory (DFT) and QMC drivers as implemented in the \tvb\ package and (2) benchmarks of Diffusion Monte Carlo (DMC) calculations for several data sets. For (1), we checked inter-package consistencies between \tvb\ and other established quantum chemistry packages. By doing so, we confirmed that DFT energies obtained by \pyscf\ are consistent with those obtained by \tvb\ within the local density approximation (LDA), and that Hartree-Fock (HF) energies obtained by \pyscf\ and \qpack\ are consistent with variational Monte Carlo energies obtained by \tvb\ with the HF wavefunctions. These validation tests constitute a further reliability check of the \tvb\ package. For (2), we benchmarked atomization energies of the Gaussian-2 set, binding energies of the S22, A24, and SCAI sets, and equilibrium lattice parameters of 12 cubic crystals using DMC calculations. We found that, for all compounds analyzed here, the DMC calculations with the LDA nodal surface give satisfactory results, i.e., consistent either with high-level computational or with experimental reference values.
\end{abstract}
\maketitle

\makeatletter
\def\Hline{
\noalign{\ifnum0=`}\fi\hrule \@height 1pt \futurelet
\reserved@a\@xhline}
\makeatother

%%%%%%%%%%%%%%%%%%%%%%%%%%%%%%%%%%%%%%%%%%%%%%%%%%%%
%  Introduction
%%%%%%%%%%%%%%%%%%%%%%%%%%%%%%%%%%%%%%%%%%%%%%%%%%%%
\section{Introduction}
\label{sec:intro}
\begin{figure}[htbp]
  \centering
  \includegraphics[width=\columnwidth]{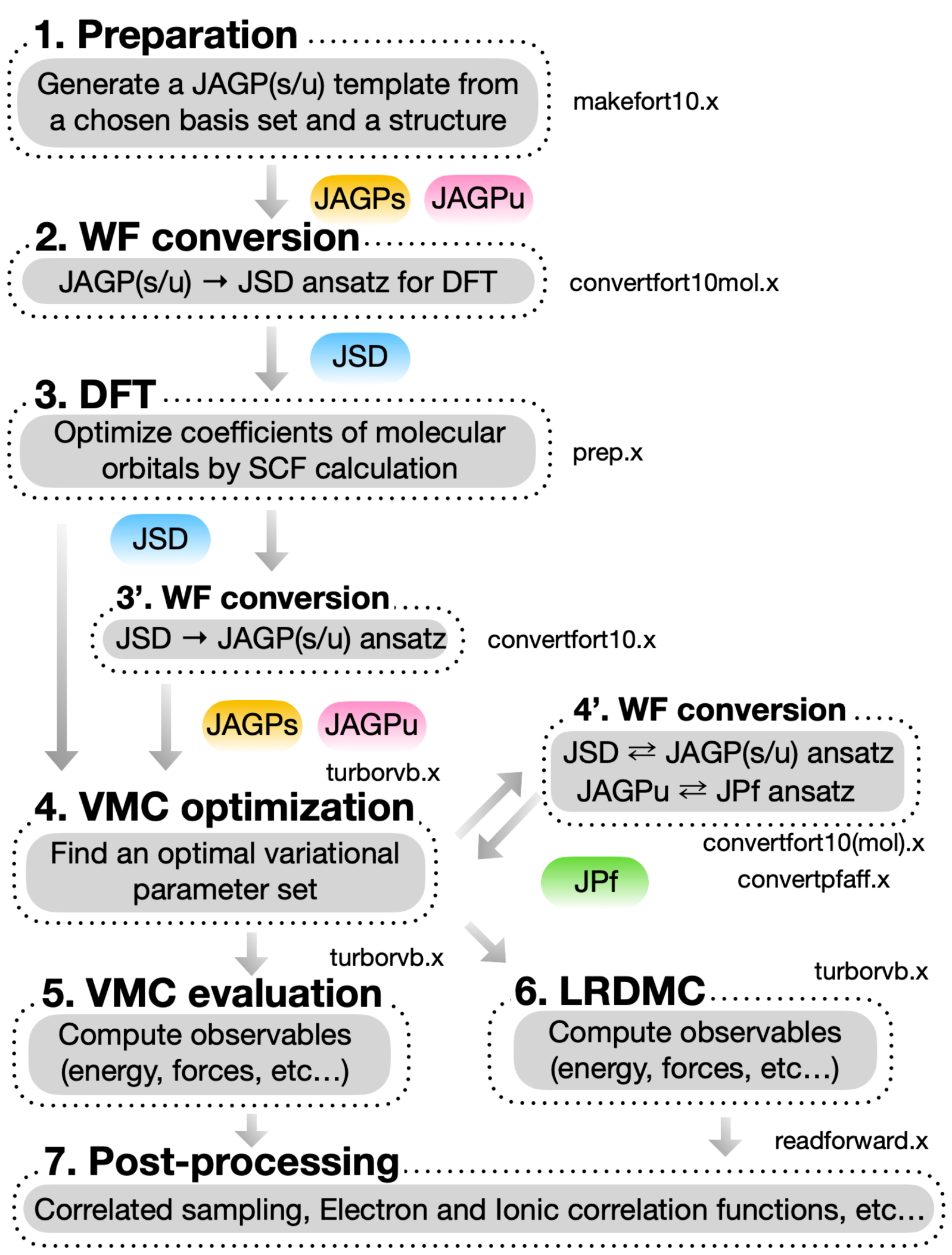}
  \caption{
    A typical workflow of a QMC calculation using \tvb;
    (1) Preparation of a JAGP(s/u) ansatz file from a chosen basis set, pseudo potentials, and a structure using {\program{makefort10.x}} binary. AGPs stands for the symmetric antisymmetrized geminal power (i.e., singlet correlation in the pairing function)~{\cite{2019GEN3}}, while AGPu stands for the broken symmetry antisymmetrized geminal power, (i.e., singlet + triplet correlations in the pairing function)~{\cite{2019GEN3}}. The generated WF is a template, i.e., it is fulfilled with random numbers; thus, one needs to initialize it, typically using DFT; 
    (2) Conversion of the generated JAGP(s/u) ansatz to JSD one using {\program{convertfort10mol.x}} binary because the subsequent DFT calculation works only with molecular orbitals (SD); 
    (3) Initializing wave functions (WFs) using the build-in DFT module ({\program{prep.x}});
    (3') If one wants to use JAGP(s/u) ansatz, one can convert the initialized JSD ansatz to a JAGP(s/u) ansatz before the VMC optimization;
    (4) Optimization of a WF at the VMC level using {\program{turborvb.x}} binary;
    (4') One can convert the optimized WF to another type of ansatz.
    Pf stands for the Pfaffian anstaz, which is the most general form of the AGP WF~{\cite{2020NAK2}}.
    (5) Computation of observables such as energy and forces at the VMC level using {\program{turborvb.x}} binary;
    (6) Computation of observables such as energy and forces at the LRDMC level using {\program{turborvb.x}};
    (7) Computation of other physical properties such as electron density.
  }
  \label{fig:workflow}
\end{figure}

In recent years, there has been a surge of interest in materials informatics and digital transformation paradigms in the materials science community, which involve utilizing information science and computational chemistry/physics techniques to design or search for novel materials. The kernel method for high-throughput electronic structure calculations is most commonly the Density Functional Theory (DFT), which has been successfully used for designing various materials {\it in silico}~{\cite{2004OGA, 2014NIS, 2017HAY, 2017WANG, 2019WANG, 2019SUN, 2021OUY, 2023KAT}}. However, DFT sometimes loses the quantitative predictive power in particular cases, such as materials at extreme conditions or with strong electronic correlation~{\cite{2014CLAY, 2016CLAY, 2018SOR, 2022LY, 2022NIK, 2023LOR}}. Instead, \emph{ab initio} quantum Monte Carlo (QMC) does not lose predictive power even for such materials because it does not rely on any exchange-correlation functionals in its formalism{~\cite{2001FOU}}. However, when it comes to \emph{ab initio} QMC applications, one of the biggest drawbacks is its complicated computational procedure. Indeed, a QMC study usually requires many involved operations, such as generating trial wave functions, variational optimizations, time-step or lattice-size extrapolation, and finite-size corrections. 
For instance, a typical workflow of a QMC calculation using the \tvb\ quantum Monte Carlo package~{\cite{2003CAS, 2020NAK2}}, is shown in Fig.~{\ref{fig:workflow}}. Automating such required tasks can offer a significant improvement in our productivity, enabling researchers to spend more time doing physics and chemistry rather than launching and monitoring jobs.
So far, many workflow management packages have been developed to achieve a more productive research activity and high-throughput calculations. Some representatives are \textsc{AiiDA}~{\cite{2020HUB}}, \textsc{AFLOW}~{\cite{2012STE}}, \textsc{Fireworks}~{\cite{2015ANU}}, and ~\textsc{atomate}~{\cite{2017MAT}}, which have been widely used for generating and/or managing material science database such as \textsc{NOMAD}~{\cite{2019CLA}} and \textsc{Materials Projects}~{\cite{2013ANU}}.
%, and \textsc{ASE}~{\cite{2017LAR}}.
%
One could immediately exploit these established workflow systems also in QMC calculations, but the combination of a QMC code with these workflow packages is not straightforward due to the complexity of the QMC calculations. Thus, to manage them, we need to implement interfaces, if possible in Python, because most of the established workflow packages are also implemented based on Python, due to its appealing features. In fact, several packages for high-throughput QMC calculations, such as \textsc{Nexus}~{\cite{2016KRO2}} and \textsc{QMC-SW}~{\cite{2019KON}}, are Python implementations. Wheeler et al. has very recently developed a new open-source Python-based package for real-space QMC, named {\program{PyQMC}}~{\cite{2023WHE}}. {\program{PyQMC}} is an all-Python package; thus, it enables one to develop algorithms and complex workflows more flexibly and user-friendly.

{\vspace{2mm}}
\tbg\ is an open-source Python package meant to manage \tvb\ calculations using a python script. In this paper, we explain the basic concepts, designs, functionalities, implemented classes, user interfaces, command-line tools of the package. \tbg\ provides Python classes and command-line tools that fully control \tvb\ jobs, which allow one to realize high-throughput QMC calculations. \tbg\ includes \pyturbo\ as a sub-package for providing users with more fundamental but more flexible building blocks to control \tvb\ jobs.
For demonstrating high-throughput QMC calculations, we also implemented another open-source python package, \tbw.
The demonstrations contain: (1) validations of DFT and QMC implementations of the \tvb\ package and (2) benchmarks of Diffusion Monte Carlo (DMC) calculations for several data sets. As per (1), we confirmed that DFT energies obtained by \pyscf~{\cite{2018SUN, 2020SUN}} are consistent with those obtained by \tvb\ within the local density approximation (LDA), and Hartree-Fock (HF) energies obtained by \pyscf\ and \qpack~{\cite{2019GAR}} are consistent with variational Monte Carlo (VMC) energies obtained by \tvb\ computed with the HF wavefunctions. The validation tests constitute a further reliability check of the \tvb\ package. As far as point (2) is concerned, we benchmarked atomization energies of the Gaussian-2 set~{\cite{1991CUR}}, binding energies of the S22~{\cite{2006JUR}}, A24~{\cite{2013REZ}}, and SCAI~{\cite{2009BER}} sets, and equilibrium lattice parameters of 12 cubic crystals. We found that, for all compounds analyzed in this study, the diffusion Monte Carlo (DMC) calculations with the LDA nodal surface give satisfactory results, i.e., consistent either with computational or with experimental reference values. 

{\vspace{2mm}}
%\michele{(HERE, WE SHOULD PROVIDE THE STRUCTURE OF THE PAPER AND A SUMMARY OF ITS CONTENT, BY REFERRING TO THE CORRESPONDING SECTIONS. IN THE PRESENT VERSION, THE READER DISCOVERS TOO LATE THAT ANOTHER IMPLEMENTATION, PYTURBO, IS PRESENTED, WHICH IS NEVER MENTIONED IN THE INTRODUCTION SO FAR. THIS SHOULD BE DONE ALREADY HERE, AS OUTLINE OF THE PAPER, SUCH AS THE READER HAS A CLEARER IDEA OF THE OVERALL STRUCTURE OF THE PAPER AND OF THE PACKAGES AS WELL.}
This paper is organized as follows: in Sec.~{\ref{sec:tbg-overview}}, we provide an overview of the \tbg\ program structure; in Sec.~{\ref{sec:pyturbo-overview}}, we provide an overview of the \pyturbo\ program structure; in Sec.~{\ref{sec:tbg-command-line}}, we describe the command-line tool and user interface implemented in \tbg; in Sec.~{\ref{sec:tbw-overview}}, we introduce \tbw, a python package for realizing high-throughput calculations using \tbg; and in Sec.~{\ref{sec:tbw-demo}}, we showcase several validation and benchmark results obtained using \tbg\ and \tbw.
%
%%%%%%%%%%%%%%%%%%%%%%%%%%%%%%%%%%%%%%%%%%%%%%%%%%%%
% Figure 1 UML class diagram of turbo-genius package
%%%%%%%%%%%%%%%%%%%%%%%%%%%%%%%%%%%%%%%%%%%%%%%%%%%%
%%%%%%%%%%%%%%%%%%%%%%%%%%%%%%%%%%%%%%%%%%%%%%%%%%%%

%%%%%%%%%%%%%%%%%%%%%%%%%%%%%%%%%%%%%%%%%%%%%%%%%%%%
%  Program overview of \tbg
%%%%%%%%%%%%%%%%%%%%%%%%%%%%%%%%%%%%%%%%%%%%%%%%%%%%
\section{\tbg: Program overview}
\label{sec:tbg-overview}
\tbg\ is implemented in \verb|Python 3| (the minimal requirement is \verb|Python|~3.7). Python was chosen because it allows seamless integration with other major workflow frameworks. The main classes of \tbg\ are mainly composed of two types of classes. One is {\program{Wavefunction}} that stores and manipulates a wavefunction information including nuclear positions and pseudopotentials. The others are classes inheriting an abstract class (\program{genius} class). The classes enable one to control \tvb\ tasks, such as generating input files, launching jobs, and analyzing outcomes. The major classes of \tbg\ are listed in Table~{\ref{tab:class-turbogenius}}. Hereafter, we will describe the functionalities of the two main classes.

%%%%%%%%%%%%%%%%%%%%%%%%%%%%%%%%%%%%%%%%%%%%%%%%%%%%
% Table 1 Major classes of the TurboGenius package
%%%%%%%%%%%%%%%%%%%%%%%%%%%%%%%%%%%%%%%%%%%%%%%%%%%%
\begin{center}
\begin{table*}[htbp]
\caption{\label{tab:class-turbogenius} Major classes of the \tbg\ package} 
\vspace{2mm}
\begin{tabular}{c|c|p{12cm}}
\Hline 
Parent Class & Class (or Method) & Description \\ 
\Hline
- & Wavefuntion & Manipulations and conversions of a \tvb\ WF file \\ 
\Hline
\multirow{6}{*}{Genius\_IO}
& DFT\_genius & Managing DFT calculations. The built-in program {\program{Prep}} is used.  \\ 
& VMCopt\_genius & Managing VMC optimizations. \\ 
& VMC\_genius & Managing single-shot VMC calculations. \\ 
& LRDMC\_genius & Managing single-shot LRDMC calculations. \\ 
& LRDMCopt\_genius & Managing LRDMC optimizations. \\ 
& Correlated\_sampling\_genius & Managing correlated sampling jobs. \\ 
\Hline
\end{tabular}
\end{table*}
\end{center}
%%%%%%%%%%%%%%%%%%%%%%%%%%%%%%%%%%%%%%%%%%%%%%%%%%%%

\subsection{Wavefunction}
\label{subsec:tbg-wavefunction}
{\program{Wavefunction}} is a class manipulating wavefunctions, such as for generating a JSD (Jastrow Slater Determinant) or JAGP (Jastrow correlated Antisymmetrized Geminal Power~{\cite{2003CAS}}) WF for a subsequent DFT initialization, and for converting a WF ansatz to another one (e.g., JSD to JAGP). 
%
% xyz file -> TurboRVB WF
The listing~{\ref{lst:turbogenius-wavefunction-generate-WF-example}} shows a script to generate a WF file for the water molecule with the cc-pVQZ and cc-pVDZ basis sets for the determinant and the Jastrow parts, respectively, accompanied with the correlation consistent effective core potentials (ccECPs~{\cite{2017BEN,2018BEN,2018ANN,2019WAN}}). The generated WF and PP files are {\program{fort.10}} and {\program{pseudo.dat}}, respectively. Notice that {\program{Wavefunction}} currently has pre-defined keywords for correlation-consistent basis sets implemented in Basis Set Exchange~{\cite{2019PRI}} for all-electron calculations, and correlation-consistent effective core potentials (ccECPs)~{\cite{2017BEN,2018BEN,2018ANN,2019WAN}} and Burkatzki-Filippi-Dolg (BFD)~{\cite{2007BUR,2008BUR}} basis sets for pseudopotential calculations. One can also use different basis set and pseudopotentials by providing them as a text {\program{list}}. The generated WF is a Slater Determinant WF with randomized MO coefficients. Indeed, a user should initialize it using the built-in DFT ({\program{prep}}) code before doing QMC calculations.
%

% TREXIO -> TurboRVB WF
{\vspace{2mm}}
The {\program{Wavefunction}} class also allows one to generate a wavefunction file from a \trexio\ file~{\cite{trexio}}. The \trexio\ library has a standard format for storing wave functions, together with a C-compatible API such that it can be easily used in any programming language, which is being developed in TREX (Targeting Real Chemical Accuracy at the Exascale) project~{\cite{trex}}. The listing~{\ref{lst:turbogenius-wavefunction-read-trexio-example}} shows a script to convert a \trexio\ file. Thus, instead of using the built-in DFT program, {\program{prep}}, one can use standard quantum chemistry packages such as GAMESS~{\cite{2020BAR}} and \pyscf~{\cite{2018SUN, 2020SUN}}, then convert it to \tvb\ WF format. The class supports both all-electron and pseudopotential WF with/without periodic boundary conditions (i.e., for both molecules and crystals), and both restricted (ROHF) and unrestricted (UHF) open-shell WFs. Notice that a restricted WF is converted to the symmetric antisymmetrized geminal power (AGPs) (i.e., singlet correlation in the pairing function)~{\cite{2019GEN3}}, while an unrestricted WF is converted to the broken symmetry antisymmetrized geminal power (AGPu), (i.e., singlet + triplet correlations in the pairing function)~{\cite{2019GEN3}}.

%%%%%%%%%%%%%%%%%%%%%%%%%%%%%%%%%%%%
\begin{lstlisting}[style=PythonStyle,linewidth=\columnwidth,caption={A python script to generate a WF file},label=lst:turbogenius-wavefunction-generate-WF-example]

# import module
from turbogenius.wavefunction import Wavefunction

# generate a WF file from water.xyz
wavefunction = Wavefunction()
wavefunction.read_from_structure(
    structure_file="water.xyz",
    det_basis_set="cc-pVQZ",
    jas_basis_set="cc-pVDZ",
    pseudo_potential="ccECP",
)
wavefunction.to_jsd() # Jastrow-Slater WF.

\end{lstlisting}
%%%%%%%%%%%%%%%%%%%%%%%%%%%%%%%%%%%%
%
%%%%%%%%%%%%%%%%%%%%%%%%%%%%%%%%%%%%
\begin{lstlisting}[style=PythonStyle,linewidth=\columnwidth,caption={A python script to read a TREXIO file},label=lst:turbogenius-wavefunction-read-trexio-example]

# import module
from turbogenius.wavefunction import Wavefunction

# Read a TREXIO File of water and convert
# it to the TurboRVB format.
wavefunction = Wavefunction()
wavefunction.from_trexio(
    trexio_filename="trexio.hdf5"
)

wavefunction.to_jsd() # Jastrow-Slater WF.
wavefunction.to_jagps() # Jastrow-AGPs WF.

\end{lstlisting}
%%%%%%%%%%%%%%%%%%%%%%%%%%%%%%%%%%%%

\subsection{GeniusIO classes}
\label{subsec:tbg-geniusio}
{\program{GeniusIO}} is a parent class of the wrappers to manage complex QMC procedures. Several classes inheriting {\program{GeniusIO}} are implemented in \tbg, as shown in Table~{\ref{tab:class-turbogenius}}. Here, to make the explanation simple, we focus on one of the classes, {\program{VMC\_genius}}. {\program{VMC\_genius}} is a class to control VMC job of \tvb. The listing~{\ref{lst:turbogenius-vmc-genius-example}} shows a Python script to compute VMC energy of the hydrogen dimer.
%%%%%%%%%%%%%%%%%%%%%%%%%%%%%%%%%%%%
\begin{lstlisting}[style=PythonStyle,linewidth=\columnwidth,caption={A python script to compute VMC energy of the hydrogen dimer},label=lst:turbogenius-vmc-genius-example]

# one needs "fort.10" (i.e., a WF file of the H2 dimer) for this calculation.

# import modules
from turbogenius.vmc_genius import VMC_genius

# (1) create a vmc_genius instance
vmc_genius = VMC_genius(
    vmcsteps=600,  # The number of MCMC steps
    num_walkers=40,  # The number of walkers
)

# (2) generate an input file
vmc_genius.generate_input(input_name="datas_vmc.input")

# (3) launch a VMC run
vmc_genius.run(input_name="datas_vmc.input", output_name="out_vmc.o")

# (4) compute energy and forces with reblocking
vmc_genius.compute_energy_and_forces(bin_block=10, warmupblocks=10)

# print vmc energy
print(f"VMC energy = {vmc_genius.energy:.5f} +- {vmc_genius.energy_error:.5f} Ha")

\end{lstlisting}
%%%%%%%%%%%%%%%%%%%%%%%%%%%%%%%%%%%%
The procedure is composed of 4 steps, i.e., (1) create a {\program{vmc\_genius}} instance, (2) generate an input file, (3) run the VMC job, and (4) post-processing (reblocking): (1) The input parameters used here are {\program{vmcsteps (int)}} and {\program{num\_walkers (int)}} specifying the total number of Markov Chain Monte Carlo (MCMC) steps and the number of walkers used in total, respectively. (2) One can generage the input file corresponding to the parameters. (3) One can launch the VMC job. (4) After the VMC job is completed, one can post-process the outcomes depending on the type of jobs. For instance, in the {\program{VMC\_genius}} class, the method {\program{compute\_energy\_and\_forces}} allows one to get the means and variances of the energy, forces, and stresses using the \tvb\ built-in scripts implementing the bootstrap and jackknife methods{~\cite{2017BEC}}, where one can use the reblocking (binning) technique to remove the autocorrelation bias.\cite{1989FLY}. The related options are {\program{bin\_block}} (int): block length and {\program{warmupblocks}} (int): the number of disregarded blocks.
%jackknife method [japanese] https://qiita.com/kaityo256/items/174347b12717ba49263d
%jackknife method [japanese] https://qiita.com/nijigen_plot/items/cb7a51b4c42349a1d0a0

{\vspace{2mm}}
The listing~{\ref{lst:turbogenius-hydrogen-workflow}} shows a Python script to compute the VMC energy of the hydrogen dimer with JSD ansatz with DFT orbitals (JDFT).  We notice that \tvb\ commands launched by \tbg\ can be specified through an environmental variable {\program{TURBOGENIUS\_QMC\_COMMAND}}. For instance, when one sets {\program{TURBOGENIUS\_QMC\_COMMAND='mpirun -np 64 turborvb-mpi.x'}}, one can launch the VMC job with 64 MPI processes. One can run a serial job by an environmental value {\program{TURBOGENIUS\_QMC\_COMMAND='turborvb-serial.x'}}.
%%%%%%%%%%%%%%%%%%%%%%%%%%%%%%%%%%%%
\begin{widetext}
\begin{lstlisting}[style=PythonStyle,linewidth=\columnwidth,caption={A python workflow to obtain the VMC energy of the Hydrogen dimer.},label=lst:turbogenius-hydrogen-workflow]

#!/usr/bin/env python

# TREXIO -> JDFT WF (fort.10) -> VMCopt (only Jastrow) -> VMC (JDFT)

# import modules
import os, shutil
from turbogenius.pyturbo.basis_set import Jas_Basis_sets
from turbogenius.wavefunction import Wavefunction
from turbogenius.vmc_opt_genius import VMCopt_genius
from turbogenius.vmc_genius import VMC_genius

# trexio filename (Hydrogen dimer)
trexio_filename = "H2_trexio.hdf5"

# Start a workflow
root_dir = os.getcwd()

# TREXIO -> TurboRVB WF
trexio_dir = os.path.join(root_dir, "01trexio")
os.makedirs(trexio_dir, exist_ok=True)
shutil.copy(
    os.path.join(root_dir, trexio_filename), os.path.join(trexio_dir, trexio_filename)
)
os.chdir(trexio_dir)

# Jastrow basis (GAMESS format)
H_jastrow_basis = """
        S  1
        1       1.873529  1.00000000
        S  1
        1       0.343709  1.00000000
        S  1
        1       0.139013  1.00000000
        P  1
        1       0.740212  1.00000000
"""

H2_jas_basis_sets = Jas_Basis_sets.parse_basis_sets_from_texts(
    [H_jastrow_basis, H_jastrow_basis], format="gamess"
)

# convert the TREXIO file to the TurboRVB WF format (i.e., fort.10)
wavefunction = Wavefunction()
wavefunction.read_from_trexio(
    trexio_filename=os.path.join(trexio_dir, trexio_filename),
    jas_basis_sets=H2_jas_basis_sets,
)

os.chdir(root_dir)

# Optimization of Jastrow factor
vmcopt_dir = os.path.join(root_dir, "02vmcopt")
os.makedirs(vmcopt_dir, exist_ok=True)
copy_files = ["fort.10", "pseudo.dat"]
for file in copy_files:
    shutil.copy(os.path.join(trexio_dir, file), os.path.join(vmcopt_dir, file))
os.chdir(vmcopt_dir)

# generate a vmcopt_genius instance
vmcopt_genius = VMCopt_genius(
    vmcoptsteps=100,
    steps=50,
    warmupblocks=0,
    num_walkers=40,
    optimizer="lr",
    learning_rate=0.35,
    regularization=0.001,
    opt_onebody=True,
    opt_twobody=True,
    opt_det_mat=False,
    opt_jas_mat=True,
    opt_det_basis_exp=False,
    opt_jas_basis_exp=False,
    opt_det_basis_coeff=False,
    opt_jas_basis_coeff=False,
)

# generate the input file and run
vmcopt_genius.generate_input(input_name="datasmin.input")
vmcopt_genius.run(input_name="datasmin.input", output_name="out_min")

# average the optimized variational parameters
vmcopt_genius.average(optwarmupsteps=5)

os.chdir(root_dir)

# VMC calculation with the optimized WF
vmc_dir = os.path.join(root_dir, "03vmc")
os.makedirs(vmc_dir, exist_ok=True)

copy_files = ["fort.10", "pseudo.dat"]
for file in copy_files:
    shutil.copy(os.path.join(vmcopt_dir, file), os.path.join(vmc_dir, file))
os.chdir(vmc_dir)

# generate a vmc_genius instance
vmc_genius = VMC_genius(
    vmcsteps=300,
    num_walkers=40,
)

# generate the input file and run
vmc_genius.generate_input(input_name="datasvmc.input")
vmc_genius.run(input_name="datasvmc.input", output_name="out_vmc")

# reblock the MCMC samples
vmc_genius.compute_energy_and_forces(bin_block=10, warmupblocks=5)

# VMC energy
energy, error = vmc_genius.energy, vmc_genius.energy_error
print(f"VMC-JDFT energy = {energy:.5f} +- {error:3f} Ha")

\end{lstlisting}
\end{widetext}
%%%%%%%%%%%%%%%%%%%%%%%%%%%%%%%%%%%%

%%%%%%%%%%%%%%%%%%%%%%%%%%%%%%%%%%%%%%%%%%%%%%%%%%%%
% pyturbo 
%%%%%%%%%%%%%%%%%%%%%%%%%%%%%%%%%%%%%%%%%%%%%%%%%%%%
\section{\pyturbo: Program overview}
\label{sec:pyturbo-overview}
\pyturbo\ is a sub-package of the \tbg\ package, which contains fundamental but flexible functionalities. The reason for implementing several classes in an independent sub-package is that there is a trade-off between flexibility and complexity (i.e., the availability of more sophisticated functionality). Indeed, we suppose advanced users will develop and exploit more elaborated procedures than the ones we expect at present. \tbg\ in its present status may not be flexible enough to support all of them. Therefore, we leave \pyturbo\ as an independent sub-package and keep its implementation as simple as possible, so that advanced users can be provided with the fundamental building blocks to fully control \tvb\ jobs in a Python environment. 
%In fact, the development \aiidaturbo\ package~{\cite{aiidaturborvb}} exploits both \tbg\ and \pyturbo\ functionalities to manage complex workflows. 
The major classes of \pyturbo\ are shown in shown in Table.~{\ref{tab:class-pyturbo}}. {\program{fort.10}} is the most fundamental file containing all the WF information except for that of pseudo potentials. The information of pseudo potential is stored in a separate file, {\program{pseudo.dat}}. {\program{IO\_fort10}} and {\program{Pseudopotentials}} are classes for manipulating a WF file and PP files, respectively. \pyturbo\ contains several classes inheriting the abstract class {\program{fortranIO}}. They are essentially Fortran90 wrappers, i.e, in a one-to-one correspondence between a \pyturbo\ class and a \tvb\ Fortran binary (e.g., {\program{Makefort10}} class in \pyturbo\ corresponds to {\program{makefort10.x}} in \tvb). The corresponding \tvb\ modules are listed in Table~{\ref{tab:modules-turborvb}}.
%
%%%%%%%%%%%%%%%%%%%%%%%%%%%%%%%%%%%%%%%%%%%%%%%%%%%%
% Table 2 Major classes of the pyturbo package
%%%%%%%%%%%%%%%%%%%%%%%%%%%%%%%%%%%%%%%%%%%%%%%%%%%%
\begin{center}
\begin{table*}[htbp]
\caption{\label{tab:class-pyturbo} Major classes of the \pyturbo\ package. Corresponding modules in \tvb\ are listed in Table~{\ref{tab:modules-turborvb}}.}
\vspace{2mm}
\begin{tabular}{c|c|p{12cm}}
\Hline 
Parent Class & Class & Description \\ 
\Hline
\multirow{2}{*}{-} 
& IO\_fort10 & Represents many-body wavefunctions (i.e., \program{fort.10}). Allows wavefunction manipulations. \\ 
& Basis\_sets & Represents basis sets. \\ 
& Pseudo\_potentials & Represents pseudo potentials. \\ 
\Hline
\multirow{10}{*}{Fortran\_IO} 
& Makefort10 & Python wrapper for a Fortran binary \program{makefort10.x} \\
& Convertfort10 & Python wrapper for a Fortran binary \program{convertfort10.x} \\
& Convertfort10mol & Python wrapper for a Fortran binary \program{convertfort10mol.x} \\
& Convertpfaff & Python wrapper for a Fortran binary \program{convertpfaff.x} \\
& Prep & Python wrapper for a Fortran binary \program{prep.x} \\
& VMCopt & Python wrapper for a Fortran binary \program{turborvb.x}, tuned for VMC optimizations. \\
& VMC & Python wrapper for a Fortran binary \program{turborvb.x}, tuned for single-shot VMC calculations. \\
& LRDMCopt & Python wrapper for a Fortran binary \program{turborvb.x}, tuned for LRDMC optimizations.\\
& LRDMC & Python wrapper for a Fortran binary \program{turborvb.x}, tuned for single-shot LRDMC calculations.\\
& Readforward & Python wrapper for a Fortran binary \program{readforward.x} \\
\Hline
\end{tabular}
\end{table*}
\end{center}
%%%%%%%%%%%%%%%%%%%%%%%%%%%%%%%%%%%%%%%%%%%%%%%%%%%%
%
%%%%%%%%%%%%%%%%%%%%%%%%%%%%%%%%%%%%%%%%%%%%%%%%%%%%
% Table 3 Main modules in TurboRVB
%%%%%%%%%%%%%%%%%%%%%%%%%%%%%%%%%%%%%%%%%%%%%%%%%%%%
\begin{center}
\begin{table*}[htbp]
\caption{\label{tab:modules-turborvb} Main modules in \tvb~{\cite{2020NAK2}}.} 
\vspace{2mm}
\begin{tabular}{c|p{15cm}}
\Hline 
 Module & Description \\ 
\Hline 
 makefort10.x             &   Generates a JSD/JAGP WF using a given basis set and structure. \\
 convertfort10mol.x       &   Adds molecular orbitals. If the number of molecular orbitals is equal to (larger than) the half of the number of electrons in a system, the resultant WF becomes JSD (JAGPn). \\
 convertfort10.x          &   Converts a JSD/JAGP/JAGPn WF to a JAGP one. It also converts an uncontracted atomic basis set to a hybrid (contracted) one using the geminal embedding scheme. \\
 convertfortpfaff.x       &   Converts a JAGP WF to JPf one. \\
 prep.x                   &   Performs a DFT calculation.                                        \\
 turborvb.x               &   Performs VMC optimization, VMC evaluation, LRDMC, structural optimization and molecular dynamics.  \\
 readforward.x            &   Performs correlated samplings, and calculates various physical properties.                               \\
\Hline 
\Hline
\end{tabular}
\end{table*}
\end{center}
%%%%%%%%%%%%%%%%%%%%%%%%%%%%%%%%%%%%%%%%%%%%%%%%%%%%
%

\subsection{IO\_fort10}
\label{subsec:pyturbo-fort10}
{\program{IO\_fort10}} is a class to manipulate the \tvb\ WF file {\program{fort.10}} and to extract particular information (e.g., the basis set for the determinant and the Jastrow parts). {\program{IO\_fort10}} internally uses the {\program{Basis\_sets}} class that can store basis-set information for both the determinant and the Jastrow parts. Other information, such as lattice vectors, atomic positions, MO coefficients, AGP and Jastrow matrix elements, are stored in the corresponding attributes implemented in {\program{IO\_fort10}}. The Atomic Simulation Environment (ASE)~{\cite{2017ASK}} package is used to read/write molecule and crystal structures for supporting various file formats.

\subsection{Basis\_sets}
The {\program{Basis\_sets}} class can store basis-set information for both the determinant and the Jastrow parts. \tbg\ supports the Gaussian-type localized atomic orbitals (GTOs):
\begin{equation}
\psi _{l, \pm m ,I}^{{\text{Gaussian}}}\left( {{\mathbf{r}};\zeta } \right) = {\left| {{\mathbf{r}} - {{\mathbf{R}}_I}} \right|^l}{\operatorname{e} ^{ - \zeta {{\left| {{\mathbf{r}} - {{\mathbf{R}}_I}} \right|}^2}}} \cdot \Re[  (-i)^{1\pm 1 \over 2} {Y_{l,m,I}}\left( {\theta ,\varphi } \right)],
\label{gaussian}
\end{equation}
where the real and the imaginary part ($m>0$)  of the spherical harmonic function $ {Y_{l,m,I}}\left( {\theta ,\varphi } \right)$ centered at ${{{\mathbf{R}}_I}}$ are taken and rewritten in Cartesian coordinates in order to work with real defined and easy to compute orbitals, $l$ is the corresponding angular momentum and $m\ge 0$ is its projection number along the $z-$quantization axis. 
For the compatibility with \trexio, the variables in the classes are defined in the exactly same way as in \trexio. One can refer to the \trexio\ documentation for the details~{\cite{trexio}}. The class also implements several parsers to write/read specific formats, such as GAMESS~{\cite{1993SCH}} and NWCHEM~{\cite{2010VAL}}.

\subsection{Pseudopotentials}
\label{subsec:pyturbo-pp}
{\program{Pseudopotentials}} class can store pseudopotential information. \pyturbo\ supports only the standard semi-local form
\begin{equation}
\hat V_{{\text{pp}}}^I\left( {{{\mathbf{r}}_i}} \right) = V_{{\text{loc}}}^I\left( {{r_{i,I}}} \right) + \sum\limits_{l = 0}^{{l_{\max }}} V_l^I\left( r_{i,I} \right) \sum\limits_{m =  - l}^l {\left| {{Y_{l,m}}} \right\rangle \left\langle {{Y_{l,m}}} \right|} 
\end{equation}
where $r_{i,I} = |{{\mathbf{r}}_i} - {{\mathbf{R}}_I}|$ is the distance between the $i$-th electron and the $I$-th ions, $l_{{\text max}}$ is the maximum angular momentum of the ion $I$, and $\sum\limits_{l = 0}^{{l_{\max }}} {\sum\limits_{m =  - l}^l {\left| {{Y_{l,m}}} \right\rangle \left\langle {{Y_{l,m}}} \right|} }$ is a projection operator on the spherical harmonics centered at the ion $I$. 
As it is now becoming a  common  practice not only in QMC, 
both the local  $V_{{\text{loc}}}^I\left( {{r_{i,I}}} \right)$ and   the non-local $V_l^I\left( r_{i,I} \right)$ functions, are expanded over a simple Gaussian basis parametrized by coefficients ({\it e.g.}, effective charge ${Z_{{\text{eff}}}}$ and other simple constants), multiplying simple powers of $r$, and a corresponding Gaussian term:
\begin{equation}
{r^2}{V_l}\left( r \right) = \sum\limits_k {{\alpha _{k,l}}{r^{\beta _{k,l}-2}}\exp \left(- {{\gamma _{k,l}}{r^2}} \right)},
\end{equation}
where ${{\alpha _{k,l}}}$, ${{\beta _{k,l}}}$ (usually small positive integers), and ${{\gamma _{k,l}}}$ are the parameters obtained by appropriate fitting. ${{\alpha _{k,l}}}$, ${{\beta _{k,l}}}$ and ${{\gamma _{k,l}}}$ are the parameters stored in the {\program{Pseudopotentials}} class.
For the compatibility with \trexio, the variables in the classes are defined in the exactly same way as in \trexio. We refer to the documentation of \trexio\ for the details~{\cite{trexio}}.

\subsection{Fortran90 wrappers}
\label{subsec:pyturbo-fortran90}
\pyturbo\ has several classes inheriting the {\program{fortranIO}} class, which are basically Fortran90 wrappers for the corresponding \tvb\ fortran binaries. The Listing~{\ref{lst:pyturbo-vmc}} shows an example to compute the VMC energy of the Hydrogen dimer (assuming the optimized WF is given by a user) using \pyturbo. The input variable of {\program{VMC}} class is {\program{namelist}} instance that is an encapsulated dictionary with {\program{keys=fortran keyword}} and {\program{value=value}} of each parameter, as shown in Listing~{\ref{lst:pyturbo-vmc}}. Indeed, the {\program{input\_parameters}} used in the Python script contains all parameters given to \tvb; thus, one can fully control \tvb\ jobs via \pyturbo.

%%%%%%%%%%%%%%%%%%%%%%%%%%%%%%%%%%%%
\begin{lstlisting}[style=PythonStyle,linewidth=\columnwidth,caption={VMC class in \pyturbo.},label=lst:pyturbo-vmc]

# pyturbo modules
from turbogenius.pyturbo.namelist import Namelist
from turbogenius.pyturbo.vmc import VMC

# input parameters of TurboRVB
input_parameters = {
    "&simulation": {
        "itestr4": 2,
        "ngen": 5000,
        "iopt": 1,
    },
    "&pseudo": {},
    "&vmc": {},
    "&readio": {},
    "&parameters": {},
}
fortran_namelist = Namelist(namelist=input_parameters)

# Generate vmc instance with the above input parameters
vmc = VMC(namelist=fortran_namelist)

# VMC run
vmc.generate_input(input_name="datasvmc0.input")
vmc.run(input_name="datasvmc0.input", output_name="out_vmc0")

# VMC run (continuation)
vmc.set_parameter("iopt", 0)
vmc.generate_input(input_name="datasvmc1.input")
vmc.run(input_name="datasvmc1.input", output_name="out_vmc1")

# VMC, check if the jobs finished correctly
flags = vmc.check_results(output_names=["out_vmc0", "out_vmc1"])

# Reblock the MCMC samples
energy, error = vmc.get_energy(init=10, bin=10)
print(f"VMC energy = {energy:.4f} +- {error:.4f}")

\end{lstlisting}
%%%%%%%%%%%%%%%%%%%%%%%%%%%%%%%%%%%%

%%%%%%%%%%%%%%%%%%%%%%%%%%%%%%%%%%%%%%%%%%%%%%%%%%%%
% turbo-genius command line
%%%%%%%%%%%%%%%%%%%%%%%%%%%%%%%%%%%%%%%%%%%%%%%%%%%%
\section{Command-line tool and user interface}
\label{sec:tbg-command-line}
\tbg\ provides a useful command-line interface, named {\program{turbogenius\_cli}}. The command-line tool can be run on a terminal by calling {\program{turbogenius}}, which is automatically installed during the setup procedure e.g., by {\program{pip}}. The command-line tool allows one to manipulate input and output files very efficiently and user-friendly. One of the most useful functions of the command-line tool is its helper, as shown in Listings~{\ref{lst:turbogenius_helper}} and \ref{lst:turbogenius_vmc_helper}. One can readily know what options are available and what each option does. This functionality is realized with the {\program{click}} package~{\cite{2022CLI}}.
\begin{lstlisting}[style=ShellStyle,caption={The helper implemented in \tbg.},label=lst:turbogenius_helper]
% turbogenius --help
Usage: turbogenius [OPTIONS] COMMAND [ARGS]...

Options:
--help  Show this message and exit.

Commands:
convertfort10        convertfort10_genius
convertfort10mol     convertfort10mol_genius
convertpfaff         readforward_genius
convertwf            convert wavefunction
correlated-sampling  correlated_sampling_genius
lrdmc                lrdmc_genius
lrdmcopt             lrdmcopt_genius
makefort10           makefort10_genius
prep                 prep_genius
vmc                  vmc_genius
vmcopt               vmcopt_genius
\end{lstlisting}
\begin{lstlisting}[style=ShellStyle,caption={The VMCopt helper implemented in \tbg.},label=lst:turbogenius_vmc_helper]
% turbogenius vmcopt --help
Usage: turbogenius vmcopt [OPTIONS]

Options: (-g, -r, and/or -post is mandatory.)
-post                 Postprocess
-r                    Run a program
-g                    Generate an input file
-vmcoptsteps INTEGER  Specify vmcoptsteps
-optwarmup INTEGER    Specify optwarmupsteps
-steps INTEGER        Specify steps per one iteration
-bin INTEGER          Specify bin_block
-warmup INTEGER       Specify warmupblocks
-nw INTEGER           Specify num_walkers
-maxtime INTEGER      Specify maxtime
-optimizer TEXT       Specify optimizer, sr or lr
-learn FLOAT          Specify learning_rate
-reg FLOAT            Specify regularization
-opt_onebody          flag for opt_onebody
-opt_twobody          flag for opt_twobody
-opt_det_mat          flag for opt_det_mat
-opt_jas_mat          flag for opt_jas_mat
-opt_det_basis_exp    flag for opt_det_basis_exp
-opt_jas_basis_exp    flag for opt_jas_basis_exp
-opt_det_basis_coeff  flag for opt_det_basis_coeff
-opt_jas_basis_coeff  flag for opt_jas_basis_coeff
-twist                flag for twisted average
-kpts INTEGER...      kpts, Specify Monkhorst-Pack grids and shifts,
                    [nkx,nky,nkz,kx,ky,kz]
-plot                 flag for plotting graph
-log TEXT             logger level, DEBUG, INFO, ERROR
--help                Show this message and exit.
%
\end{lstlisting}
%

%%%%%%%%%%%%%%%%%%%%%%%%%%%%%%%%%%%%%%%%%%%%%%%%%%%%
%  TurboWorkflow
%%%%%%%%%%%%%%%%%%%%%%%%%%%%%%%%%%%%%%%%%%%%%%%%%%%%
\section{\tbw: A python package for realizing high-throughput calculations using \tbg}
\label{sec:tbw-overview}
Since \tbg\ is able to fully control \tvb\ jobs, one can implement workflows by combining it with a file/job managing package. To demonstrate it, as a proof of concept, we developed an open-source python package, \tbw, that enables one to compose simple workflows by combining \tbg\ with a job managing python package, \tbf. We notice that one can exploit other established workflow packages such as AiiDA~{\cite{2020HUB}} and Fireworks~{\cite{2015ANU}} as job-managing systems. Combining established workflow managers with \tbg\ is an intriguing future work. The main classes of \tbw\ are summarized in Table~{\ref{tab:class-turboworkflows}}. Each workflow class inherits the parent {\program{Workflow}} class with options useful for a QMC calculation. For instance, in the {\program{VMC\_workflow}, one can specify a target accuracy (i.e., statistical error) of the VMC calculation. The {\program{VMC\_workflow} first submits an initial VMC run to a machine with the specified MPI and OpenMP processes to get a stochastic error bar per Monte Carlo step. Since the error bar is inversely proportional to the square root of the number of Monte Carlo samplings, the necessary steps to achieve the target accuracy is readily estimated by the initial run. The {\program{VMC\_workflow} then submits subsequent production VMC runs with the estimated necessary number of steps. Similar functionalities are also implemented in other workflow scripts such as {\program{VMCopt\_workflow}, {\program{LRDMC\_workflow}, and {\program{LRDMCopt\_workflow}. Figure~{\ref{SI-fig:uml-vmc-workflow}} shows a Unified Modeling Language (UML) diagram of the {\program{VMC\_workflow}}. The users can also define their own workflows inheriting the {\program{Workflow}} class.

\vspace{1mm}
\tbw\ can solve the dependencies of a given set of workflows and manage sequential jobs. The {\program{Launcher}} allows one to pass values and files obtained by a workflow to another workflow by using the {\program{Variable}} class, as shown in the listing~{\ref{lst:turboworkflows-water-workflow}}. As shown in the listing~{\ref{lst:turboworkflows-water-workflow}}, the {\program{Launcher}} class accepts workflows as a list, solves the dependencies of the workflows, and submits independent sequential jobs simultaneously. {\program{Launcher}} realizes this feature by the so-called topological ordering of a Directed Acyclic Graph (DAG) and the built-in python module, {\program{asyncio}}.
The listing~{\ref{lst:turboworkflows-water-workflow}} shows a \tbw\ workflow script to perform a sequential job, \pyscf\ $\rightarrow$ \trexio\ $\rightarrow$ \tvb\ WF (JSD ansatz) $\rightarrow$ VMC optimization (Jastrow factor optimization) $\rightarrow$ VMC $\rightarrow$ LRDMC (lattice space $\rightarrow$ 0). Finally, we get the extrapolated LRDMC energy of the water dimer.
Figure~{\ref{SI-fig:uml-launcher}} shows the UML diagram corresponding to the script shown in the listing~{\ref{lst:turboworkflows-water-workflow}}. 
If one wants to manipulate files or values in a more complex way, one should define a new workflow class inheriting the {\program{Workflow}} class and pass it into the {\program{Encapsulated\_Workflow}} class.

%%%%%%%%%%%%%%%%%%%%%%%%%%%%%%%%%%%%%%%%%%%%%%%%%%%%
% Major classes of the turboworkflows package
%%%%%%%%%%%%%%%%%%%%%%%%%%%%%%%%%%%%%%%%%%%%%%%%%%%%
\begin{center}
\begin{table*}[htbp]
\caption{\label{tab:class-turboworkflows} Major classes of the \tbw\ package.}
\vspace{2mm}
\begin{tabular}{c|c|p{12cm}}
\Hline 
Parent Class & Class & Description \\ 
\Hline
\multirow{3}{*}{-} 
& Workflow & Abstract class for implementing workflows \\ 
& Encapsulated\_Workflow & Encapsulated workflow class including input and output file handlings \\ 
& Launcher & A class for managing Encapsulated\_Workflow instances, e.g., solving dependency. \\ 
\Hline
\multirow{1}{*}{Workflow} 
%& Makefort10\_workflow & A workflow implementation based on \program{Makefort10\_genius}  \\
%& Convertfort10mol\_workflow & A workflow implementation based on \program{Convertfort10mol\_genius}  \\
%& Convertfort10\_workflow & A workflow implementation based on \program{Convertfort10\_genius}  \\
& DFT\_workflow & A workflow implementation based on \program{DFT\_genius}  \\
& VMC\_workflow & A workflow implementation based on \program{VMC\_genius}  \\
& VMCopt\_workflow & A workflow implementation based on \program{VMCopt\_genius}  \\
& LRDMC\_workflow & A workflow implementation based on \program{LRDMC\_genius}  \\
& LRDMCopt\_workflow & A workflow implementation based on \program{LRDMCopt\_genius}  \\
\Hline
\multirow{1}{*}{Workflow} 
& PySCF\_workflow & A workflow implementation based on \pyscf  \\
& TREXIO\_workflow & A workflow implementation based on \trexio  \\
\Hline
\end{tabular}
\end{table*}
\end{center}
%%%%%%%%%%%%%%%%%%%%%%%%%%%%%%%%%%%%%%%%%%%%%%%%%%%%
%
%%%%%%%%%%%%%%%%%%%%%%%%%%%%%%%%%%%%%%%%%%%%%%%%%%%%%%%%%%%%%%%%%%%%%%%%%
% python workflow example and result.
%%%%%%%%%%%%%%%%%%%%%%%%%%%%%%%%%%%%%%%%%%%%%%%%%%%%%%%%%%%%%%%%%%%%%%%%%
\begin{widetext}
\begin{lstlisting}[style=PythonStyle,linewidth=\columnwidth,caption={A python workflow to obtain the extrapolated LRDMC energy ($a \to 0$) of the water molecule.},label=lst:turboworkflows-water-workflow]

#!/usr/bin/env python

# python packages
import os
import shutil

# turboworkflows packages
from turboworkflows.workflow_trexio import TREXIO_convert_to_turboWF
from turboworkflows.workflow_vmc import VMC_workflow
from turboworkflows.workflow_vmcopt import VMCopt_workflow
from turboworkflows.workflow_lrdmc_ext import LRDMC_ext_workflow
from turboworkflows.workflow_encapsulated import Encapsulated_Workflow
from turboworkflows.workflow_lanchers import Launcher, Variable

# dictionary of Jastrow basis (GAMESS format)
jastrow_basis_dict = {
    "H": """
        S  1
        1       1.873529  1.00000000
        S  1
        1       0.802465  1.00000000
        S  1
        1       0.147217  1.00000000
        """,
    "O": """
        S  1
        1       1.686633  1.00000000
        S  1
        1       0.237997  1.00000000
        S  1
        1       0.125346  1.00000000
        P  1
        1       1.331816  1.00000000
        """,
}

# Convert from a TREXIO file to a WF with TurboRVB format
trexio_workflow = Encapsulated_Workflow(
    label="trexio-workflow",
    dirname="trexio-workflow",
    input_files=["water.hdf5"],
    workflow=TREXIO_convert_to_turboWF(
        trexio_filename="water.hdf5",
        jastrow_basis_dict=jastrow_basis_dict,
    ),
)

# VMC optimization of Jastrow factor
# One-, two-, and three-body Jastrow are optimized
vmcopt_workflow = Encapsulated_Workflow(
    label="vmcopt-workflow",
    dirname="vmcopt-workflow",
    input_files=[
        Variable(label="trexio-workflow", vtype="file", name="fort.10"),
        Variable(label="trexio-workflow", vtype="file", name="pseudo.dat"),
    ],
    workflow=VMCopt_workflow(
        # cluster information
        cores=1,
        openmp=1,
        # vmc optimization, parameters
        vmcopt_max_continuation=2,
        vmcopt_num_walkers=40,
        vmcopt_target_error_bar=7.5e-3,
        vmcopt_trial_optsteps=10,
        vmcopt_trial_steps=50,
        vmcopt_production_optsteps=40,
        vmcopt_optwarmupsteps_ratio=0.8,
        vmcopt_bin_block=1,
        vmcopt_warmupblocks=0,
        vmcopt_optimizer="lr",
        vmcopt_learning_rate=0.35,
        vmcopt_regularization=0.001,
        vmcopt_onebody=True,
        vmcopt_twobody=True,
        vmcopt_det_mat=False,
        vmcopt_jas_mat=True,
        vmcopt_det_basis_exp=False,
        vmcopt_jas_basis_exp=False,
        vmcopt_det_basis_coeff=False,
        vmcopt_jas_basis_coeff=False,
        vmcopt_maxtime=172000,
    ),
)

# VMC calculation with the optimized WF.
vmc_workflow = Encapsulated_Workflow(
    label="vmc-workflow",
    dirname="vmc-workflow",
    input_files=[
        Variable(label="vmcopt-workflow", vtype="file", name="fort.10"),
        Variable(label="vmcopt-workflow", vtype="file", name="pseudo.dat"),
    ],
    workflow=VMC_workflow(
        # cluster information
        cores=1,
        openmp=1,
        # vmc parameters
        vmc_max_continuation=2,
        vmc_num_walkers=40,
        vmc_target_error_bar=5.0e-3,
        vmc_trial_steps=150,
        vmc_bin_block=10,
        vmc_warmupblocks=5,
        vmc_maxtime=172000,
    ),
)
# LRDMC calculations with the optimized WF
# LRDMC energies are computed with a = 0.20, 0.30, and 0.40, and then, extrapolated to a->0 limit.
lrdmc_ext_workflow = Encapsulated_Workflow(
    label="lrdmc-ext-workflow",
    dirname="lrdmc-ext-workflow",
    input_files=[
        Variable(label="vmc-workflow", vtype="file", name="fort.10"),
        Variable(label="vmc-workflow", vtype="file", name="pseudo.dat"),
    ],
    workflow=LRDMC_ext_workflow(
        # cluster information
        cores=1,
        openmp=1,
        # lrdmc, parameters
        lrdmc_max_continuation=2,
        lrdmc_num_walkers=40,
        lrdmc_target_error_bar=5.0e-3,
        lrdmc_trial_steps=150,
        lrdmc_bin_block=10,
        lrdmc_warmupblocks=5,
        lrdmc_correcting_factor=10,
        lrdmc_trial_etry=Variable(label="vmc-workflow", vtype="value", name="energy"),
        lrdmc_alat_list=[-0.20, -0.30, -0.40, -0.50],
        lrdmc_nonlocalmoves="tmove",
        lrdmc_maxtime=172000,
    ),
)

# add the workflows to the Launcher class
cworkflows_list = [
    trexio_workflow,
    vmcopt_workflow,
    vmc_workflow,
    lrdmc_ext_workflow,
]
launcher = Launcher(cworkflows_list=cworkflows_list)

# Launch the jobs
launcher.launch()


\end{lstlisting}
\end{widetext}
%%%%%%%%%%%%%%%%%%%%%%%%%%%%%%%%%%%%%%%%%%%%%%%%%%%%%%%%%%%%%%%%%%%%%%%%%

\section{Demonstrations of high-throughput calculations}
\label{sec:tbw-demo}
In this section, we will show several results obtained using \tbg\ and \tbw. Earlier versions of \tbg\ was used for computing phonon dispersion calculations~{\cite{2021NAK1}}, equation of state calculations~{\cite{2021NAK1}}, potential energy surfaces of dimers~{\cite{2019NAK, 2022NAK1}}, binding energies of molecules~{\cite{2019NAK, 2022NAK1, 2023RAG}}, hydrogen liquids~{\cite{2022TIR1}}, and hydrogen solids~{\cite{2023LOR}}. However, they were not fully performed using a python script. The current versions of \tbg\ and \tbw\ allow one to fully control \tvb\ calculations using a python script. In this paper, we show two types of demonstrations, validations of the \tvb\ package and benchmarking QMC calculations using \tvb. The summary of the demonstrations are shown in Tables~{\ref{tab:demonstration-validation-list}} and ~{\ref{tab:demonstration-benchmark-list}}. 

%%%%%%%%%%%%%%%%%%%%%%%%%%%%%%%%%%%%%%%%%%%%%%%%%%%%
% Demonstration-validation-list
%%%%%%%%%%%%%%%%%%%%%%%%%%%%%%%%%%%%%%%%%%%%%%%%%%%%
\begin{center}
\begin{table*}[htbp]
\caption{\label{tab:demonstration-validation-list} Summary of the validation tests using \tbg\ and \tbw.}
\vspace{2mm}
\begin{tabular}{c|cc|p{10cm}}
\Hline 
Data set       &  Target~(\tvb) & Ref.~(\pyscf) & Purpose of the validation \\ 
\Hline 
38 molecules   &    LDA-DFT (\program{prep})    &    LDA-DFT     & Consistency check between the DFT module of \pyscf\ and that of \tvb\ for open-boundary condition systems (molecules). The detail is written in Sec.~{\ref{subsubsec:tbw-demo-LDA}}. \\ 
\hline
10 crystals     &    LDA-DFT (\program{prep})    &    LDA-DFT     & Consistency check between the DFT module of \pyscf\ and that of \tvb\ for periodic-boundary condition systems with insulating electronic states (without the smearing technique). $k$ = 4$\times$4$\times$4 was used. The detail is written in Sec.~{\ref{subsubsec:tbw-demo-LDA}}. \\ 
\hline
4 crystals     &    LDA-DFT (\program{prep})    &    LDA-DFT     & Consistency check between the DFT module of \pyscf\ and that of \tvb\ for periodic-boundary condition systems with metallic electronic states (with the smearing technique). $k$ = 4$\times$4$\times$4 was used. The detail is written in Sec.~{\ref{subsubsec:tbw-demo-LDA}}. \\ 
\Hline
100 molecules  &     VMC (\program{turborvb})     &    RHF/ROHF    &  Consistency check between HF calculations done by \pyscf\ and VMC calculations without Jastrow factor done by \tvb\ for open-boundary condition systems (molecules). RHF and ROHF were used for spin-unpolarized and spin-polarized systems, respectively. 
The same consistency check was done between \tvb\ and \qpack. The detail is written in sec.~{\ref{subsubsec:tbw-demo-QMC}}.
\\ 
\hline
49 molecules  &   VMC (\program{turborvb})    &    UHF         &  Consistency check between HF calculations done by \pyscf\ and VMC calculations without Jastrow factor done by \tvb\ for open-boundary condition systems (molecules) with spin-polarized states (UHF was used). The detail is written in Sec.~{\ref{subsubsec:tbw-demo-QMC}}. \\
\Hline
9 crystals &     VMC (\program{turborvb})     &    RHF/ROHF    &  Consistency check between HF calculations done by \pyscf\ and VMC calculations without Jastrow factor done by \tvb\ for periodic-boundary condition systems. RHF and ROHF were used for spin-unpolarized and spin-polarized systems, respectively. $k$ = $\Gamma$, $k$ = (0.25, 0.25, 0.25), and $k$ = 4$\times$4$\times$4 were tested. The detail is written in Sec.~{\ref{subsubsec:tbw-demo-QMC}}.\\ 
\Hline
\end{tabular}
\end{table*}
\end{center}

%%%%%%%%%%%%%%%%%%%%%%%%%%%%%%%%%%%%%%%%%%%%%%%%%%%%
% Demonstration-benchmark-list
%%%%%%%%%%%%%%%%%%%%%%%%%%%%%%%%%%%%%%%%%%%%%%%%%%%%
\begin{center}
\begin{table*}[htbp]
\caption{\label{tab:demonstration-benchmark-list} A summary of the benchmarks using \tbg\ and \tbw.}
\vspace{2mm}
\begin{tabular}{c|p{14cm}}
\Hline
Data set  & Description of the benchmark test \\ 
\Hline
G2-set    & Benchmarking the atomization energies of 55 molecules. They were computed using LRDMC at the $a \rightarrow 0$ limit. The JSD ansatz with the LDA-PZ nodal surface was employed. The references are experimental values. The detail is written in Sec.~{\ref{subsubsec:tbw-demo-benchmark-complex}}. \\ 
\hline
S22-set   & Benchmarking the binding energies of 22 complex systems. They were computed using LRDMC at the $a \rightarrow 0$ limit. The JSD ansatz with the LDA-PZ nodal surface was employed. The references are CCSD(T) values. The detail is written in Sec.~{\ref{subsubsec:tbw-demo-benchmark-complex}}. \\
\hline
A24-set   & Benchmarking the binding energies of 24 complex systems. They were computed using LRDMC at the $a \rightarrow 0$ limit. The JSD ansatz with the LDA-PZ nodal surface was employed. The references are CCSD(T) values. The detail is written in Sec.~{\ref{subsubsec:tbw-demo-benchmark-complex}}. \\
\hline
SCAI-set  & Benchmarking the binding energies of 24 complex systems. They were computed using LRDMC at the $a \rightarrow 0$ limit. The JSD ansatz with the LDA-PZ nodal surface was employed. The references are CCSD(T) values. The detail is written in Sec.~{\ref{subsubsec:tbw-demo-benchmark-complex}}. \\
\Hline
CO dimer  & Benchmarking the equilibrium bond length and harmonic vibrational frequency of the CO dimer. They were estimated from potential energy surfaces with the JSD (LDA-PZ nodal surface) and JAGP ansatz. The references are experimental values. The details is written in Sec.~{\ref{subsec:tbw-demo-benchmark-CO-dimer}}. \\
\Hline
Cubic crystals & Benchmarking the equilibrium lattice parameters 12 cubic crystals. They were estimated by fitting the equation of states obtained by LRDMC. The JSD ansatz with the LDA-PZ nodal surface was employed. The references are experimental values. The detail is written in Sec.~{\ref{subsec:tbw-demo-benchmark-EOS}}. \\
\Hline
\end{tabular}
\end{table*}
\end{center}
%%%%%%%%%%%%%%%%%%%%%%%%%%%%%%%%%%%%%%%%%%%%%%%%%%%%

\subsection{Validations of QMC implementations}
Validation of scientific softwares, such as checking consistency among softwares that implement the same theory as employed in this study (i.e., {\it inter-software} test), is an important step to ensure one's software reliability. The widespread use of validation tests is also important to ensure the trustability of numerical simulations in general. Such validation tests have been performed not-so-widely in the \emph{ab initio} QMC community, since QMC requires complex computational procedures, as mentioned in the introduction. \tbg\ and \tbw\ enable one to do the tests much more easily and efficiently. In this paper, we report the results of {\it inter-software} tests for the \tvb\ package (v1.0.0). We checked {\it inter-package} consistencies between \tvb\ and other established quantum chemistry packages, such as \pyscf~{\cite{2018SUN,2020SUN}} and \qpack~{\cite{2019GAR}}. The details are written in Secs.~{\ref{subsubsec:tbw-demo-LDA}} and ~{\ref{subsubsec:tbw-demo-QMC}}. Notice that the data and Python scripts to reproduce the validation tests are available from our public repositories (see Sec.~{\ref{sec:data}}).

\subsubsection{Validation of DFT module: LDA (\pyscf) v.s. LDA (\tvb-prep)}
\label{subsubsec:tbw-demo-LDA}
Using the implemented workflows, we have checked the consistency of the DFT-LDA calculations among packages. DFT energies should be consistent among packages as far as the same basis sets and ECPs are used even though those DFT codes employ different implementation schemes. For the reference calculations, we used the \pyscf\ package (v2.0.1)~{\cite{2018SUN, 2020SUN}} with the Perdew-Zunger (PZ81) local density approximation (PZ-LDA)~{\cite{1981PER}}. For the test sets, we have chosen (1) 38 molecules with singlet spins, (2) 10 insulating crystals, where both orthorhombic and non-orthorhombic cells are included, and (3) 4 metallic crystals. For the crystals, we employed the $k$ = 4 $\times$ 4 $\times$ 4 (i.e., twisted average) grid such that the reciprocal grid includes both real and complex points. For the real-space grid, which is needed for the numerical integration employed in the built-in DFT module ({\program{prep}}), we employed 0.05 Bohr for the molecules and insulating crystals, while 0.03 Bohr for the metallic crystals. We employed the Fermi-Dirac smearing method with 0.01~Ha for the metals. Figures.~{\ref{SI-fig:sanity-check-molecules-LDA}}, {\ref{SI-fig:sanity-check-crystals-LDA-pbc-kgrid-insulator}}, and {\ref{SI-fig:sanity-check-crystals-LDA-pbc-kgrid-metal}} show the consistencies between \pyscf\ and \tvb-{\program{prep}} LDA calculations for all the above cases. The very slight differences come from the implementations (i.e., \tvb-prep module employs the numerical integration to compute the overlap and Hamiltonian matrix elements). The corresponding numbers are shown in Tables.~{\ref{SI-tab:sanity-check-molecules-LDA}}, {\ref{SI-tab:sanity-check-crystals-LDA-pbc-kgrid-insulator}}, and {\ref{SI-tab:sanity-check-crystals-LDA-pbc-kgrid-metal}}. The consistencies between the packages show that the implementations of DFT calculations in the \tvb\ package is correctly done.

\subsubsection{Validation of \tvb\ QMC module: HF (\pyscf) v.s. VMC (\tvb\ w/o Jastrow)}
\label{subsubsec:tbw-demo-QMC}
One of the most prominent features of \tbg\ is the functionality to convert a \trexio\ file to a \tvb\ WF because it allows one to use any quantum chemistry or DFT packages to generate trial WFs as long as they employ localized basis sets. We have carefully verified the implementation of the converter by confirming the consistency between HF and VMC energies without Jastrow factor, both for molecules (open systems) and crystals (periodic systems). 
More specifically, we computed the HF energies of 100 molecules and 9 crystals (such that both orthorhombic and non-orthorhombic cells are included) by \pyscf\ v2.0.1, where the ccECP~{\cite{2017BEN,2018BEN,2018ANN,2019WAN}} with accompanied basis sets were employed. The obtained \pyscf\ checkpoint files were converted to \tvb\ WFs via \trexio\ files using the converter implemented in \tbg\ package. Then, we computed VMC energies of the \tvb\ WFs {\emph{without}} Jastrow factor. The \pyscf\ to \tvb\ conversion via \trexio\ supports both restricted (i.e., ROHF) and unrestricted (i.e., UHF) open-shell WFs. We tested the former implementation using the 100 molecules, while the latter using the 49 molecules. For the 100 molecules, the same consistency checks were done between HF calculations by \qpack~(v2.1.2) and \tvb. For the crystals, we tested both single-$k$ and multi-$k$ (i.e., twisted average) calculations. For the single-$k$ tests, $k$=(0.00, 0.00, 0.00) and (0.25, 0.25, 0.25) were used. For the twisted average tests, we employed $k$ = 4 $\times$ 4 $\times$ 4 such that the grid includes both real and complex points. Notice that, in the twisted average case, we compared the VMC energies with the averages of the HF energies obtained at each $k$-point \emph{independently}, since the VMC is independently done for each $k$ point (i.e., MCMC is done for each $k$.) We did not use metals for the VMC validation tests. This is because, for metals, the HF energy obtained with the smearing technique is not consistent with the VMC ones  due to the fact that the orbital contributions above the Fermi energy are truncated when converting the DFT orbitals to the single Slater-Determinant ansatz. The HF and VMC energies should be consistent within the statistical errors (i.e., within $3 \sigma$) as long as the conversions are done correctly. 
The results of the validation tests are shown in Figs.{\ref{SI-fig:sanity-check-molecules-RHF}}-{\ref{SI-fig:sanity-check-crystals-k-twist}}. The corresponding values are shown in Tables~{\ref{SI-tab:sanity-check-molecules-RHF}}-{\ref{SI-tab:sanity-check-crystals-k-twist}}. The results show that the HF and VMC energies are consistent within the statistical errors (i.e., within $3 \sigma$), implying that the implementations of the WF converter and VMC calculations in the \tvb\ package are correctly done.

\subsection{Benchmarking of QMC calculations}
Benchmarking a theory with several typical systems is a task as important as the validation test, with the aim at examining the accuracy of the theory. Such benchmark calculations can also be performed efficiently using \tbg\ and \tbw. In this work, we benchmarked atomization energies of the Gaussian-2 set~{\cite{1991CUR}}, binding energies of the S22~{\cite{2006JUR}}, A24~{\cite{2013REZ}}, and SCAI~{\cite{2009BER}} sets, and equilibrium lattice parameters of 12 crystals via equation of states calculations. We found that, for all the compounds, the diffusion Monte Carlo calculations with the PZ-LDA nodal surface give satisfactory results, i.e., consistent either with CCSD(T) or experimental values. The details of the results are reported in the following parts.

\subsubsection{G2, S22, A24, and SCAI benchmark sets: atomization energy and binding energy calculations}
\label{subsubsec:tbw-demo-benchmark-complex}
We report the result of the benchmark tests using the G2~{\cite{1991CUR}}, S22~{\cite{2006JUR}}, A24~{\cite{2013REZ}}, and SCAI~{\cite{2009BER}} sets. The G2-set targets atomization energies of 55 molecules, while the other sets benchmark binding energies of complex systems. The benchmark calculations were performed by \tbw. Our python workflow launched a sequential job for each atom and molecule, \pyscf\ $\rightarrow$ \trexio\ $\rightarrow$ \tvb\ WF (Jastrow Slater determinant ansatz) $\rightarrow$ VMC optimization (Jastrow factor) $\rightarrow$ VMC $\rightarrow$ LRDMC (lattice space $\rightarrow$ 0). Finally, we got extrapolated LRDMC energies of atoms and molecules of the benchmark sets. The quality of the Jastrow optimization did not affect the final extrapolated LRDMC energy because the determinant localization approximation (DLA)~{\cite{2019ZEN}} was employed. Indeed, the workflow was fully automatic and fully reproducible since the determinant part which determines the nodal surface was obtained deterministically, and the Jastrow factor, which was obtained by stochastic optimization, did not affect the extrapolated FN energy. The geometries of the G2 set were taken from previous benchmark studies~{\cite{1997CUR, 2008FEL, 2005OEI}}, while those of the other dataset were taken from Benchmark Energy and Geometry DataBase (BEGDB)~{\cite{2008REZ}}

%%%%%%%%%%%%%%%%%%%%%%%%%%%%%%%%%%%%%%%%%%%%%%%%%%%%
% G2-set
%%%%%%%%%%%%%%%%%%%%%%%%%%%%%%%%%%%%%%%%%%%%%%%%%%%%
\begin{figure*}[htbp]
  \centering
  \includegraphics[width=\textwidth]{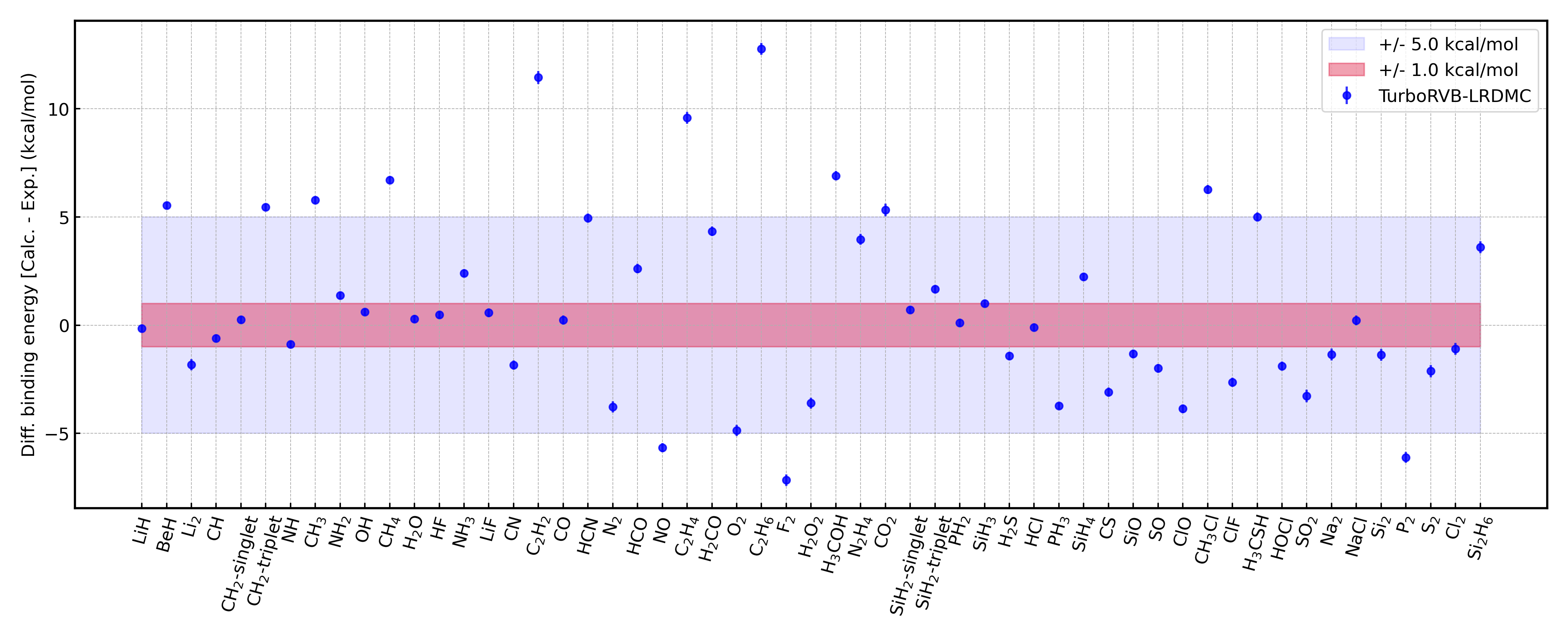}
  \caption{Deviation of the LRDMC atomization energies obtained in this study from the experimentally obtained values. Corrections for zero-point energies and relativistic effects have been included before computing the differences between the LRDMC and experimental values~{\cite{2010NEM}}. 
  The blue and red bounds represent discrepancies $\pm$~5 kcal/mol and $\pm$~1 kcal/mol, respectively. DFT calculations with PZ-LDA~{\cite{1981PER}} exchange-correlation functional were used to generate the trial WFs. The cc-pVQZ basis set with the accompanied ccECP pseudo potentials~{\cite{2017BEN,2018BEN,2018ANN,2019WAN}} were employed for the DFT calculations.}

  \label{fig:G2-set}
\end{figure*}
%%%%%%%%%%%%%%%%%%%%%%%%%%%%%%%%%%%%%%%%%%%%%%%%%%%%
%
%%%%%%%%%%%%%%%%%%%%%%%%%%%%%%%%%%%%%%%%%%%%%%%%%%%%
% S22-set
%%%%%%%%%%%%%%%%%%%%%%%%%%%%%%%%%%%%%%%%%%%%%%%%%%%%
\begin{figure*}[htbp]
  \centering
  \includegraphics[width=\textwidth]{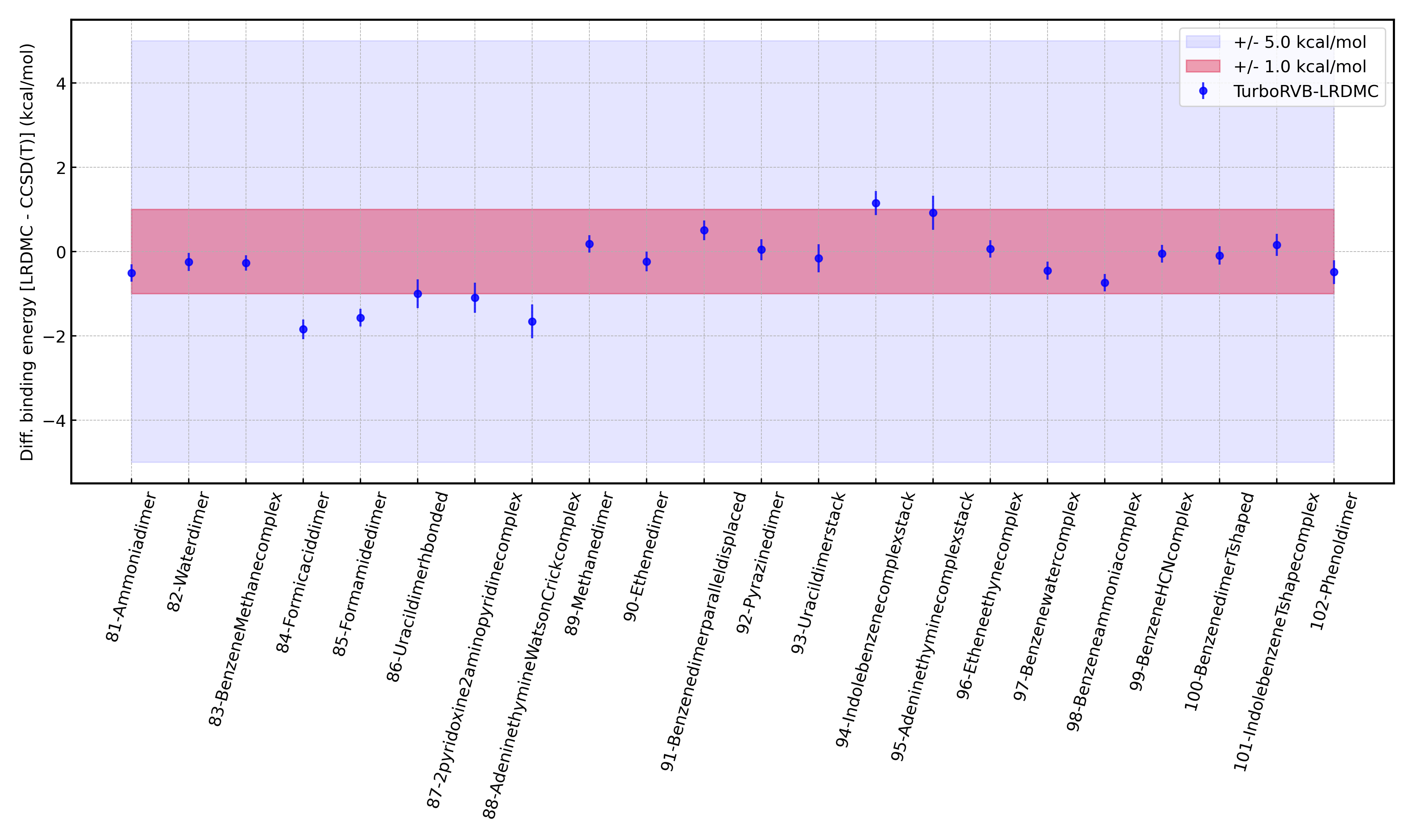}
  \caption{The binding energy comparison of the S22 benchmark set. The differences between the LRDMC values obtained in this study and the reference CCSD(T) values are plotted. The blue and red bounds represent discrepancies $\pm$~5 kcal/mol and $\pm$~1 kcal/mol, respectively. DFT calculations with PZ-LDA~{\cite{1981PER}} exchange-correlation functional were used to generate the trial WFs. The cc-pVQZ basis set with the accompanied ccECP pseudo potentials~{\cite{2017BEN,2018BEN,2018ANN,2019WAN}} were employed for the DFT calculations.}
  \label{fig:S22-set}
\end{figure*}
%%%%%%%%%%%%%%%%%%%%%%%%%%%%%%%%%%%%%%%%%%%%%%%%%%%%
%
%%%%%%%%%%%%%%%%%%%%%%%%%%%%%%%%%%%%%%%%%%%%%%%%%%%%
% A24-set
%%%%%%%%%%%%%%%%%%%%%%%%%%%%%%%%%%%%%%%%%%%%%%%%%%%%
\begin{figure*}[htbp]
  \centering
  \includegraphics[width=\textwidth]{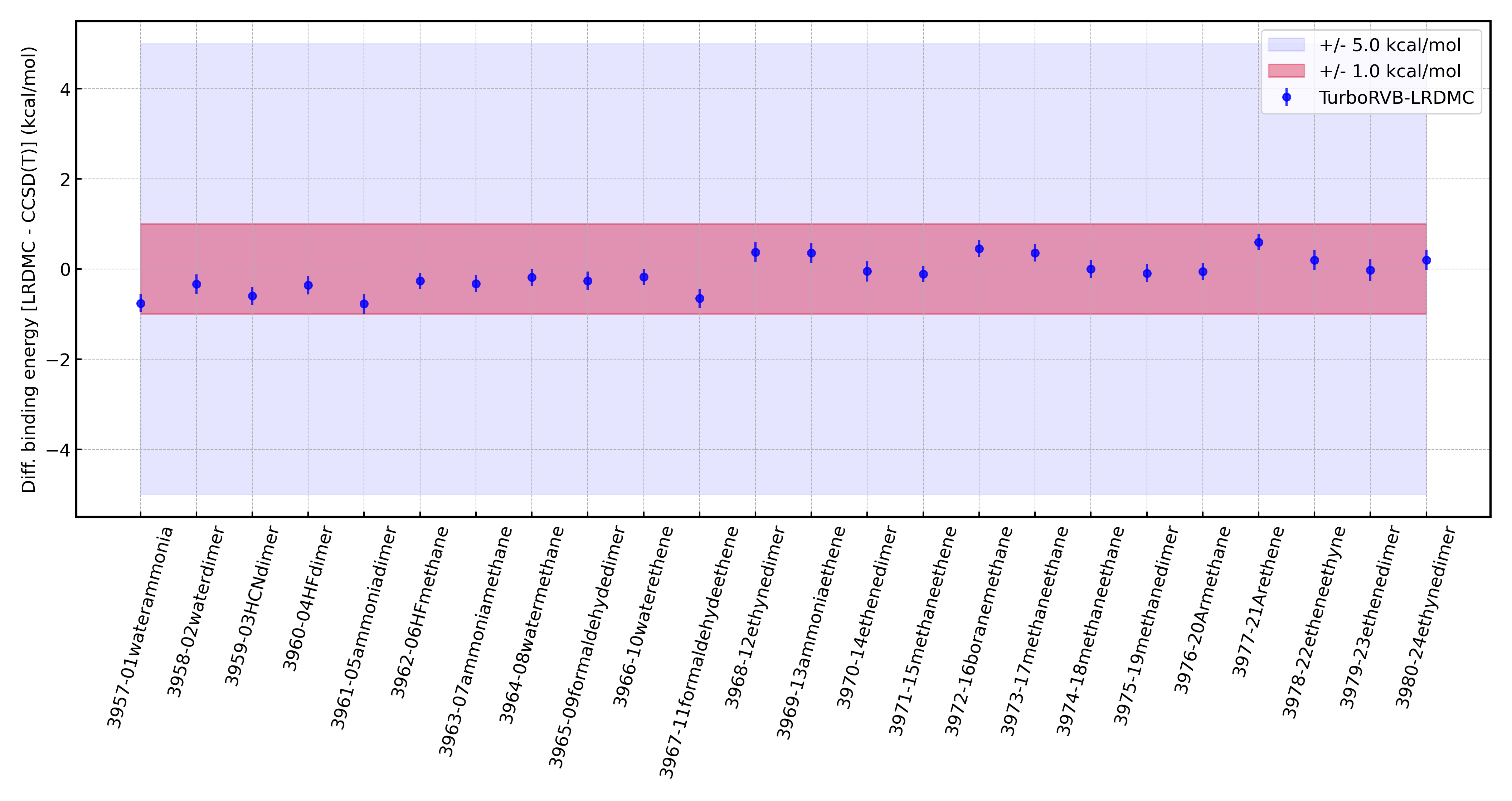}
  \caption{The binding energy comparison of the A24 benchmark set. The differences between the LRDMC values obtained in this study and the reference CCSD(T) values are plotted. The blue and red bounds represent discrepancies $\pm$~5 kcal/mol and $\pm$~1 kcal/mol, respectively. DFT calculations with PZ-LDA~{\cite{1981PER}} exchange-correlation functional were used to generate the trial WFs. The cc-pVQZ basis set with the accompanied ccECP pseudo potentials~{\cite{2017BEN,2018BEN,2018ANN,2019WAN}} were employed for the DFT calculations.}
  \label{fig:A24-set}
\end{figure*}
%%%%%%%%%%%%%%%%%%%%%%%%%%%%%%%%%%%%%%%%%%%%%%%%%%%%
%
%%%%%%%%%%%%%%%%%%%%%%%%%%%%%%%%%%%%%%%%%%%%%%%%%%%%
% SCAI-set
%%%%%%%%%%%%%%%%%%%%%%%%%%%%%%%%%%%%%%%%%%%%%%%%%%%%
\begin{figure*}[htbp]
  \centering
  \includegraphics[width=\textwidth]{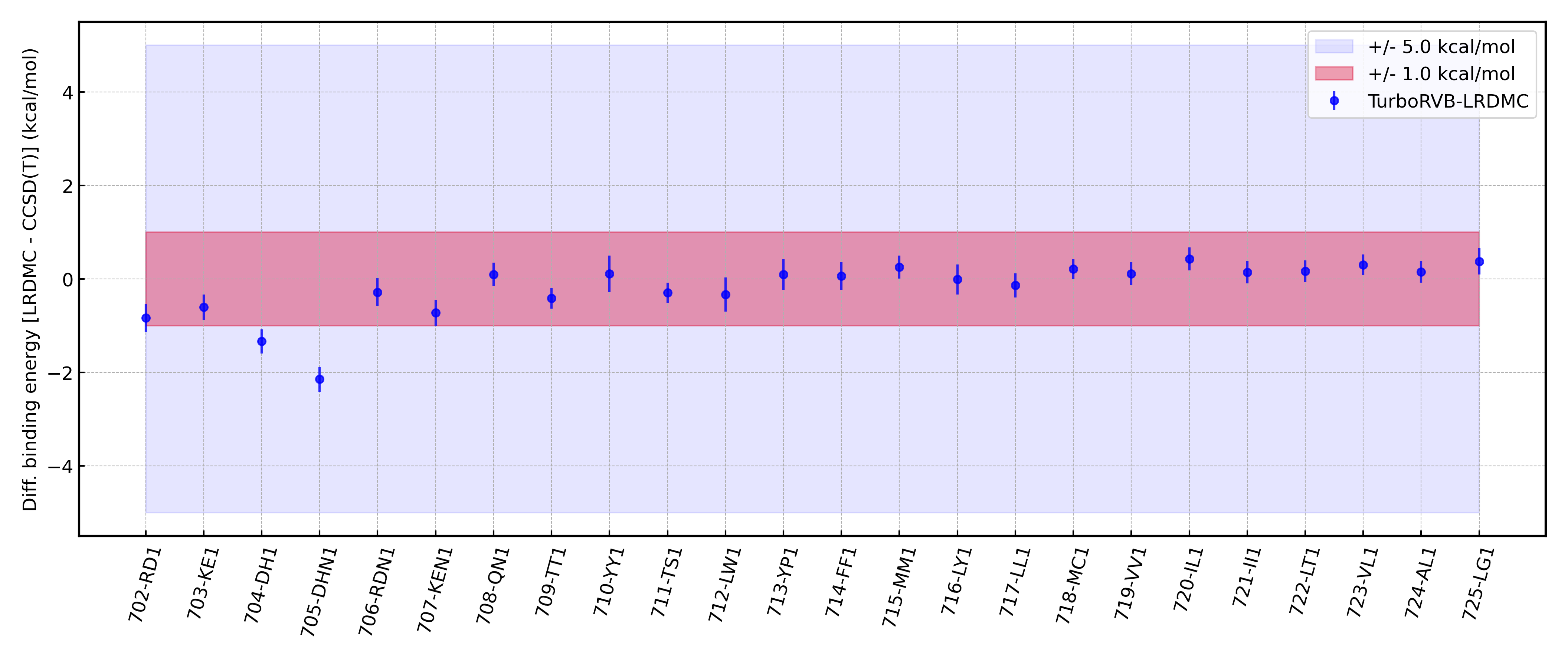}
  \caption{The binding energy comparison of the SCAI benchmark set. The differences between the LRDMC values obtained in this study and the reference CCSD(T) values are plotted. The blue and red bounds represent discrepancies $\pm$~5 kcal/mol and $\pm$~1 kcal/mol, respectively. DFT calculations with PZ-LDA~{\cite{1981PER}} exchange-correlation functional were used to generate the trial WFs. The cc-pVQZ basis set with the accompanied ccECP pseudo potentials~{\cite{2017BEN,2018BEN,2018ANN,2019WAN}} were employed for the DFT calculations.}
  \label{fig:SCAI-set}
\end{figure*}
%%%%%%%%%%%%%%%%%%%%%%%%%%%%%%%%%%%%%%%%%%%%%%%%%%%%
%

{\vspace{2mm}}
% G2-set
Figure.~{\ref{fig:G2-set}} shows the benchmark result for the G2-set~{\cite{1991CUR}}. The corresponding numbers are shown in Table~{\ref{SI-tab:G2-set}}. We computed the atomization energies of the 55 molecules included in the G2-set by pseudo-potential LRDMC calculations with the JDFT ansatz. We employed the cc-pVQZ basis set with the accompanied ccECP~{\cite{2017BEN,2018BEN,2018ANN,2019WAN}} pseudo potentials.
For the DFT calculations, we used the \pyscf\ package (v2.0.1)~{\cite{2018SUN, 2020SUN}} with PZ-LDA~{\cite{1981PER}} exchange-correlation functional.
% JDFT ansatz
The obtained Mean absolute deviation (MAD) with the JDFT ansatz is 3.242 kcal/mol, which is consistent with previous studies. Nemec et al.~{\cite{2010NEM}} used all-electron DMC on Slater determinant (SD) WF to obtain the binding energies of the G2 set to an MAD of 3.2 kcal/mol. Grossman~{\cite{2002GRO} reported a similar accuracy, a MAD of 2.9 kcal/mol, in binding energies by DMC pseudopotential calculations. Abhishek et al.~{\cite{2023RAG} reported a MAD of 3.2 kcal/mol by all-electron LRDMC calculations with the JDFT ansatz. 
% JAGP ansatz
Recently, Abhishek et al.~{\cite{2023RAG} also reported a MAD of 1.6~kcal/mol by all-electron LRDMC calculations with the JAGPs ansatz. \tbg\ and \tbw\ support the same binding energy calculations with the JAGP ansatz also. However, to reproduce their results, large Jastrow basis sets and very careful optimizations are needed; otherwise, optimizations could be stuck at local mimina. The fully automatic optimization of WFs with many variational parameters is not established yet. This should be solved in the near future for realizing robust high-throughput QMC calculations with optimized (beyond-DFT) nodal surfaces.
%Instead, the obtained MAD with the JAGPs ansatz, ~ {\color{red}X.XXX} kcal/mol, is much better than that with the JDFT ansatz, indicating that the JAGPs ansatz considers electron correlation the the JDFT ansatz misses, and it reaches the chemical accuracy. Notice that the nodal surfaces of the JAGPs ansatz were optimized at the VMC level using the linear method. Abhishek et al.~{\cite{2023ABH} also reported a MAD of 1.6~kcal/mol by all-electron LRDMC calculations with the JAGPs ansatz.

% S22-set
The S22 dataset was developed by Hobza et al. for testing interaction energies for small complex systems~{\cite{2006JUR}}. Figure.~{\ref{fig:S22-set}} shows the benchmark result for the S22-set. The corresponding numbers are shown in Tab.~{\ref{SI-tab:S22-set}}. 
We employed the cc-pVQZ basis set with the accompanied ccECP~{\cite{2017BEN,2018BEN,2018ANN,2019WAN}} pseudo potentials. For the DFT calculations, we used the \pyscf\ package (v2.0.1)~{\cite{2018SUN, 2020SUN}} with PZ-LDA~{\cite{1981PER}} exchange-correlation functional.
The obtained MAD is 0.610 kcal/mol. A subset of the S22-benchmark was studied by Dubecky et. al.~{\cite{2013DUB, 2014DUB}}. They extracted a subset of the S22 benchmark test, ammonia dimer, water dimer, methane dimer, ethene dimer, ethene--ethyne, benzene--water, benzene--methane, and benzene dimer (T-shape).  They employed the ECPs with the corresponding basis sets (aug-TVZ) developed by Burkatzki et al.~{\cite{2007BUR}}. They used the B3LYP exchange-correlation functional for generating the trial wave functions. Their obtained binding energies, -3.10(6), -5.15(8), -0.44(5), -1.47(9), -1.56(8), -3.53(13), -1.30(13), and -2.88(16) kcal/mol are very closed to ours, -3.68(21), -5.26(21), -0.35(20), -1.75(23), -1.47(20), -3.73(21), -1.77(18) and -2.83(22) kcal/mol for ammonia dimer, water dimer, methane dimer, ethene dimer, ethene--ethyne, benzene--water, benzene--methane, and benzene dimer (T-shape), respectively. The full set of the S22-benchmark was studied by Korth et. al.~{\cite{2008KOR}}. They used the guidance functions of the Slater-Jastrow type with Hartree-Fock determinants and Schmidt-Moskowitz type correlation functions.~{\cite{1990SCH}}. They used quadruple $\zeta$ valence GTO basis sets fully optimized for the ECPs developed by Ovcharenko et al.~{\cite{2001OVC}}. Their obtained MAD (0.68 kcal/mol) is very close to ours (0.61 kcal/mol).

% A24-set
The A24 dataset is a set of non-covalent systems large enough to include various types of interactions~{\cite{2013REZ}}. The dataset was intended for testing accuracy of computational methods which are used as a benchmark in larger model systems. Figure.~{\ref{fig:A24-set}} shows the benchmark result for the A24-set. The corresponding numbers are shown in Table~{\ref{SI-tab:A24-set}}. We employed the cc-pVQZ basis set with the accompanied ccECP~{\cite{2017BEN,2018BEN,2018ANN,2019WAN}} pseudo potentials. For the DFT calculations, we used the \pyscf\ package (v2.0.1)~{\cite{2018SUN, 2020SUN}} with PZ-LDA~{\cite{1981PER}} exchange-correlation functional. The obtained MAD is 0.315 kcal/mol. The full set of the A24-benchmark was studied by Dubecky et. al.~{\cite{2014DUB}}. They investigated the effects of the basis set, Jastrow factor, and optimization protocols. They finally obtained MAD of 0.15 kcal/mol with the single-determinant trial wave functions of Slater-Jastrow type using B3LYP orbitals and aug-TZV basis sets accompanied with the ECPs developed by Burkatzki et al.~{\cite{2007BUR}}. Their MAD (0.15 kcal/mol) is very closed to the value reported in this study (0.315 kcal/mol).

% SCAI-set
The SCAI dataset is developed to benchmark interactions between amino acid side chains~{\cite{2009BER}}. The dataset contains a representative set of 24 of the 400 (i.e., 20 $\times$ 20) possible interacting side chain pairs. Figure.~{\ref{fig:SCAI-set}} shows the benchmark result for the SCAI-set. The corresponding numbers are shown in Table.~{\ref{SI-tab:SCAI-set}}. We employed the cc-pVQZ basis set with the accompanied ccECP~{\cite{2017BEN,2018BEN,2018ANN,2019WAN}} pseudo potentials. For the DFT calculations, we used the \pyscf\ package (v2.0.1)~{\cite{2018SUN, 2020SUN}} with PZ-LDA~{\cite{1981PER}} exchange-correlation functional. The obtained MAD is ~ 0.402 kcal/mol, which is as small as the S22 and A24 benchmark sets. To the best of our knowledge, no one has benchmarked the SCAI data set using QMC. Among the systems, 704-DH1 and 705-DHN1 show the largest deviations, -1.33(26) kcal/mol and -2.15(26) kcal/mol, respectively.
% http://www.hirokawa-shoten.co.jp/storage/ad0fdf6c5552abadd9dd8700b1e6b01427fd9558f667bdce8ac68ed5508e3707.pdf

%JSCH-set
%tobe?

%L7-set
%tobe?

\subsubsection{Potential Energy Surface (PES) calculations}
\label{subsec:tbw-demo-benchmark-CO-dimer}
Potential Energy Surface (PES) calculations are often computed for dimers in QMC for benchmarking, i.e., for comparing binding energies, equilibrium bond lengths, and harmonic frequencies with experimental values, and for checking if the obtained forces and pressures are biased or not~{\cite{2003ASS, 2014MOR, 2019GEN3, 2021VAN, 2021TII, 2022NAK1, 2022LAR}}. To reduce the computation costs of PES calculations and avoid being trapped at local minima, Jastrow factors are usually optimized at a certain bond length and copied to WFs with other bond lengths which will then be optimized with a better starting point. \tbw\ automatizes these procedure and, for a good point, solves the dependency automatically. An example workflow is shown in Listing~{\ref{SI-lst:turboworkflows-CO-workflow}}, where the initial Jastrow optimization procedure is defined for a bond length (1.10 \AA) and the optimized Jastrow factors are copied to WFs at other bond lengths and optimized again. The point is the {\program{Value}} instance that defines workflows that should be completed before. The workflow for the PES calculation is showin in Fig.~{\ref{fig:workflow-PES-calculation}}. The PESs were computed at both the VMC and LRDMC levels. Two ansatz were employed in this study, JDFT and JAGPs. The JDFT ansatz were optimized according to the procedure described above using the linear method~{\cite{2007UMR}} at the VMC level and then they were converted to the JAGPs ansatz. The JAGPs were further optimized using the linear method at the VMC level. The optimized JDFT and JAGP ansatz were used for the subsequent LRDMC calculations. The T-move approach~{\cite{2006CAS}} with the lattice discretization~=~0.30 Bohr was employed for the LRDMC calculations. The obtained PESs are shown in Fig.~{\ref{fig:PES-CO}}. The left-hand side of Fig.~{\ref{fig:PES-CO}} shows that the PESs obtained with JDFT and JAGPs ansatz at the VMC level, while the righthand of Fig.~{\ref{fig:PES-CO}} shows that the PESs obtained with JDFT and JAGPs ansatz at the LRDMC level.
Table~{\ref{tab:PES-CO}} summarizes the equilibrium bond length and the harmonic frequency obtained from the VMC and LRDMC calculations and those obtained from experiments.
%The equilibrium bond lengths obtained at the VMC level are 1.1150(2) \AA\ and 1.1186(2) \AA\ with JDFT and JAGPs ansatz, respectively. Those obtained at the LRDMC level are 1.1223(2) \AA\ and 1.1240(2) \AA\ with JDFT and JAGPs ansatz, respectively. The harmonic frequencies obtained at the VMC level are  2272(3) cm$^{-1}$and  2233(3) cm$^{-1}$ with JDFT and JAGPs ansatz, respectively. Those obtained at the LRDMC level are 2212(3) cm$^{-1}$ and 2194(3) cm$^{-1}$ with JDFT and JAGPs ansatz, respectively. The experimental equilibrium bond length and harmonic frequency are 1.128323 \AA\ and 2169.81358 cm$^{-1}$, respectively~{\cite{2013HUB}}.Indeed, 
The LRDMC calculation with the JAGPs ansatz gives the closest values to the experiment, as expected. The remaining discrepancies can be solved using a larger determinant and Jastrow basis sets as they both affect the quality of the nodal surfaces. Such a benchmark test for other molecules is an intersting future work. The workflow will be also useful for benchmarking new ansatz and algorithms implemented in \tvb.

%%%%%%%%%%%%%%%%%%%%%%%%%%%%%%%%%%%%%%%%%%%%%%%%%%%%
% PES of CO dimer
%%%%%%%%%%%%%%%%%%%%%%%%%%%%%%%%%%%%%%%%%%%%%%%%%%%%
\begin{figure*}[htbp]
  \centering
  \includegraphics[width=\textwidth]{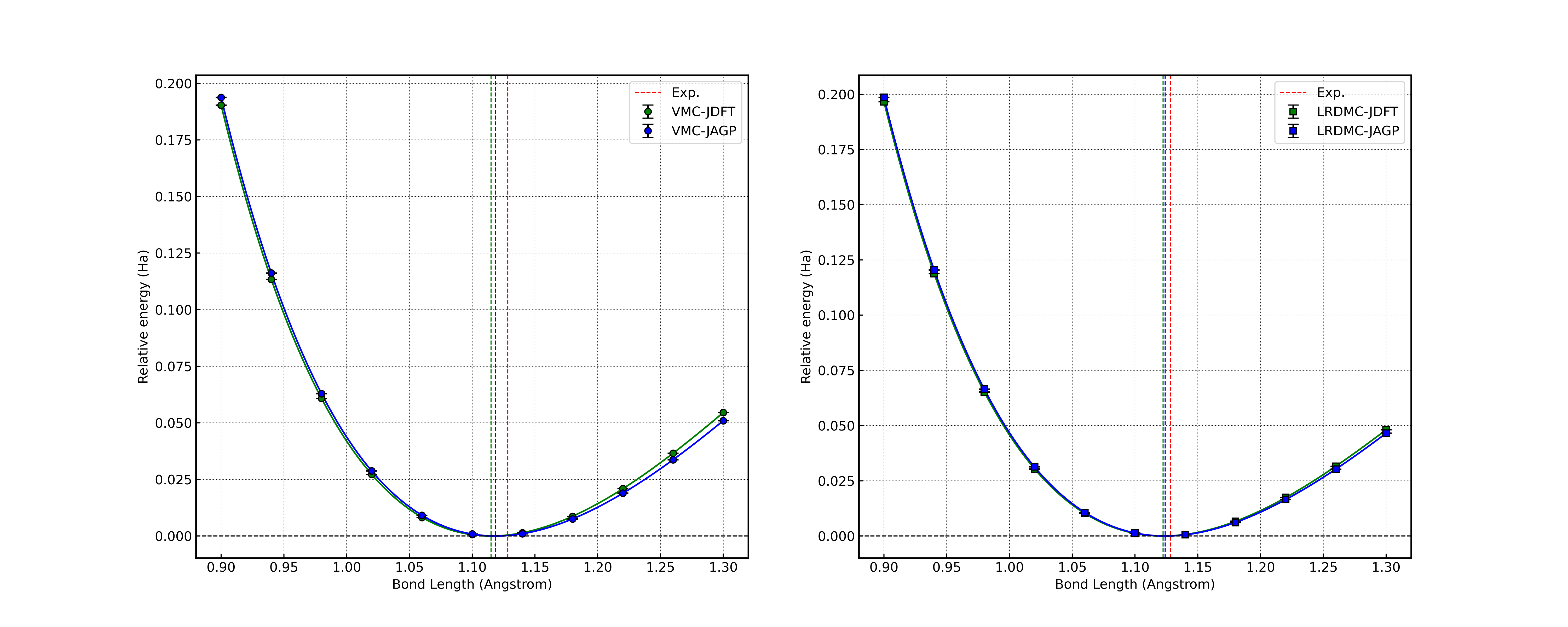}
  \caption{PESs of the CO dimer computed by a python workflow implemented using \tbw. The left and right PESs were computed at the VMC and LRDMC levels, respectively. The green and blue vertical broken lines represent the equilibrium distances obtained from the JSD (with the LDA-PZ nodal surface) and JAGP PESs, respectively. The red vertical broken line shows the experimental equilibrium distance.}
  \label{fig:PES-CO}
\end{figure*}
%%%%%%%%%%%%%%%%%%%%%%%%%%%%%%%%%%%%%%%%%%%%%%%%%%%%

%%%%%%%%%%%%%%%%%%%%%%%%%%%%%%%%%%%%%%%%%%%%%%%%%%%%
% Workflows of the PES calculation
%%%%%%%%%%%%%%%%%%%%%%%%%%%%%%%%%%%%%%%%%%%%%%%%%%%%
\begin{figure*}[htbp]
  \centering
  \includegraphics[width=\textwidth]{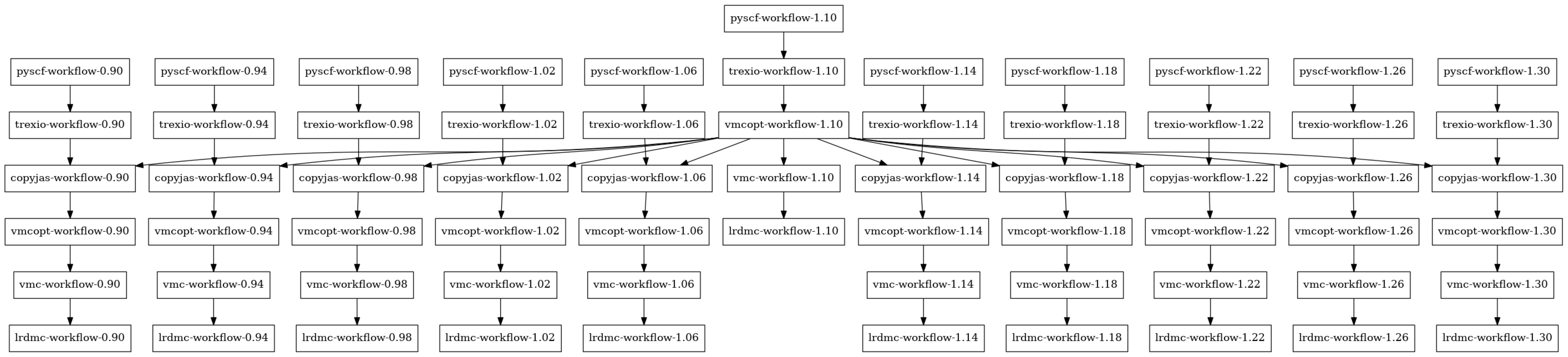}
  \caption{Dependencies of the PES calculation for the CO dimer. The dependencies are automatically solved by the {\program{launcher}} implemented in \tbw.}
  \label{fig:workflow-PES-calculation}
\end{figure*}
%%%%%%%%%%%%%%%%%%%%%%%%%%%%%%%%%%%%%%%%%%%%%%%%%%%%

%%%%%%%%%%%%%%%%%%%%%%%%%%%%%%%%%%%%%%%%%%%%%%%%%%%%
% Table. template 
\begin{center}
\begin{table}[hbtp]
\caption{\label{tab:PES-CO}
Equilibrium bond distances $r_{\rm eq}$ (\AA) and harmonic frequencies $\omega$ (cm$^{-1}$) of the CO dimer obtained with the JDFT and JAGPs ansatz both at the VMC and LRDMC levels. }
\begin{tabular}{c|c|cc}
\Hline
Method & Ansatz & $r_{{\rm eq}}$ (\AA) & $\omega$ (cm$^{-1}$) \\
\Hline
\multirow{2}{*}{VMC} 
  &  JDFT  &  1.1150(2)   &  2272(3)   \\
  &  JAGPs &  1.1186(2)   &  2233(3)   \\  
\Hline
\multirow{2}{*}{LRDMC} 
  &  JDFT   &  1.1223(2)  &  2212(3)   \\
  &  JAGPs  &  1.1240(2)  &  2194(3)   \\
\Hline
\multirow{1}{*}{Exp.} 
  &  -  & 1.128323~{\footnotemark[1]} & 2169.81358~{\footnotemark[1]} \\
\Hline
\end{tabular}
\footnotetext[1]{These values are taken from Ref.~\onlinecite{2013HUB}.}
\end{table}
\end{center}
%%%%%%%%%%%%%%%%%%%%%%%%%%%%%%%%%%%%%%%%%%%%%%%%%%%%

\subsubsection{Benchmarking Equations of State in Solids}
\label{subsec:tbw-demo-benchmark-EOS}
We report the Equation of States (EOSs) of 12 solids computed at the VMC and LRDMC levels by \tbw. Such a benchmark has been done for several crystals to test the accuracy of QMC calculations~{\cite{2013SHU, 2016SAN}}. Table~{\ref{tab:eos-qmc}} shows the Crystallography Open Database(COD)-IDs~{\cite{2009GRA, 2012GRA}} of the 12 crystals computed in this demonstration. The 12 crystals were chosen because Ref.~{\onlinecite{2012HAO}} summarizes experimental lattice parameters at 0~K with the zero-point energy subtracted, which are directly comparable with our results.

{\vspace{2mm}}
First of all, we carefully checked the basis-set convergence since \pyscf, which generates trial wavefunctions for the subsequent \tvb\ calculations, employs the localized basis set also for the periodic systems. In other words, the convergence is sometimes difficult to be achieved unlike the plane-wave one. To check the basis-set convergence, we compared the EOSs of the 12 crystals obtained by \pyscf\ (localized basis-set) and QuantumEspresso (Plane-Wave basis set with a 800~Ryd cutoff) with the same ccECP pseudopotentials~{\cite{2017BEN,2018BEN,2018ANN,2019WAN}}. The basis set convergence check using 1$\times$1$\times$1 supercell is shown in Fig.~{\ref{SI-fig:eos-basis-check-s-1-1-1}} and the finally chosen basis sets are listed in Table~{\ref{tab:eos-qmc}}. We found that, to achieve the convergence, a large basis set (e.g., V5Z) is often needed. Notice that, in the localized basis sets, orbitals whose exponent is smaller than 0.10 were cut to avoid the numerical instability (i.e., linear-dependency~{\cite{2021NAK1}}). The consistency holds also for the 2$\times$2$\times$2 supercells as shown in Fig.~{\ref{SI-fig:eos-basis-check-s-2-2-2}}. The slow convergence with respect to the basis set size comes from the fact that the provided basis sets were tuned using molecules, not solids. Indeed, the exponents are not suitable for solids. Therefore, to achieve a better convergence, we recommend to use basis sets optimized for solids~{\cite{2022YE}}.

{\vspace{2mm}}
We have confirmed that the 2$\times$2$\times$2 supercell with $k$=2$\times$2$\times$2 twisted average is large enough to mitigate the one-body finite size effect for Diamond (Fig.~{\ref{SI-fig:eos-onebody-check-diamond}}). It should be common among all the compounds we are studying because they are all insulators with similar lattice parameters. We have not checked whether 2$\times$2$\times$2 supercells are large enough to mitigate the two-body error, but we can assume so because the EOS calculations depend on the relative energies. In fact, Ref.~{\onlinecite{2007MAE}} shows that larger supercells do not change the lattice parameter.

{\vspace{2mm}}
Figure~{\ref{fig:eos-qmc} shows the EOSs of the 12 crystals computed at the VMC and LRDMC levels with the LDA-PZ nodal surfaces with the converged basis sets. The DFT calculations with the XC=LDA-PZ~{\cite{1981PER}}, PBE~{\cite{1996PER}}, and PBEsol~{\cite{2008PER}} were performed using \qe\ with the Ultra-soft PPs provided by the PS-library Project (v.1.0.0)~{\cite{2014AND}}. The 2$\times$2$\times$2 supercells and $k$=2$\times$2$\times$2 meshes were employed for all the calculations. The obtained PESs were fitted by the Vinet function~\cite{1987VIN}. Table~{\ref{tab:eos-qmc}} shows the lattice parameters ($a_{0}$) obtained from the Vinet fittings and the available experimental values~{\cite{2012HAO}}. We evaluated the performance of the methods by mean absolute error (MAE: $\frac{1}{M} (\Sigma{|a_{\rm calc.} - a_{\rm exp.}|})$) and mean absolute relative error (MARE: $\frac{1}{M} (\Sigma{\frac{|a_{\rm calc.} - a_{\rm exp.}|}{a_{\rm exp.}}} \times 100)$), where $M$ is the sample size (i.e., $M = 12$ in this work). Fig.~{\ref{fig:vol-percentage-error}} shows percentage errors in the calculated lattice parameters compared to experimental ones. 

{\vspace{2mm}}
On one hand, our results show that the accuracy of the VMC calculations on the lattice parameter is strongly dependent on crystals. For instance, the estimated equilibrium lattice parameter of Diamond is well consistent with the experimental value both at the DFT and VMC levels, while that of NaCl was severely underestimated. The MARE for the VMC calculations is 0.5087(75) \%, which is slightly better than that for the DFT-LDA calculations, while worse than that of the DFT-PBEsol calculations. The results imply the VMC calculations (with the small Jastrow factor employed in this study) is sometimes not sufficient to get accurate EOSs.
%In fact, Santana et al.~{\cite{2016SAN}} reported that, for CaO, the DMC calculation with the LDA nodal surface gives a highly accurate structural properties, while our VMC calculation underestimates it. 

%
{\vspace{2mm}}
On the other hand, our results show that the DMC calculations with the LDA-PZ nodal surfaces are much less dependent on the choices of XCs for generating the trial wave functions, showing the accuracy of the methods. The conclusion is in line with the benchmark calculations presented in Refs.~{\onlinecite{2013SHU}} and ~{\onlinecite{2016SAN}}, showing that DMC is highly accurate in describing the structural properties of a broad range of solids and that these structural properties are rather insensitive to the given nodal surfaces. The MARE for the DMC calculations is 0.229(11) \%, which is the best result among the methods tested in this study. The remaining discrepancy between the experiments and calculations could come either from the fixed nodal surface (i.e., the LDA-PZ is used in this study), from the qualities of the PPs (ccECP), or from their related non-local properties (i.e., in case of T-moves, the quality of the Jastrow factor can still affect the results). In these regards, more comprehensive benchmark tests on the EOS calculations using \tbw\ will be an interesting perspective.

%
%%%%%%%%%%%%%%%%%%%%%%%%%%%%%%%%%%%%%%%%%%%%%%%%%%%%
% EOS, Comparison, Exp. VMC, DMC, DFT
%%%%%%%%%%%%%%%%%%%%%%%%%%%%%%%%%%%%%%%%%%%%%%%%%%%%
\begin{figure*}[htbp]
  \centering
  \includegraphics[width=\textwidth]{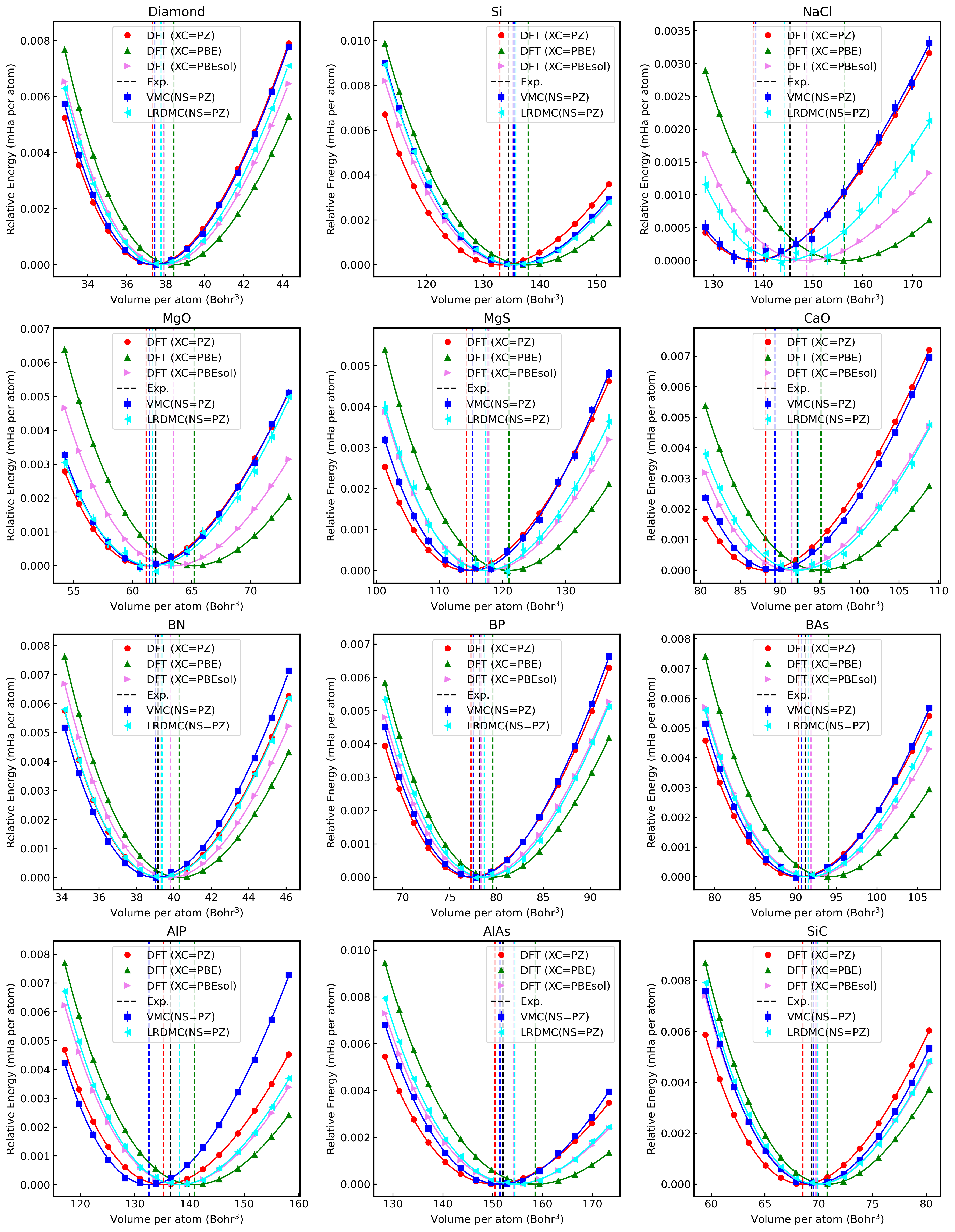}
  \caption{Results of the equation of state calculations for the 12 crystals by DFT, VMC, and LRDMC. The dot points are obtained values, and the solid lines are the curves obtained by a fit of the Vinet equation to calculations. The VMC and DMC points have statistical errors. In the plots, the error bars represent 2$\sigma$. The experimental equilibrium volumes and those obtained by a fit of the Vinet equation to calculations are plotted as vertical broken lines. Notice that the finite temperature thermal expansion and zero point energy were corrected in the experimental values~{\cite{2012HAO}}.}
  \label{fig:eos-qmc}
\end{figure*}
%%%%%%%%%%%%%%%%%%%%%%%%%%%%%%%%%%%%%%%%%%%%%%%%%%%%
\begin{figure}[htbp]
  \centering
  \includegraphics[width=\columnwidth]{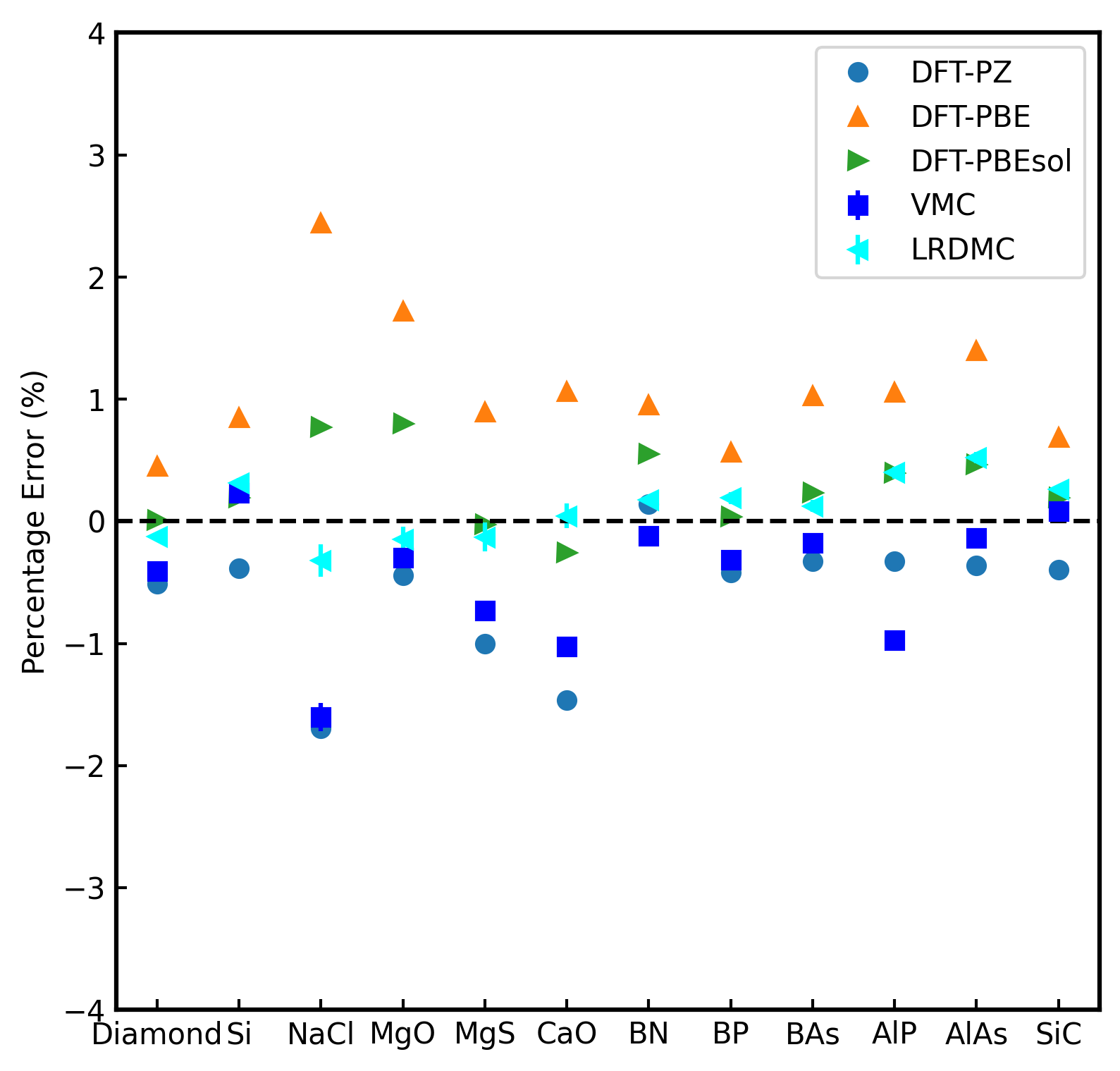}
  \caption{Percentage errors in the obtained lattice parameters as compared to the experimental values~{\cite{2012HAO}}. In the plots, the error bars represent 2$\sigma$. The positive (negative) percentage indicates that the calculation overestimates (underestimates) the lattice parameter.}
  \label{fig:vol-percentage-error}
\end{figure}
%%%%%%%%%%%%%%%%%%%%%%%%%%%%%%%%%%%%%%%%%%%%%%%%%%%%
% Table. template 
\begin{center}
\begin{table*}[hbtp]
\caption{\label{tab:eos-qmc}
The columns 1-5 contain, compounds, crystal type (A4, diamond; B1, CsCl; B3, zinc blende), CODID, basis set, and ECP. The columns 6-10 contain the equilibrium lattice parameters obtained from the Vinet fitting to DFT, VMC, and LRDMC calculations performed in this paper. The VMC and DMC results include statistical errors (1$\sigma$) in the individual calculations. The column 11 contains the experimental values, where the finite temperature thermal expansion and zero point energy were subtracted in the work of Hao et al.~{\cite{2012HAO}}. The lattice parameters are given in \AA. The two statistics are shown in the table, mean absolute error (MAE: $\frac{1}{M} (\Sigma{|a_{\rm calc.} - a_{\rm exp.}|})$) and mean absolute relative error (MARE: $\frac{1}{M} (\Sigma{\frac{|a_{\rm calc.} - a_{\rm exp.}|}{a_{\rm exp.}}} \times 100)$), where $M$ is the sample size (i.e., $M = 12$ in this work).
}
\begin{tabular}{ccc|cc|cccccc}
\Hline
\multicolumn{3}{c|}{System} & \multicolumn{2}{c|}{Basis set and ECP} & \multicolumn{6}{c}{Equilibrium lattice parameter (\AA)} \\
\Hline
 Compound & Crystal type &   CODID &    Basis set &   ECP & DFT (PZ) & DFT (PBE) & DFT (PBEsol) &         VMC &       LRDMC &  Exp. \\
\Hline
  Diamond &           A4 & 2101499 & ccecp-ccpvtz & ccECP &   3.5368 &    3.5711 &       3.5552 & 3.54043(45) & 3.55055(49) & 3.555 \\
       Si &           A4 & 1526655 & ccecp-ccpv5z & ccECP &   5.4011 &    5.4681 &       5.4323 &  5.4346(11) &  5.4390(11) & 5.422 \\
     NaCl &           B1 & 1000041 & ccecp-ccpv5z & ccECP &   5.4707 &    5.7009 &       5.6080 &  5.4758(32) &  5.5472(37) & 5.565 \\
      MgO &           B1 & 1000053 & ccecp-ccpvqz & ccECP &   4.1695 &    4.2602 &       4.2214 &  4.1756(12) &  4.1819(21) & 4.188 \\
      MgS &           B1 & 8104342 & ccecp-ccpv5z & ccECP &   5.1361 &    5.2346 &       5.1867 &  5.1502(16) &  5.1814(29) & 5.188 \\
      CaO &           B1 & 1011094 & ccecp-ccpvqz & ccECP &   4.7111 &    4.8320 &       4.7688 &  4.7318(11) &  4.7832(23) & 4.781 \\
       BN &           B3 & 9008834 & ccecp-ccpvtz & ccECP &   3.5991 &    3.6282 &       3.6139 & 3.58974(51) & 3.60033(55) & 3.594 \\
       BP &           B3 & 1541726 & ccecp-ccpvqz & ccECP &   4.5080 &    4.5528 &       4.5287 & 4.51271(56) &  4.5356(11) & 4.527 \\
      BAs &           B3 & 9008833 & ccecp-ccpvqz & ccECP &   4.7486 &    4.8130 &       4.7752 &  4.7556(11) &  4.7700(12) & 4.764 \\
      AlP &           B3 & 9008831 & ccecp-ccpv5z & ccECP &   5.4322 &    5.5077 &       5.4714 &  5.3969(12) &  5.4718(14) & 5.450 \\
     AlAs &           B3 & 9008830 & ccecp-ccpv5z & ccECP &   5.6287 &    5.7281 &       5.6753 &  5.6412(13) &  5.6784(13) & 5.649 \\
      SiC &           B3 & 1010995 & ccecp-ccpvtz & ccECP &   4.3309 &    4.3780 &       4.3565 & 4.35153(60) & 4.35948(62) & 4.348 \\
\Hline
MAE (\AA) &            - &       - &            - &     - &   0.0307 &    0.0537 &       0.0158 & 0.02560(39) & 0.01147(53) &     - \\
MARE (\%) &            - &       - &            - &     - &   0.6219 &    1.0947 &       0.3271 &  0.5087(75) &   0.229(11) &     - \\
\Hline
\end{tabular}
\end{table*}
\end{center}
%%%%%%%%%%%%%%%%%%%%%%%%%%%%%%%%%%%%%%%%%%%%%%%%%%%%
%

%%%%%%%%%%%%%%%%%%%%%%%%%%%%%%%%%%%%%%%%%%%%%%%%%%%%
%  Conclusions
%%%%%%%%%%%%%%%%%%%%%%%%%%%%%%%%%%%%%%%%%%%%%%%%%%%%
\section{Conclusions}
\label{sec:conclusion}
In this paper, we describe the features of the recently developed \tbg\ and demonstrate its applications. \tbg\ is a collection of python wrappers for the \emph{ab initio} QMC code, \tvb. The users can combine the modules implemented in \tbg\ with their python scripts to manage QMC tasks in a python script. \tbw, which is implemented using \tbg, is a python package realizing QMC workflows. \tbw\ enables one to run sequential QMC calculations fully automatically and manage file and job transfers from/to cluster machines. As demonstrated in this paper, these Python packages are particularly helpful in performing validations of the methods and algorithms and in conducting benchmark calculations for various materials. In terms of future works, for example, generating a materials database with the \emph{ab initio} QMC would be a very intriguing work, considering the recent successes of the DFT-based materials databases. As mentioned in the introduction, the importance of data provenance and data curation in the field of materials science has increased thanks to the development of information science and technology. In this regard, an accurate QMC database can be utilized, for instance, for the construction of machine learning potentials with accuracy exceeding DFT-based ones and for training machine learning exchange-correlation functionals. A package for high-throughput electronic structure calculations is an infrastructure technology in the materials science community. Thus, it should be continuously developed and maintained as an open-source package for the long-term perspective.

\vspace{1mm}
\begin{acknowledgments}
% NIMS-PC
K.N. is grateful for computational resources from the Numerical Materials Simulator at National Institute for Materials Science (NIMS).
% FUGAKU
K.N. and M.C. are grateful for computational resources of the supercomputer Fugaku provided by RIKEN through the HPCI System Research Projects (Project IDs: hp200164, hp210038, hp220060, and hp230030).
% K.N. financial support
K.N. acknowledges financial support from the JSPS Overseas Research Fellowships, from Grant-in-Aid for Early Career Scientists (Grant No.~JP21K17752), from Grant-in-Aid for Scientific Research (Grant No.~JP21K03400), and from MEXT Leading Initiative for Excellent Young Researchers (Grant No.~JPMXS0320220025).
% TREXIO implementation support
The authors acknowledge fruitful discussion with E. Posenitskiy and A. Scemama about the implementation of the WF converter (from \trexio\ to \tvb).
% Quantum Package support
The author thank A. Scemama for providing HF energies computed by \qpack\ for the validation tests in this work.
% TREX ack.
This work is supported by the European Centre of Excellence in Exascale Computing TREX - Targeting Real Chemical Accuracy at the Exascale. This project has received funding from the European Union’s Horizon 2020 - Research and Innovation program - under grant agreement no. 952165.

\vspace{2mm}
We dedicate this paper to the memory of Prof. Sandro Sorella (SISSA), who passed away during the collaboration. He has been one of the most influential contributors to the Quantum Monte Carlo community. In particular, he deeply inspired this work, with the development of his \emph{ab initio} QMC code, \tvb.
\end{acknowledgments}

\section{Data availability} \label{sec:data}
The {\program{TREXIO}} files used for the validation tests are available from our ZENODO repository [\url{https://doi.org/10.5281/zenodo.8382156}] and from NIMS Materials Data Repository (MDR) [\url{https://doi.org/10.48505/nims.4231}]. Several sample Python scripts for the validations tests are available from our GitHub repository [{\url{https://github.com/kousuke-nakano/turbotutorials}}].

\section{Code availability and reliability} \label{sec:code}
\tbg, \tbf, and \tbw\ are available from our GitHub repositories, [\url{https://github.com/kousuke-nakano/turbogenius}], [\url{https://github.com/kousuke-nakano/turbofilemanager}], and [\url{https://github.com/kousuke-nakano/turboworkflows}], respectively. To ensure the reliability of the \tbg\ package, we have adopted standard continuous integration and deployment (CD/CI) practices. Specifically, we have prepared unit tests as well as regression tests that are executed automatically using GitHub actions whenever changes are pushed to the repository. These tests cover many functionalities in the packages; Thus, they help us with identifying any potential issues and/or bugs in the packages. As open-source projects, we encourage contributions from anyone interested in the development of these packages. The QMC kernel, \tvb, is also available from the GitHub repository [\url{https://github.com/sissaschool/turborvb}].

\section{Conflict of interest} \label{sec:interests}
The authors declare no conflict of interest.

%%%%%%%%%%%%%%%%%%%%%%%%%%%%%%%
\bibliographystyle{apsrev4-2}
\bibliography{./references.bib}
%%%%%%%%%%%%%%%%%%%%%%%%%%%%%%%

\end{document}

\title{Supplementary Information for \tbg: Python suite for high-throughput calculations of {\emph{ab initio}} quantum Monte Carlo methods}
\author{Kousuke Nakano}
\email{kousuke\_1123@icloud.com}
\affiliation{International School for Advanced Studies (SISSA), Via Bonomea 265, 34136, Trieste, Italy}
\affiliation{Center for Basic Research on Materials, National Institute for Materials Science (NIMS), Tsukuba, Ibaraki 305-0047, Japan}
\author{Oto Kohul\'{a}k}
\affiliation{International School for Advanced Studies (SISSA), Via Bonomea 265, 34136, Trieste, Italy}
\affiliation{Laboratoire de Chimie et Physique Quantiques (LCPQ), Universit{\'e} de Toulouse (UPS) and CNRS, Toulouse, France}
\author{Abhishek Raghav}
\affiliation{International School for Advanced Studies (SISSA), Via Bonomea 265, 34136, Trieste, Italy}
\affiliation{Institut de Min{\'e}ralogie, de Physique des Mat{\'e}riaux et de Cosmochimie (IMPMC), Sorbonne Universit{\'e}, CNRS UMR 7590, IRD UMR 206, MNHN, 4 Place Jussieu, 75252 Paris, France}
\author{Michele Casula}
\affiliation{Institut de Min{\'e}ralogie, de Physique des Mat{\'e}riaux et de Cosmochimie (IMPMC), Sorbonne Universit{\'e}, CNRS UMR 7590, IRD UMR 206, MNHN, 4 Place Jussieu, 75252 Paris, France}
\author{Sandro Sorella}
\affiliation{International School for Advanced Studies (SISSA), Via Bonomea 265, 34136, Trieste, Italy}

\makeatletter
\renewcommand{\refname}{}
\renewcommand*{\citenumfont}[1]{#1}
\renewcommand*{\bibnumfmt}[1]{[#1]}
\makeatother

\setcounter{table}{0}
\setcounter{equation}{0}
\setcounter{figure}{0}
\renewcommand{\thepage}{S\arabic{page}}
\renewcommand{\thetable}{S-\Roman{table}}
\renewcommand{\thefigure}{S-\arabic{figure}}
\renewcommand{\theequation}{S-\arabic{equation}}
\renewcommand{\thelstlisting}{S-\arabic{lstlisting}}

\newcommand{\tvb}{\textsc{TurboRVB}} 
\newcommand{\tbw}{\textsc{TurboWorkflows}} 
\newcommand{\tbg}{\textsc{TurboGenius}} 
\newcommand{\pyturbo}{\textsc{PyTurbo}}
\newcommand{\tbf}{\textsc{TurboFilemanager}}
\newcommand{\aiidaturbo}{\textsc{AiiDA-TurboRVB}}
\newcommand{\pyscf}{\textsc{PySCF}}
\newcommand{\qe}{\textsc{Quantum Espresso}}
\newcommand{\trexio}{\textsc{TREXIO}}
\newcommand{\program}{\texttt}

%%%%%%%%%%%%%%%%%%%%%%%%%%%%%%%%%%%%%%%%%%%%%%%%%%%%
\date{\today}
\begin{abstract}
This file includes the supplementary information for the paper titled ``\tbg: Python suite for high-throughput calculations of the ab-initio quantum Monte Carlo methods".
\end{abstract}
\maketitle
\thispagestyle{empty}
\addtocounter{page}{-1}
\newpage

%%%%%%%%%%%%%%%%%%%%%%%%%%%%%%%%%%%%%%%%%%%%%%%%%%%%
% UML activity diagram of VMC workflow
%%%%%%%%%%%%%%%%%%%%%%%%%%%%%%%%%%%%%%%%%%%%%%%%%%%%
\begin{figure*}[htbp]
  \centering
  \includegraphics[width=\textwidth]{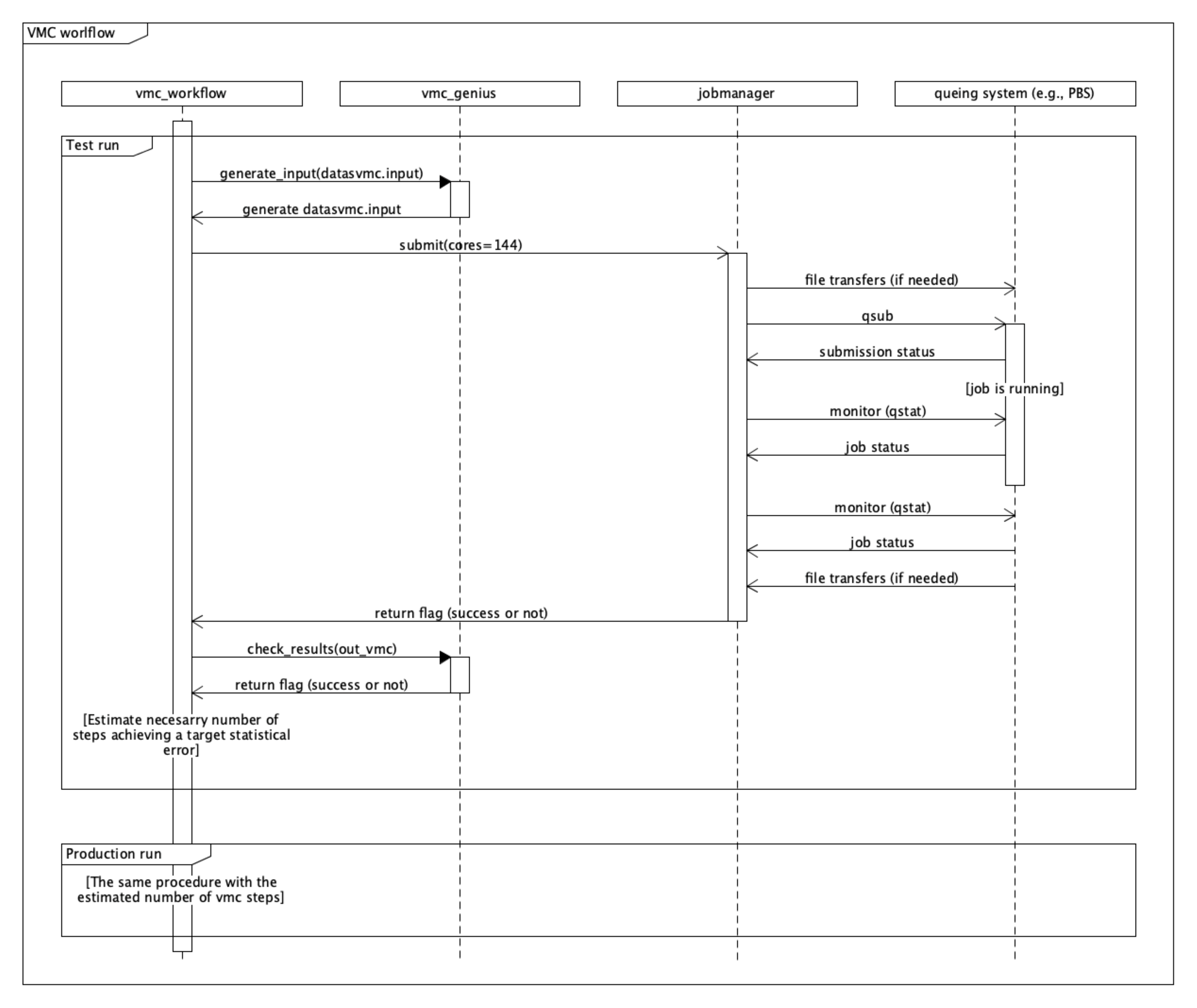}
  \caption{UML activity diagram of the VMC\_workflow class.}
  \label{fig:uml-vmc-workflow}
\end{figure*}
%%%%%%%%%%%%%%%%%%%%%%%%%%%%%%%%%%%%%%%%%%%%%%%%%%%%

%%%%%%%%%%%%%%%%%%%%%%%%%%%%%%%%%%%%%%%%%%%%%%%%%%%%
% UML activity diagram of Launcher
%%%%%%%%%%%%%%%%%%%%%%%%%%%%%%%%%%%%%%%%%%%%%%%%%%%%
\begin{figure*}[htbp]
  \centering
  \includegraphics[width=\textwidth]{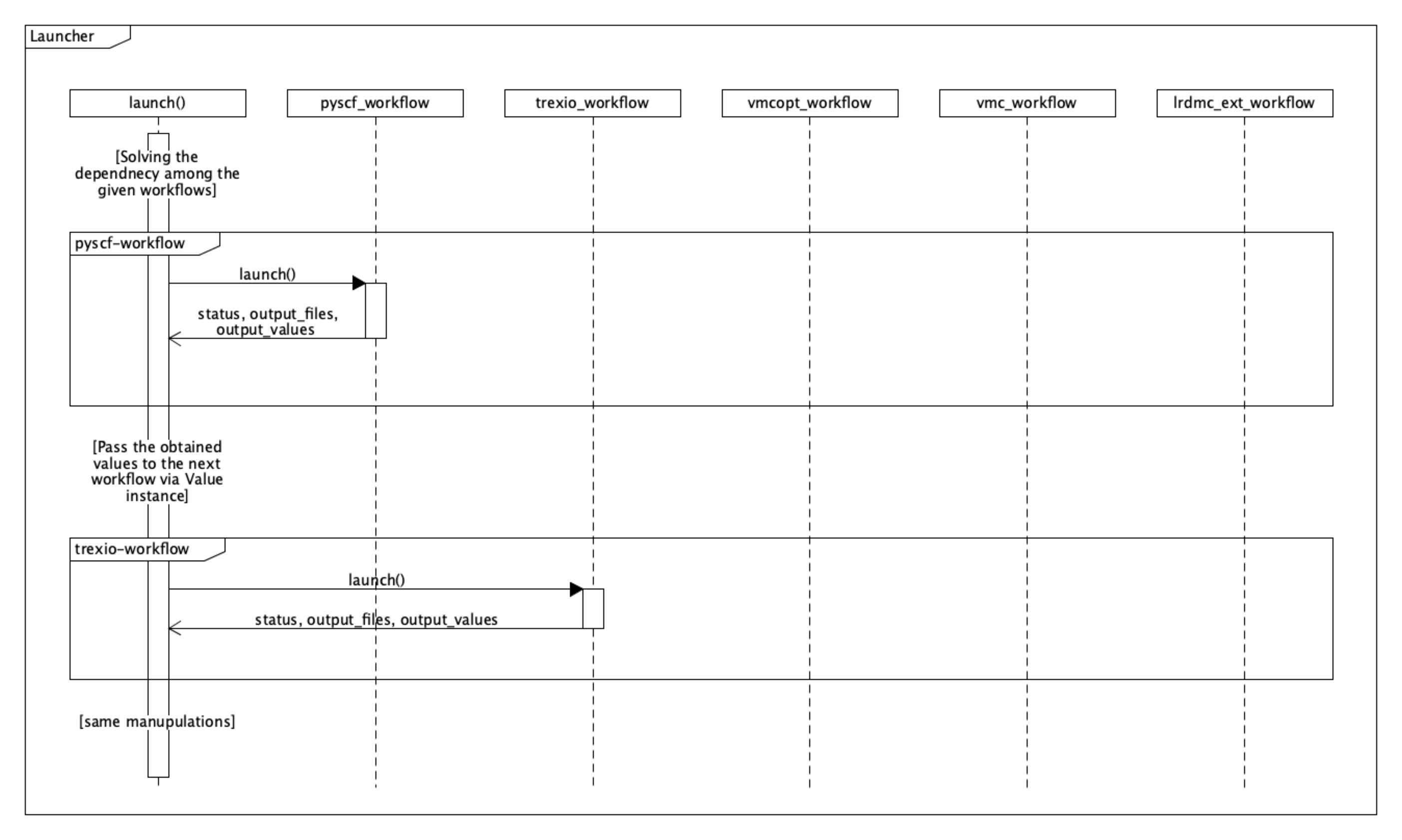}
  \caption{UML activity diagram of the Launcher class.}
  \label{fig:uml-launcher}
\end{figure*}
%%%%%%%%%%%%%%%%%%%%%%%%%%%%%%%%%%%%%%%%%%%%%%%%%%%%

\newpage

% python workflow example and result.
\begin{widetext}
\begin{lstlisting}[style=PythonStyle,linewidth=\columnwidth,caption={A python workflow to to obtain PES of the carbon monoxide.},label=lst:turboworkflows-CO-workflow]

#!/usr/bin/env python

# python packages
import os, sys
import numpy as np

# turboworkflows packages
from turboworkflows.workflow_encapsulated import eWorkflow
from turboworkflows.workflow_lrdmc import LRDMC_workflow
from turboworkflows.workflow_vmc import VMC_workflow
from turboworkflows.workflow_pyscf import PySCF_workflow
from turboworkflows.workflow_trexio import TREXIO_convert_to_turboWF
from turboworkflows.workflow_vmcopt import VMCopt_workflow
from turboworkflows.workflow_collections import Jastrowcopy_workflow
from turboworkflows.workflow_lanchers import Launcher, Variable

from ase import Atoms
max_d=1.30
min_d=0.90
num_d=11

d_list=[f'{np.round(a,2):.2f}' for a in np.linspace(min_d, max_d, num_d)]
d_ref=d_list[int(len(d_list) / 2)]

root_dir=os.getcwd()
result_dir=os.path.join(os.getcwd(), "results")
os.makedirs(result_dir, exist_ok=True)
os.chdir(result_dir)

cworkflows_list=[]

# jastrow basis dictionary
from jastrow_basis import jastrow_basis_dict

for d in d_list:
    CO_molecule = Atoms('CO', positions=[(0, 0, -float(d)/2), (0, 0, float(d)/2)])
    CO_molecule.write(f'CO_{d}.xyz')

    pyscf_workflow = eWorkflow(
        label=f'pyscf-workflow-{d}',
        dirname=f'pyscf-workflow-{d}',
        input_files=[f'CO_{d}.xyz'],
        workflow=PySCF_workflow(
            ## structure file
            structure_file=f'CO_{d}.xyz',
            ## cluster machine
            server_machine_name="my-cluster-machine",
            cores=64,
            openmp=64,
            queue="DEFAULT",
            version="stable",
            sleep_time=10,
            jobpkl_name="job_manager",
            ## pyscf
            pyscf_rerun=False,
            pyscf_pkl_name="pyscf_genius",
            charge=0,
            spin=0,
            basis="ccecp-ccpvqz",
            ecp="ccecp",
            scf_method="DFT",
            dft_xc="LDA_X,LDA_C_PZ",
            pyscf_output="out.pyscf",
            pyscf_chkfile="pyscf.chk",
        )
    )

    cworkflows_list.append(pyscf_workflow)

    trexio_workflow = eWorkflow(
        label=f'trexio-workflow-{d}',
        dirname=f'trexio-workflow-{d}',
        input_files=[Variable(label=f'pyscf-workflow-{d}', vtype='file', name='trexio.hdf5')],
        workflow=TREXIO_convert_to_turboWF(
            trexio_filename="trexio.hdf5",
            twist_average=False,
            jastrow_basis_dict=jastrow_basis_dict,
            max_occ_conv=0,
            mo_num_conv=-1,
            trexio_rerun=False,
            trexio_pkl_name="trexio_genius"
        )
    )
    cworkflows_list.append(trexio_workflow)

    if d!=d_ref:
        copyjas_workflow = eWorkflow(
            label=f'copyjas-workflow-{d}',
            dirname=f'copyjas-workflow-{d}',
            input_files=[Variable(label=f'trexio-workflow-{d}', vtype='file', name='fort.10'), Variable(label=f'vmcopt-workflow-{d_ref}', vtype='file', name='fort.10'), Variable(label=f'trexio-workflow-{d}', vtype='file', name='pseudo.dat')],
            rename_input_files=["fort.10", "fort.10_new", "pseudo.dat"],
            workflow=Jastrowcopy_workflow(
                jastrowcopy_fort10_to="fort.10",
                jastrowcopy_fort10_from="fort.10_new",
            )
        )
        cworkflows_list.append(copyjas_workflow)

        vmcopt_input_files = [Variable(label=f'copyjas-workflow-{d}', vtype='file', name='fort.10'), Variable(label=f'copyjas-workflow-{d}', vtype='file', name='pseudo.dat')]

    else:
        vmcopt_input_files = [Variable(label=f'trexio-workflow-{d}', vtype='file', name='fort.10'), Variable(label=f'trexio-workflow-{d}', vtype='file', name='pseudo.dat')]

    vmcopt_workflow = eWorkflow(
        label=f'vmcopt-workflow-{d}',
        dirname=f'vmcopt-workflow-{d}',
        input_files=vmcopt_input_files,
        workflow=VMCopt_workflow(
            ## cluster machine
            server_machine_name="my-cluster-machine",
            cores=1536,
            openmp=1,
            queue="LARGE",
            version="stable",
            sleep_time=7200,
            jobpkl_name="job_manager",
            ## vmcopt
            vmcopt_max_continuation=2,
            vmcopt_pkl_name="vmcopt_genius",
            vmcopt_target_error_bar=1.0e-3,
            vmcopt_trial_optsteps=50,
            vmcopt_trial_steps=50,
            vmcopt_production_optsteps=1200,
            vmcopt_optwarmupsteps_ratio=0.8,
            vmcopt_bin_block=1,
            vmcopt_warmupblocks=0,
            vmcopt_optimizer="lr",
            vmcopt_learning_rate=0.35,
            vmcopt_regularization=0.001,
            vmcopt_onebody=True,
            vmcopt_twobody=True,
            vmcopt_det_mat=False,
            vmcopt_jas_mat=True,
            vmcopt_det_basis_exp=False,
            vmcopt_jas_basis_exp=False,
            vmcopt_det_basis_coeff=False,
            vmcopt_jas_basis_coeff=False,
            vmcopt_num_walkers = -1,
            vmcopt_maxtime=172000,
        )
    )
    cworkflows_list.append(vmcopt_workflow)

    vmc_workflow = eWorkflow(
        label=f'vmc-workflow-{d}',
        dirname=f'vmc-workflow-{d}',
        input_files=[Variable(label=f'vmcopt-workflow-{d}', vtype='file', name='fort.10'),
                    Variable(label=f'vmcopt-workflow-{d}', vtype='file', name='pseudo.dat')],
        workflow=VMC_workflow(
            ## cluster machine
            server_machine_name="my-cluster-machine",
            cores=1536,
            openmp=1,
            queue="LARGE",
            version="stable",
            sleep_time=3600,
            jobpkl_name="job_manager",
            ## vmc
            vmc_max_continuation=2,
            vmc_pkl_name="vmc_genius",
            vmc_target_error_bar=7.0e-5,
            vmc_trial_steps= 150,
            vmc_bin_block = 10,
            vmc_warmupblocks = 5,
            vmc_num_walkers = -1,
            vmc_maxtime=172000,
        )
    )
    cworkflows_list.append(vmc_workflow)

    lrdmc_workflow = eWorkflow(
        label=f'lrdmc-workflow-{d}',
        dirname=f'lrdmc-workflow-{d}',
        input_files=[Variable(label=f'vmc-workflow-{d}', vtype='file', name='fort.10'),
                    Variable(label=f'vmc-workflow-{d}', vtype='file', name='pseudo.dat')],
        workflow=LRDMC_workflow(
            ## cluster machine
            server_machine_name="my-cluster-machine",
            cores=1536,
            openmp=1,
            queue="LARGE",
            version="stable",
            sleep_time=3600,
            jobpkl_name="job_manager",
            ## lrdmc
            lrdmc_max_continuation=2,
            lrdmc_pkl_name="lrdmc_genius",
            lrdmc_target_error_bar=7.0e-5,
            lrdmc_trial_steps=150,
            lrdmc_bin_block=10,
            lrdmc_warmupblocks=5,
            lrdmc_correcting_factor=10,
            lrdmc_trial_etry=Variable(label=f'vmc-workflow-{d}', vtype='value', name='energy'),
            lrdmc_alat=-0.25,
            lrdmc_nonlocalmoves="dla",
            lrdmc_num_walkers=-1,
            lrdmc_maxtime=172000,
        )
    )

    cworkflows_list.append(lrdmc_workflow)

launcher=Launcher(cworkflows_list=cworkflows_list, dependency_graph_draw=True)
launcher.launch()

\end{lstlisting}
\end{widetext}

\newpage

%%%%%%%%%%%%%%%%%%%%%%%%%%%%%%%%%%%%%%%%%%%%%%%%%%%%
\begin{table}[htpb]
 \caption{The comparison of the LDA-PZ energies obtained by \pyscf\ and those obtained by \tvb\ with the same basis sets for 38 molecules. The \trexio\ files used for these calculations are available from our public repository (See Sec.~\ref{M-sec:data} in the main text for the details.)}
 \label{tab:sanity-check-molecules-LDA}
 \centering
\begin{tabular}{cc|cc|c}
\hline
 Molecules & Basis set & LDA energy (Ha) [\pyscf] & LDA energy (Ha) [\tvb] & Difference (Ha) \\
\hline
          H$_{2}$ & ccecp-ccpv5z &               -1.052985 &                  -1.052981 &       +0.000004 \\
              LiH & ccecp-ccpvqz &               -0.766956 &                  -0.766955 &       +0.000001 \\
         Li$_{2}$ & ccecp-ccpvqz &               -0.424899 &                  -0.424898 &       +0.000000 \\
         CH$_{4}$ & ccecp-ccpvqz &               -7.999578 &                  -7.999577 &       +0.000002 \\
         H$_{2}$O & ccecp-ccpvqz &              -17.149230 &                 -17.149231 &       -0.000002 \\
               HF & ccecp-ccpvqz &              -24.789625 &                 -24.789628 &       -0.000003 \\
         NH$_{3}$ & ccecp-ccpvqz &              -11.649519 &                 -11.649520 &       -0.000001 \\
              LiF & ccecp-ccpvqz &              -24.488407 &                 -24.488394 &       +0.000013 \\
   C$_{2}$H$_{2}$ & ccecp-ccpvqz &              -12.387838 &                 -12.387840 &       -0.000002 \\
               CO & ccecp-ccpvqz &              -21.599002 &                 -21.599006 &       -0.000004 \\
              HCN & ccecp-ccpvqz &              -16.080629 &                 -16.080632 &       -0.000003 \\
          N$_{2}$ & ccecp-ccpvqz &              -19.782359 &                 -19.782358 &       +0.000001 \\
   C$_{2}$H$_{4}$ & ccecp-ccpvqz &              -13.619256 &                 -13.619257 &       -0.000001 \\
        H$_{2}$CO & ccecp-ccpvqz &              -22.767856 &                 -22.767858 &       -0.000002 \\
   C$_{2}$H$_{6}$ & ccecp-ccpvqz &              -14.834373 &                 -14.834374 &       -0.000001 \\
          F$_{2}$ & ccecp-ccpvqz &              -48.234293 &                 -48.234293 &       -0.000000 \\
   H$_{2}$O$_{2}$ & ccecp-ccpvqz &              -33.029733 &                 -33.029733 &       +0.000001 \\
       H$_{3}$COH & ccecp-ccpvqz &              -23.966120 &                 -23.966118 &       +0.000001 \\
   N$_{2}$H$_{4}$ & ccecp-ccpvqz &              -22.090553 &                 -22.090549 &       +0.000003 \\
         CO$_{2}$ & ccecp-ccpvqz &              -37.641550 &                 -37.641554 &       -0.000004 \\
SiH$_{2}$-singlet & ccecp-ccpvqz &               -4.949041 &                  -4.949041 &       +0.000001 \\
         H$_{2}$S & ccecp-ccpvqz &              -11.301548 &                 -11.301539 &       +0.000009 \\
              HCl & ccecp-ccpvqz &              -15.489788 &                 -15.489782 &       +0.000005 \\
         PH$_{3}$ & ccecp-ccpvqz &               -8.264345 &                  -8.264338 &       +0.000007 \\
        SiH$_{4}$ & ccecp-ccpvqz &               -6.189380 &                  -6.189385 &       -0.000005 \\
               CS & ccecp-ccpvqz &              -15.691692 &                 -15.691695 &       -0.000003 \\
              SiO & ccecp-ccpvqz &              -19.851557 &                 -19.851567 &       -0.000011 \\
       CH$_{3}$Cl & ccecp-ccpvqz &              -22.317235 &                 -22.317234 &       +0.000001 \\
              ClF & ccecp-ccpvqz &              -39.027754 &                 -39.027747 &       +0.000007 \\
       H$_{3}$CSH & ccecp-ccpvqz &              -18.134103 &                 -18.134100 &       +0.000003 \\
             HOCl & ccecp-ccpvqz &              -31.405201 &                 -31.405195 &       +0.000006 \\
         SO$_{2}$ & ccecp-ccpvqz &              -42.089335 &                 -42.089337 &       -0.000002 \\
         Na$_{2}$ & ccecp-ccpvqz &               -0.399062 &                  -0.399064 &       -0.000002 \\
             NaCl & ccecp-ccpvqz &              -15.171508 &                 -15.171512 &       -0.000005 \\
          P$_{2}$ & ccecp-ccpvqz &              -13.030309 &                 -13.030299 &       +0.000010 \\
         Cl$_{2}$ & ccecp-ccpvqz &              -29.774661 &                 -29.774649 &       +0.000012 \\
  Si$_{2}$H$_{6}$ & ccecp-ccpvqz &              -11.238912 &                 -11.238919 &       -0.000007 \\
         Ne$_{2}$ & ccecp-ccpv6z &              -69.728581 &                 -69.728586 &       -0.000004 \\
\hline
\end{tabular}
\end{table}
\newpage
%%%%%%%%%%%%%%%%%%%%%%%%%%%%%%%%%%%%%%%%%%%%%%%%%%%%
\begin{table}[htpb]
 \caption{The comparison of the LDA-PZ energies obtained by \pyscf\ and those obtained by \tvb\ with the same basis sets for 10 crystals with twisted average ($k$ = 4 $\times$ 4 $\times$ 4). The \trexio\ files used for these calculations are available from our public repository (See Sec.~\ref{M-sec:data} in the main text for the details.)}
 \label{tab:sanity-check-crystals-LDA-pbc-kgrid-insulator}
 \centering
\begin{tabular}{cccc|cc|c}
\hline
  CODID & Crystals & Crystalsystem &  Spacegroup & LDA energy (Ha) [\pyscf] & LDA energy (Ha) [\tvb] & Difference (Ha) \\
\hline
1011097 & alpha-SiO$_{2}$ &      trigonal &         152 &             -108.021418 &                -108.021413 &       +0.000005 \\
1526655 &              Si &         cubic &         227 &              -31.213676 &                 -31.213692 &       -0.000016 \\
2101499 &         Diamond &         cubic &         227 &              -45.398692 &                 -45.398686 &       +0.000007 \\
9008997 &            h-BN &     hexagonal &         194 &              -25.627859 &                 -25.627856 &       +0.000003 \\
9011660 &            w-BN &     hexagonal &         186 &              -25.640240 &                 -25.640238 &       +0.000002 \\
1534043 & Si$_{3}$N$_{4}$ &         cubic &         220 &             -206.216212 &                -206.216225 &       -0.000013 \\
1010954 &    hp-SiO$_{2}$ &         cubic &         227 &             -287.772846 &                -287.772842 &       +0.000005 \\
9008667 &             LiF &         cubic &         225 &              -98.395453 &                 -98.395446 &       +0.000007 \\
1000053 &             MgO &         cubic &         225 &              -67.925608 &                 -67.925614 &       -0.000006 \\
9008830 &            AlAs &         cubic &         216 &              -33.435721 &                 -33.435736 &       -0.000015 \\
\hline
\end{tabular}
\end{table}
%%%%%%%%%%%%%%%%%%%%%%%%%%%%%%%%%%%%%%%%%%%%%%%%%%%%
\begin{table}[h]
 \caption{The comparison of the LDA-PZ energies obtained by \pyscf\ and those obtained by \tvb\ with the same basis sets for 4 crystals with twisted average ($k$ = 4 $\times$ 4 $\times$ 4). The \trexio\ files used for these calculations are available from our public repository (See Sec.~\ref{M-sec:data} in the main text for the details.)}
 \label{tab:sanity-check-crystals-LDA-pbc-kgrid-metal}
 \centering
\begin{tabular}{cccc|cc|c}
\hline
 CODID & Crystals & Crystalsystem &  Spacegroup & LDA energy (Ha) [\pyscf] & LDA energy (Ha) [\tvb] & Difference (Ha) \\
\hline
  57408 &   Li-fcc &         cubic &         225 &              -29.573978 &                 -29.573972 &       +0.000006 \\
 109012 &  Li-cI16 &         cubic &         220 &             -118.109248 &                -118.109234 &       +0.000014 \\
8104233 &       Na &         cubic &         229 &              -94.770109 &                 -94.770104 &       +0.000005 \\
2300202 &       Fe &         cubic &         229 &             -246.587802 &                -246.587783 &       +0.000019 \\
\hline
\end{tabular}
\end{table}
%%%%%%%%%%%%%%%%%%%%%%%%%%%%%%%%%%%%%%%%%%%%%%%%%%%%

\newpage
\begin{longtable}{cccc|cc|c}
\caption{The comparison of the RHF(ROHF) energies obtained by \pyscf\ and the VMC energies obtained by \tvb\ with the WFs converted from the corresponding \pyscf\ checkpoints files via \trexio\ for the 100 molecules. The error bars refer to $3 \sigma$ of the VMC calculations. The \trexio\ files used for these calculations are available from our public repository (See Sec.~\ref{M-sec:data} in the main text for the details.)}
\label{tab:sanity-check-molecules-RHF} \\
\hline
Molecules &  Charge &  Spin &    Basis set & HF energy (Ha) [\pyscf] & HF energy (Ha) [\tvb] & Difference (Ha) \\
\hline 
\endfirsthead

% head
\hline
Molecules &  Charge &  Spin &    Basis set & HF energy (Ha) [\pyscf] & HF energy (Ha) [\tvb] & Difference (Ha) \\
\hline 
\endhead

% foot
\hline\multicolumn{7}{r}{continued.} \\
\endfoot

% lastfoot
\hline\multicolumn{7}{r}{end.} \\
\endlastfoot

% table
                H &       0 &     1 & ccecp-ccpvqz &              -0.500000 &             -0.499992(18) &   +0.000008(18) \\
               He &       0 &     0 & ccecp-ccpvqz &              -2.861679 &             -2.861677(39) &   +0.000003(39) \\
               Li &       0 &     1 & ccecp-ccpv5z &              -0.196853 &            -0.1968528(15) &  -0.0000001(15) \\
               Be &       0 &     0 & ccecp-ccpv6z &              -0.961893 &             -0.961896(31) &   -0.000003(31) \\
                B &       0 &     1 & ccecp-ccpv6z &              -2.539305 &             -2.539324(32) &   -0.000019(32) \\
                C &       0 &     2 & ccecp-ccpv6z &              -5.314319 &             -5.314307(33) &   +0.000012(33) \\
                N &       0 &     3 & ccecp-ccpv6z &              -9.633866 &             -9.633874(31) &   -0.000007(31) \\
                O &       0 &     2 & ccecp-ccpv6z &             -15.689783 &            -15.689774(34) &   +0.000009(34) \\
                F &       0 &     1 & ccecp-ccpv6z &             -23.937918 &            -23.937922(37) &   -0.000004(37) \\
               Ne &       0 &     0 & ccecp-ccpv6z &             -34.708819 &            -34.708814(28) &   +0.000005(28) \\
               Na &       0 &     1 & ccecp-ccpvqz &              -0.186205 &            -0.1862031(56) &  +0.0000019(56) \\
               Mg &       0 &     0 & ccecp-ccpvqz &              -0.788392 &             -0.788372(27) &   +0.000020(27) \\
               Al &       0 &     1 & ccecp-ccpvqz &              -1.877084 &             -1.877101(27) &   -0.000017(27) \\
               Si &       0 &     2 & ccecp-ccpvqz &              -3.672547 &             -3.672553(30) &   -0.000006(30) \\
                P &       0 &     3 & ccecp-ccpvqz &              -6.340966 &             -6.340973(31) &   -0.000007(31) \\
                S &       0 &     2 & ccecp-ccpvqz &              -9.920494 &             -9.920504(29) &   -0.000010(29) \\
               Cl &       0 &     1 & ccecp-ccpvqz &             -14.691247 &            -14.691263(34) &   -0.000016(34) \\
               Ar &       0 &     0 & ccecp-ccpvqz &             -20.779663 &            -20.779668(26) &   -0.000005(26) \\
                K &       0 &     1 & ccecp-ccpvqz &             -27.934622 &            -27.934614(28) &   +0.000009(28) \\
               Ca &       0 &     0 & ccecp-ccpvqz &             -36.349734 &            -36.349743(35) &   -0.000009(35) \\
               Sc &       0 &     1 & ccecp-ccpvtz &             -46.122158 &            -46.122171(31) &   -0.000013(31) \\
               Ti &       0 &     2 & ccecp-ccpvtz &             -57.598094 &            -57.598110(29) &   -0.000016(29) \\
                V &       0 &     3 & ccecp-ccpvtz &             -70.898419 &            -70.898434(34) &   -0.000015(34) \\
               Cr &       0 &     6 & ccecp-ccpvtz &             -86.048478 &            -86.048479(28) &   -0.000001(28) \\
               Mn &       0 &     5 & ccecp-ccpvdz &            -103.120673 &           -103.120673(35) &   -0.000000(35) \\
               Fe &       0 &     4 & ccecp-ccpvtz &            -122.572205 &           -122.572210(39) &   -0.000005(39) \\
               Co &       0 &     3 & ccecp-ccpvqz &            -144.264994 &           -144.264997(55) &   -0.000003(55) \\
               Ni &       0 &     2 & ccecp-ccpvtz &            -168.272851 &           -168.272814(51) &   +0.000038(51) \\
               Cu &       0 &     1 & ccecp-ccpvtz &            -195.337049 &           -195.337048(69) &   +0.000001(69) \\
               Zn &       0 &     0 & ccecp-ccpvtz &            -225.275021 &           -225.275017(70) &   +0.000005(70) \\
               Ga &       0 &     1 & ccecp-ccpvqz &              -1.984210 &             -1.984208(32) &   +0.000001(32) \\
               Ge &       0 &     2 & ccecp-ccpvqz &              -3.664901 &             -3.664907(26) &   -0.000006(26) \\
               As &       0 &     3 & ccecp-ccpvqz &              -6.065876 &             -6.065870(38) &   +0.000006(38) \\
               Se &       0 &     2 & ccecp-ccpvqz &              -9.149929 &             -9.149927(36) &   +0.000002(36) \\
               Br &       0 &     1 & ccecp-ccpvqz &             -13.121474 &            -13.121470(29) &   +0.000004(29) \\
               Kr &       0 &     0 & ccecp-ccpvqz &             -18.228060 &            -18.228079(28) &   -0.000019(28) \\
               H- &      -1 &     0 & ccecp-ccpvqz &              -0.476733 &             -0.476736(37) &   -0.000003(37) \\
              He+ &       1 &     1 & ccecp-ccpvqz &              -1.999787 &             -1.999809(30) &   -0.000022(30) \\
              He- &      -1 &     1 & ccecp-ccpvqz &              -2.251807 &             -2.251804(28) &   +0.000004(28) \\
              Li- &      -1 &     0 & ccecp-ccpvqz &              -0.187667 &             -0.187681(29) &   -0.000013(29) \\
              LiH &       0 &     0 & ccecp-ccpvqz &              -0.749666 &             -0.749652(31) &   +0.000015(31) \\
              BeH &       0 &     1 & ccecp-ccpvqz &              -1.544205 &             -1.544220(30) &   -0.000015(30) \\
         Li$_{2}$ &       0 &     0 & ccecp-ccpvqz &              -0.399560 &             -0.399556(27) &   +0.000004(27) \\
               CH &       0 &     1 & ccecp-ccpvqz &              -5.906189 &             -5.906203(35) &   -0.000014(35) \\
 CH$_{2}$-singlet &       0 &     0 & ccecp-ccpvqz &              -6.523371 &             -6.523377(29) &   -0.000007(29) \\
 CH$_{2}$-triplet &       0 &     2 & ccecp-ccpvqz &              -6.564203 &             -6.564222(27) &   -0.000019(27) \\
               NH &       0 &     2 & ccecp-ccpvqz &             -10.212125 &            -10.212108(33) &   +0.000017(33) \\
         CH$_{3}$ &       0 &     1 & ccecp-ccpvqz &              -7.205901 &             -7.205875(45) &   +0.000026(45) \\
         NH$_{2}$ &       0 &     1 & ccecp-ccpvqz &             -10.821853 &            -10.821870(31) &   -0.000017(31) \\
               OH &       0 &     1 & ccecp-ccpvqz &             -16.300322 &            -16.300335(30) &   -0.000013(30) \\
         CH$_{4}$ &       0 &     0 & ccecp-ccpvqz &              -7.847115 &             -7.847102(34) &   +0.000013(34) \\
         H$_{2}$O &       0 &     0 & ccecp-ccpvqz &             -16.944934 &            -16.944924(32) &   +0.000009(32) \\
               HF &       0 &     0 & ccecp-ccpvqz &             -24.597194 &            -24.597211(32) &   -0.000017(32) \\
         NH$_{3}$ &       0 &     0 & ccecp-ccpvqz &             -11.460900 &            -11.460893(30) &   +0.000007(30) \\
              LiF &       0 &     0 & ccecp-ccpvqz &             -24.281560 &            -24.281576(25) &   -0.000016(25) \\
               CN &       0 &     1 & ccecp-ccpvqz &             -15.088891 &            -15.088881(38) &   +0.000010(38) \\
   C$_{2}$H$_{2}$ &       0 &     0 & ccecp-ccpvqz &             -12.115888 &            -12.115899(27) &   -0.000010(27) \\
               CO &       0 &     0 & ccecp-ccpvqz &            -144.264994 &           -144.264997(55) &   -0.000003(55) \\
              HCN &       0 &     0 & ccecp-ccpvqz &             -15.781632 &            -15.781644(30) &   -0.000012(30) \\
          N$_{2}$ &       0 &     0 & ccecp-ccpvqz &             -19.463663 &            -19.463643(38) &   +0.000020(38) \\
              HCO &       0 &     1 & ccecp-ccpvqz &             -21.805381 &            -21.805384(32) &   -0.000002(32) \\
               NO &       0 &     1 & ccecp-ccpvqz &             -25.412957 &            -25.412974(30) &   -0.000018(30) \\
   C$_{2}$H$_{4}$ &       0 &     0 & ccecp-ccpvqz &             -13.329730 &            -13.329719(33) &   +0.000011(33) \\
        H$_{2}$CO &       0 &     0 & ccecp-ccpvqz &             -22.431967 &            -22.431958(27) &   +0.000009(27) \\
          O$_{2}$ &       0 &     2 & ccecp-ccpvqz &             -31.420428 &            -31.420430(32) &   -0.000002(32) \\
   C$_{2}$H$_{6}$ &       0 &     0 & ccecp-ccpvqz &             -14.526751 &            -14.526748(29) &   +0.000003(29) \\
          F$_{2}$ &       0 &     0 & ccecp-ccpvqz &             -47.824583 &            -47.824585(30) &   -0.000002(30) \\
   H$_{2}$O$_{2}$ &       0 &     0 & ccecp-ccpvqz &             -32.606154 &            -32.606174(32) &   -0.000020(32) \\
       H$_{3}$COH &       0 &     0 & ccecp-ccpvqz &             -23.609361 &            -23.609348(31) &   +0.000012(31) \\
   N$_{2}$H$_{4}$ &       0 &     0 & ccecp-ccpvqz &             -21.708394 &            -21.708387(31) &   +0.000007(31) \\
         CO$_{2}$ &       0 &     0 & ccecp-ccpvqz &             -37.113177 &            -37.113187(32) &   -0.000010(32) \\
SiH$_{2}$-singlet &       0 &     0 & ccecp-ccpvqz &              -4.852829 &             -4.852828(33) &   +0.000001(33) \\
SiH$_{2}$-triplet &       0 &     2 & ccecp-ccpvqz &              -4.845542 &             -4.845530(31) &   +0.000012(31) \\
         PH$_{2}$ &       0 &     1 & ccecp-ccpvqz &              -7.508716 &             -7.508714(30) &   +0.000001(30) \\
        SiH$_{3}$ &       0 &     1 & ccecp-ccpvqz &              -5.463594 &             -5.463585(31) &   +0.000009(31) \\
         H$_{2}$S &       0 &     0 & ccecp-ccpvqz &             -11.132951 &            -11.132948(32) &   +0.000004(32) \\
              HCl &       0 &     0 & ccecp-ccpvqz &             -15.320450 &            -15.320478(35) &   -0.000028(35) \\
         PH$_{3}$ &       0 &     0 & ccecp-ccpvqz &              -8.116063 &             -8.116043(30) &   +0.000020(30) \\
        SiH$_{4}$ &       0 &     0 & ccecp-ccpvqz &              -6.086629 &             -6.086614(38) &   +0.000015(38) \\
               CS &       0 &     0 & ccecp-ccpvqz &             -15.402549 &            -15.402531(39) &   +0.000018(39) \\
              SiO &       0 &     0 & ccecp-ccpvqz &             -19.550999 &            -19.551009(28) &   -0.000010(28) \\
               SO &       0 &     2 & ccecp-ccpvqz &             -25.695157 &            -25.695170(38) &   -0.000013(38) \\
              ClO &       0 &     1 & ccecp-ccpvqz &             -30.394450 &            -30.394447(38) &   +0.000003(38) \\
       CH$_{3}$Cl &       0 &     0 & ccecp-ccpvqz &             -21.994015 &            -21.994012(33) &   +0.000003(33) \\
              ClF &       0 &     0 & ccecp-ccpvqz &             -38.652034 &            -38.652008(33) &   +0.000027(33) \\
       H$_{3}$CSH &       0 &     0 & ccecp-ccpvqz &             -17.809387 &            -17.809395(28) &   -0.000007(28) \\
             HOCl &       0 &     0 & ccecp-ccpvqz &             -31.018384 &            -31.018365(28) &   +0.000019(28) \\
         SO$_{2}$ &       0 &     0 & ccecp-ccpvqz &             -41.493266 &            -41.493266(36) &   -0.000000(36) \\
         Na$_{2}$ &       0 &     0 & ccecp-ccpvqz &              -0.372231 &             -0.372232(28) &   -0.000001(28) \\
             NaCl &       0 &     0 & ccecp-ccpvqz &             -14.993974 &            -14.993958(38) &   +0.000016(38) \\
         Si$_{2}$ &       0 &     0 & ccecp-ccpvqz &              -7.320352 &             -7.320365(30) &   -0.000013(30) \\
          P$_{2}$ &       0 &     0 & ccecp-ccpvqz &             -12.742485 &            -12.742494(23) &   -0.000009(23) \\
          S$_{2}$ &       0 &     2 & ccecp-ccpvqz &             -19.920105 &            -19.920101(35) &   +0.000005(35) \\
         Cl$_{2}$ &       0 &     0 & ccecp-ccpvqz &             -29.423830 &            -29.423800(37) &   +0.000030(37) \\
  Si$_{2}$H$_{6}$ &       0 &     0 & ccecp-ccpvqz &             -11.021000 &            -11.021016(31) &   -0.000017(31) \\
          K$_{2}$ &       0 &     0 & ccecp-ccpvqz &             -55.865096 &            -55.865107(33) &   -0.000011(33) \\
         Ne$_{2}$ &       0 &     0 & ccecp-ccpv6z &             -69.417548 &            -69.417569(55) &   -0.000021(55) \\
         O$_{2}$+ &       1 &     1 & ccecp-ccpvqz &             -30.940484 &            -30.940494(26) &   -0.000010(26) \\
        O$_{2}$++ &       2 &     0 & ccecp-ccpvqz &             -29.978460 &            -29.978464(27) &   -0.000003(27) \\
         O$_{2}$- &      -1 &     3 & ccecp-ccpvqz &             -31.164295 &            -31.164306(27) &   -0.000011(27) \\
\hline
\end{longtable}
\newpage
%%%%%%%%%%%%%%%%%%%%%%%%%%%%%%%%%%%%%%%%%%%%%%%%%%%%
\begin{longtable}{cccc|cc|c}
\caption{The comparison of the UHF energies obtained by \pyscf\ and the VMC energies obtained by \tvb\ with the WFs converted from the corresponding \pyscf\ checkpoints files via \trexio\ for the 49 molecules. The error bars refer to $3 \sigma$ of the VMC calculations. The \trexio\ files used for these calculations are available from our public repository (See Sec.~\ref{M-sec:data} in the main text for the details.)}
\label{tab:sanity-check-molecules} \\
\hline
Molecules &  Charge &  Spin &    Basis set & HF energy (Ha) [\pyscf] & HF energy (Ha) [\tvb] & Difference (Ha) \\
\hline 
\endfirsthead

% head
\hline
Molecules &  Charge &  Spin &    Basis set & HF energy (Ha) [\pyscf] & HF energy (Ha) [\tvb] & Difference (Ha) \\
\hline 
\endhead

% foot
\hline\multicolumn{7}{r}{continued.} \\
\endfoot

% lastfoot
\hline\multicolumn{7}{r}{end.} \\
\endlastfoot

% table
                H &       0 &     1 & ccecp-ccpvqz &              -0.500000 &             -0.500011(19) &   -0.000011(19) \\
               Li &       0 &     1 & ccecp-ccpv5z &              -0.196853 &            -0.1968520(20) &  +0.0000008(20) \\
                B &       0 &     1 & ccecp-ccpv6z &              -2.543510 &             -2.543520(37) &   -0.000010(37) \\
                C &       0 &     2 & ccecp-ccpv6z &              -5.319856 &             -5.319848(35) &   +0.000009(35) \\
                N &       0 &     3 & ccecp-ccpv6z &              -9.638600 &             -9.638594(31) &   +0.000005(31) \\
                O &       0 &     2 & ccecp-ccpv6z &             -15.696937 &            -15.696941(30) &   -0.000004(30) \\
                F &       0 &     1 & ccecp-ccpv6z &             -23.943069 &            -23.943065(28) &   +0.000004(28) \\
               Na &       0 &     1 & ccecp-ccpvqz &              -0.186205 &            -0.1862063(70) &  -0.0000013(70) \\
               Al &       0 &     1 & ccecp-ccpvqz &              -1.881133 &             -1.881133(29) &   +0.000000(29) \\
               Si &       0 &     2 & ccecp-ccpvqz &              -3.677013 &             -3.677011(28) &   +0.000002(28) \\
                P &       0 &     3 & ccecp-ccpvqz &              -6.341271 &             -6.341273(29) &   -0.000001(29) \\
                S &       0 &     2 & ccecp-ccpvqz &              -9.926822 &             -9.926832(28) &   -0.000010(28) \\
               Cl &       0 &     1 & ccecp-ccpvqz &             -14.697582 &            -14.697592(26) &   -0.000011(26) \\
                K &       0 &     1 & ccecp-ccpvqz &             -27.934701 &            -27.934716(33) &   -0.000016(33) \\
               Sc &       0 &     1 & ccecp-ccpvtz &             -46.124357 &            -46.124365(30) &   -0.000008(30) \\
               Ti &       0 &     2 & ccecp-ccpvtz &             -57.614825 &            -57.614826(28) &   -0.000001(28) \\
                V &       0 &     3 & ccecp-ccpvtz &             -70.892874 &            -70.892876(29) &   -0.000002(29) \\
               Cr &       0 &     6 & ccecp-ccpvtz &             -86.048802 &            -86.048803(31) &   -0.000002(31) \\
               Mn &       0 &     5 & ccecp-ccpvdz &            -101.744686 &           -101.744694(34) &   -0.000008(34) \\
               Fe &       0 &     4 & ccecp-ccpvtz &            -122.654507 &           -122.654504(40) &   +0.000002(40) \\
               Co &       0 &     3 & ccecp-ccpvqz &            -144.271266 &           -144.271273(55) &   -0.000007(55) \\
               Ni &       0 &     2 & ccecp-ccpvtz &            -168.486062 &           -168.486075(67) &   -0.000012(67) \\
               Cu &       0 &     1 & ccecp-ccpvtz &            -195.337588 &           -195.337586(65) &   +0.000002(65) \\
               Ga &       0 &     1 & ccecp-ccpvqz &              -1.987373 &             -1.987371(33) &   +0.000002(33) \\
               Ge &       0 &     2 & ccecp-ccpvqz &              -3.668481 &             -3.668463(26) &   +0.000018(26) \\
               As &       0 &     3 & ccecp-ccpvqz &              -6.066117 &             -6.066098(37) &   +0.000019(37) \\
               Se &       0 &     2 & ccecp-ccpvqz &              -9.155003 &             -9.155017(32) &   -0.000014(32) \\
               Br &       0 &     1 & ccecp-ccpvqz &             -13.126572 &            -13.126580(28) &   -0.000008(28) \\
              He+ &       1 &     1 & ccecp-ccpvqz &              -1.999787 &             -1.999782(28) &   +0.000005(28) \\
              He- &      -1 &     1 & ccecp-ccpvqz &              -2.253591 &             -2.253590(29) &   +0.000001(29) \\
              BeH &       0 &     1 & ccecp-ccpvqz &              -1.544551 &             -1.544556(39) &   -0.000005(39) \\
               CH &       0 &     1 & ccecp-ccpvqz &              -5.910949 &             -5.910948(28) &   +0.000002(28) \\
 CH$_{2}$-triplet &       0 &     2 & ccecp-ccpvqz &              -6.570105 &             -6.570116(30) &   -0.000011(30) \\
               NH &       0 &     2 & ccecp-ccpvqz &             -10.220726 &            -10.220740(39) &   -0.000014(39) \\
         CH$_{3}$ &       0 &     1 & ccecp-ccpvqz &              -7.210633 &             -7.210627(33) &   +0.000006(33) \\
         NH$_{2}$ &       0 &     1 & ccecp-ccpvqz &             -10.827123 &            -10.827132(35) &   -0.000009(35) \\
               OH &       0 &     1 & ccecp-ccpvqz &             -16.305559 &            -16.305564(28) &   -0.000006(28) \\
               CN &       0 &     1 & ccecp-ccpvqz &             -15.107056 &            -15.107047(29) &   +0.000009(29) \\
              HCO &       0 &     1 & ccecp-ccpvqz &             -21.811238 &            -21.811259(39) &   -0.000021(39) \\
               NO &       0 &     1 & ccecp-ccpvqz &             -25.421073 &            -25.421057(27) &   +0.000017(27) \\
          O$_{2}$ &       0 &     2 & ccecp-ccpvqz &             -31.444494 &            -31.444481(27) &   +0.000013(27) \\
SiH$_{2}$-triplet &       0 &     2 & ccecp-ccpvqz &              -4.847604 &             -4.847610(31) &   -0.000006(31) \\
         PH$_{2}$ &       0 &     1 & ccecp-ccpvqz &              -7.514646 &             -7.514636(23) &   +0.000010(23) \\
        SiH$_{3}$ &       0 &     1 & ccecp-ccpvqz &              -5.465299 &             -5.465287(30) &   +0.000012(30) \\
               SO &       0 &     2 & ccecp-ccpvqz &             -25.712539 &            -25.712538(32) &   +0.000001(32) \\
              ClO &       0 &     1 & ccecp-ccpvqz &             -30.402361 &            -30.402377(30) &   -0.000016(30) \\
          S$_{2}$ &       0 &     2 & ccecp-ccpvqz &             -19.935972 &            -19.935957(33) &   +0.000015(33) \\
         O$_{2}$+ &       1 &     1 & ccecp-ccpvqz &             -30.951269 &            -30.951288(25) &   -0.000019(25) \\
         O$_{2}$- &      -1 &     3 & ccecp-ccpvqz &             -31.191031 &            -31.191034(29) &   -0.000003(29) \\
\hline
\end{longtable}
\newpage
%%%%%%%%%%%%%%%%%%%%%%%%%%%%%%%%%%%%%%%%%%%%%%%%%%%%
\begin{table}[htpb]
 \caption{The comparison of the RHF(ROHF) energies obtained by \pyscf\ and the VMC energies obtained by \tvb\ with the WFs converted from the corresponding \pyscf\ checkpoints files via \trexio\ for 9 crystals at $k$ = $\Gamma$. The error bars refer to $3 \sigma$ of the VMC calculations. The \trexio\ files used for these calculations are available from our public repository (See Sec.~\ref{M-sec:data} in the main text for the details.)}
 \label{tab:sanity-check-crystals-k-gamma}
 \centering
\begin{tabular}{cccccc|cc|c}
\hline
  CODID &         Crystal & Crystalsystem &  Spacegroup &  Charge &  Spin & HF energy (Ha) [pySCF] & HF energy (Ha) [\tvb] & Difference (Ha) \\
\hline
1011097 & alpha-SiO$_{2}$ &      trigonal &         152 &       0 &     0 &            -106.707420 &            -106.70747(21) &    -0.00005(21) \\
1526655 &              Si &         cubic &         227 &       0 &     0 &             -30.097653 &             -30.09766(15) &    -0.00001(15) \\
2101499 &         Diamond &         cubic &         227 &       0 &     0 &             -43.928111 &             -43.92810(15) &    +0.00001(15) \\
9008997 &            h-BN &     hexagonal &         194 &       0 &     0 &             -24.883084 &             -24.88315(13) &    -0.00007(13) \\
9011660 &            w-BN &     hexagonal &         186 &       0 &     0 &             -24.341137 &             -24.34112(15) &    +0.00002(15) \\
9008667 &             LiF &         cubic &         225 &       0 &     0 &             -97.762562 &             -97.76255(22) &    +0.00001(22) \\
1000053 &             MgO &         cubic &         225 &       0 &     0 &             -66.958279 &             -66.95830(16) &    -0.00003(16) \\
1528336 &               O &    monoclinic &          12 &       0 &     4 &             -62.927947 &             -62.92795(19) &    -0.00001(19) \\
9008830 &            AlAs &         cubic &         216 &       0 &     0 &             -32.418408 &             -32.41850(15) &    -0.00009(15) \\
\hline
\end{tabular}
\end{table}
%%%%%%%%%%%%%%%%%%%%%%%%%%%%%%%%%%%%%%%%%%%%%%%%%%%%
\begin{table}[htpb]
 \caption{The comparison of the RHF(ROHF) energies obtained by \pyscf\ and the VMC energies obtained by \tvb\ with the WFs converted from the corresponding \pyscf\ checkpoints files via \trexio\ for 9 crystals at $k$ = (0.25, 0.25, 0.25). The error bars refer to $3 \sigma$ of the VMC calculations. The \trexio\ files used for these calculations are available from our public repository (See Sec.~\ref{M-sec:data} in the main text for the details.)}
 \label{tab:sanity-check-crystals-k-complex}
 \centering
\begin{tabular}{cccccc|cc|c}
\hline
  CODID &         Crystal & Crystalsystem &  Spacegroup &  Charge &  Spin & HF energy (Ha) [\pyscf] & HF energy (Ha) [\tvb] & Difference (Ha) \\
\hline
1011097 & alpha-SiO$_{2}$ &      trigonal &         152 &       0 &     0 &            -106.737229 &            -106.73730(24) &    -0.00007(24) \\
1526655 &              Si &         cubic &         227 &       0 &     0 &             -30.474552 &             -30.47466(19) &    -0.00010(19) \\
2101499 &         Diamond &         cubic &         227 &       0 &     0 &             -44.575966 &             -44.57597(14) &    -0.00000(14) \\
9008997 &            h-BN &     hexagonal &         194 &       0 &     0 &             -25.494781 &             -25.49480(13) &    -0.00001(13) \\
9011660 &            w-BN &     hexagonal &         186 &       0 &     0 &             -25.754344 &             -25.75435(14) &    -0.00001(14) \\
9008667 &             LiF &         cubic &         225 &       0 &     0 &             -97.719914 &             -97.71994(25) &    -0.00003(25) \\
1000053 &             MgO &         cubic &         225 &       0 &     0 &             -67.174487 &             -67.17452(18) &    -0.00003(18) \\
1528336 &               O &    monoclinic &          12 &       0 &     4 &             -62.950168 &             -62.95008(16) &    +0.00009(16) \\
9008830 &            AlAs &         cubic &         216 &       0 &     0 &             -32.697346 &             -32.69733(16) &    +0.00002(16) \\
\hline
\end{tabular}
\end{table}
%%%%%%%%%%%%%%%%%%%%%%%%%%%%%%%%%%%%%%%%%%%%%%%%%%%%
\begin{table}[htpb]
 \caption{The comparison of the RHF(ROHF) energies obtained by \pyscf\ and the VMC energies obtained by \tvb\ with the WFs converted from the corresponding \pyscf\ checkpoints files via \trexio\ for 9 crystals with twisted average ($k$ = 4 $\times$ 4 $\times$ 4). The error bars refer to $3 \sigma$ of the VMC calculations. The \trexio\ files used for these calculations are available from our public repository (See Sec.~\ref{M-sec:data} in the main text for the details.)}
 \label{tab:sanity-check-crystals-k-twist}
 \centering
\begin{tabular}{cccccc|cc|c}
\hline
  CODID &         Crystal & Crystalsystem &  Spacegroup &  Charge &  Spin & HF energy (Ha) [\pyscf] & HF energy (Ha) [\tvb] & Difference (Ha) \\
\hline
1011097 & alpha-SiO$_{2}$ &      trigonal &         152 &       0 &     0 &            -106.730695 &            -106.73079(34) &    -0.00010(34) \\
1526655 &              Si &         cubic &         227 &       0 &     0 &             -30.496624 &            -30.496662(96) &   -0.000038(96) \\
2101499 &         Diamond &         cubic &         227 &       0 &     0 &             -44.593443 &             -44.59338(19) &    +0.00007(19) \\
9008997 &            h-BN &     hexagonal &         194 &       0 &     0 &             -25.372483 &            -25.372477(92) &   +0.000006(92) \\
9011660 &            w-BN &     hexagonal &         186 &       0 &     0 &             -25.436420 &            -25.436378(98) &   +0.000041(98) \\
9008667 &             LiF &         cubic &         225 &       0 &     0 &             -97.719191 &             -97.71901(36) &    +0.00018(36) \\
1000053 &             MgO &         cubic &         225 &       0 &     0 &             -67.178425 &             -67.17845(26) &    -0.00003(26) \\
1528336 &               O &    monoclinic &          12 &       0 &     4 &             -62.958163 &             -62.95806(14) &    +0.00010(14) \\
9008830 &            AlAs &         cubic &         216 &       0 &     0 &             -32.711201 &             -32.71116(12) &    +0.00004(12) \\
\hline
\end{tabular}
\end{table}
%%%%%%%%%%%%%%%%%%%%%%%%%%%%%%%%%%%%%%%%%%%%%%%%%%%%
\newpage

%%%%%%%%%%%%%%%%%%%%%%%%%%%%%%%%%%%%%%%%%%%%%%%%%%%%
\begin{figure}[htbp]
  \centering
  \includegraphics[width=\textwidth]{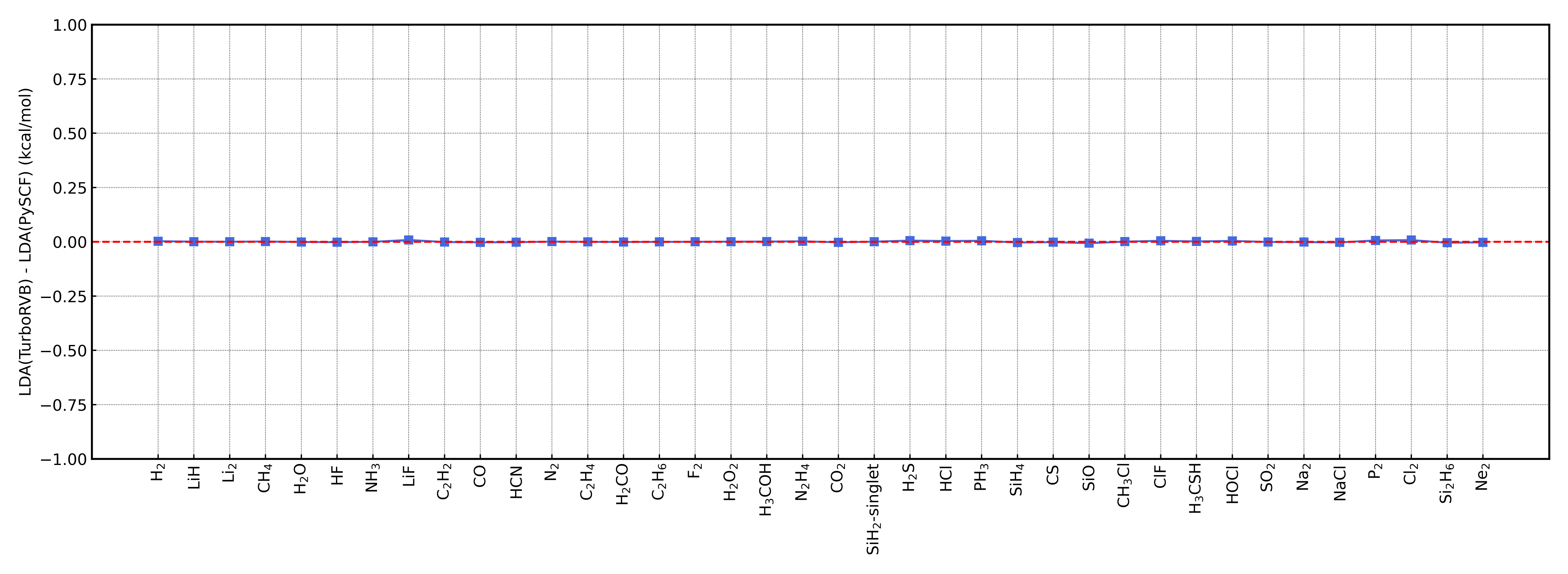}
  \caption{The comparison of the LDA-PZ energies obtained by \pyscf\ and those energies obtained by \tvb-prep with the same basis sets for 38 molecules.}
  \label{fig:sanity-check-molecules-LDA}
\end{figure}
%%%%%%%%%%%%%%%%%%%%%%%%%%%%%%%%%%%%%%%%%%%%%%%%%%%%
\begin{figure}[htbp]
  \centering
  \includegraphics[width=0.5\columnwidth]{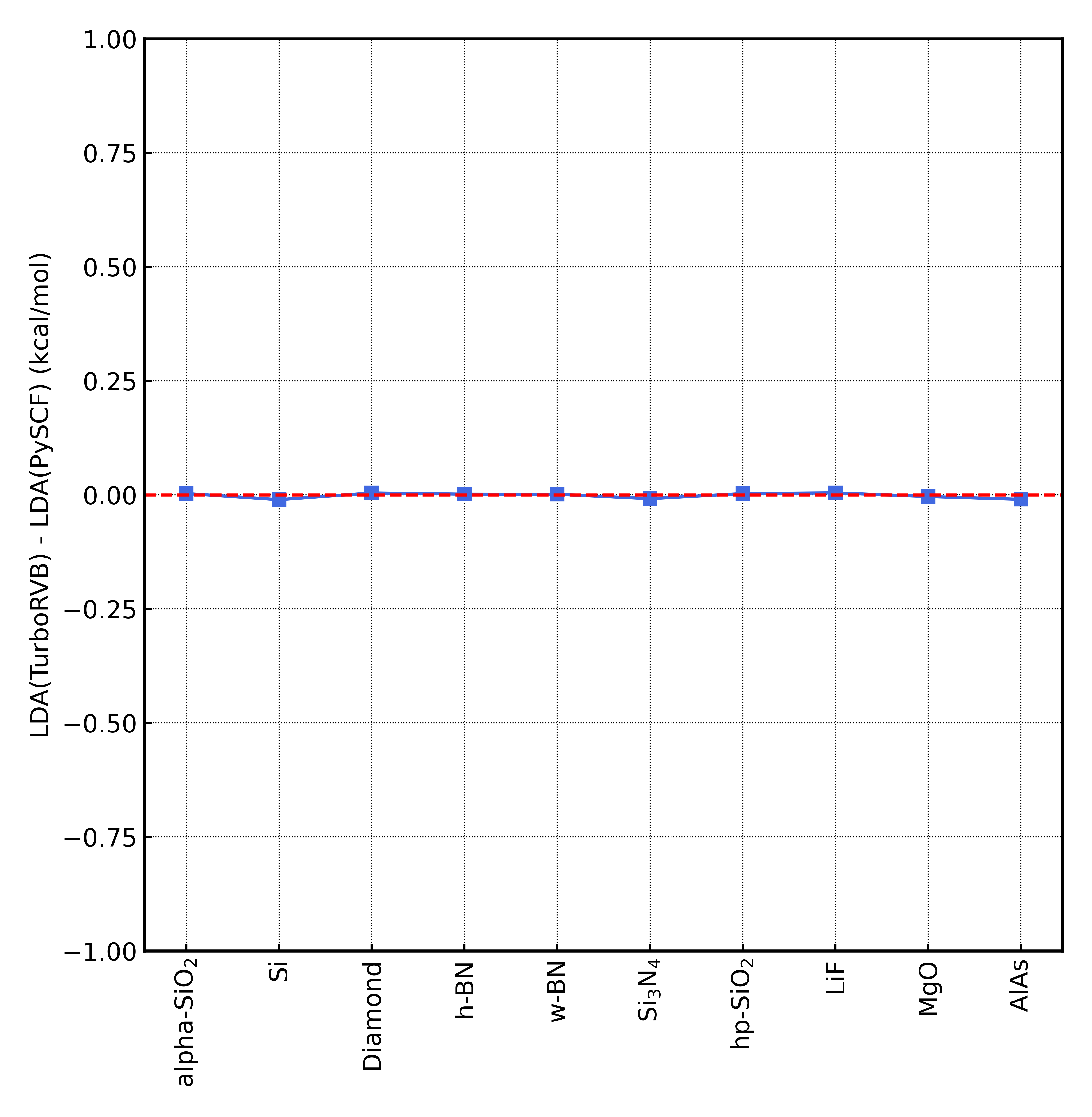}
  \caption{The comparison of the LDA-PZ energies obtained by \pyscf\ and those energies obtained by \tvb-prep with the same basis sets for 10 crystals.}
  \label{fig:sanity-check-crystals-LDA-pbc-kgrid-insulator}
\end{figure}
%%%%%%%%%%%%%%%%%%%%%%%%%%%%%%%%%%%%%%%%%%%%%%%%%%%%
\begin{figure}[htbp]
  \centering
  \includegraphics[width=0.5\columnwidth]{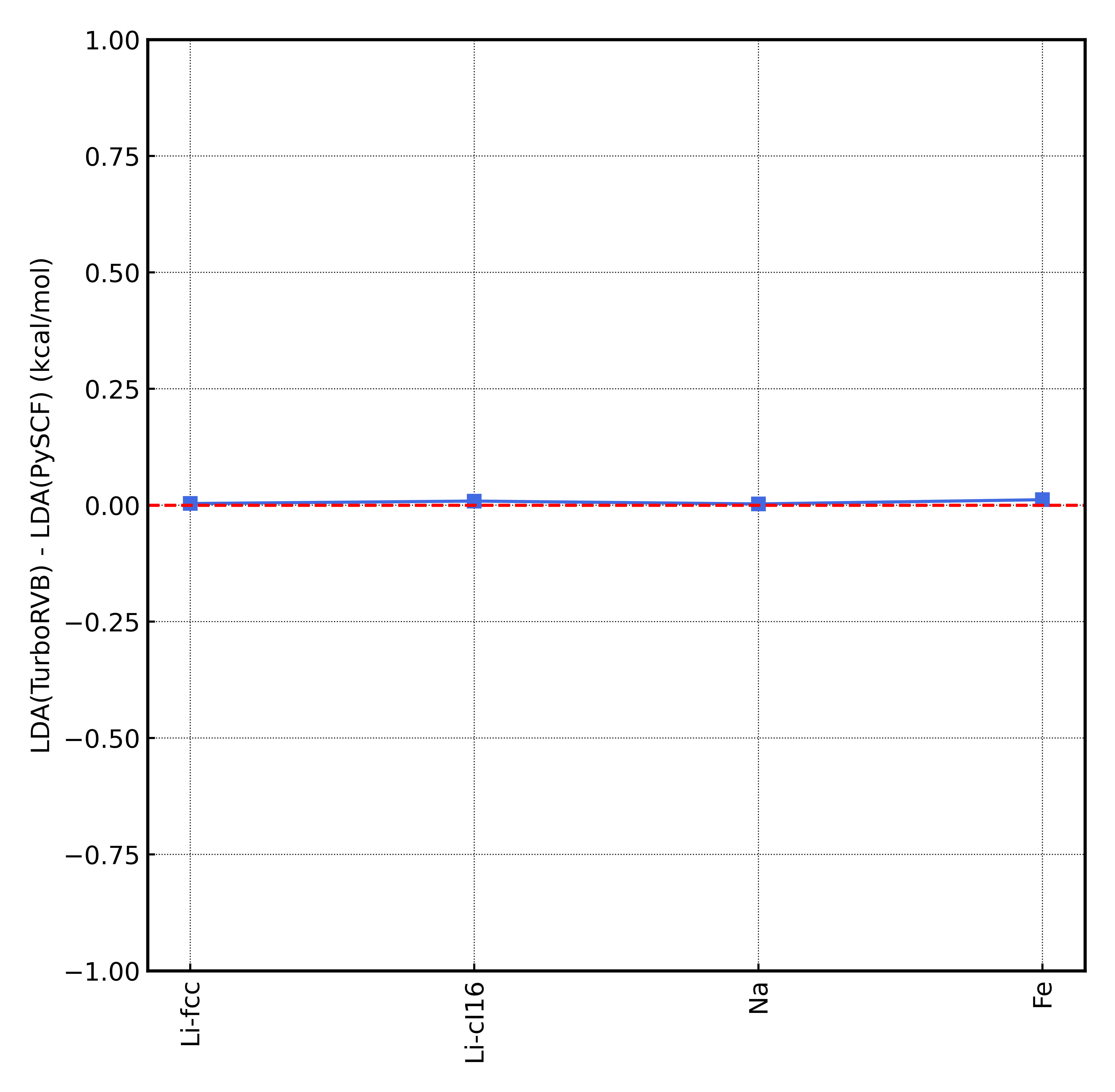}
  \caption{The comparison of the LDA-PZ energies obtained by \pyscf\ and those energies obtained by \tvb-prep with the same basis sets for 4 crystals.}
  \label{fig:sanity-check-crystals-LDA-pbc-kgrid-metal}
\end{figure}
%%%%%%%%%%%%%%%%%%%%%%%%%%%%%%%%%%%%%%%%%%%%%%%%%%%%
\begin{figure}[htbp]
  \centering
  \includegraphics[width=\textwidth]{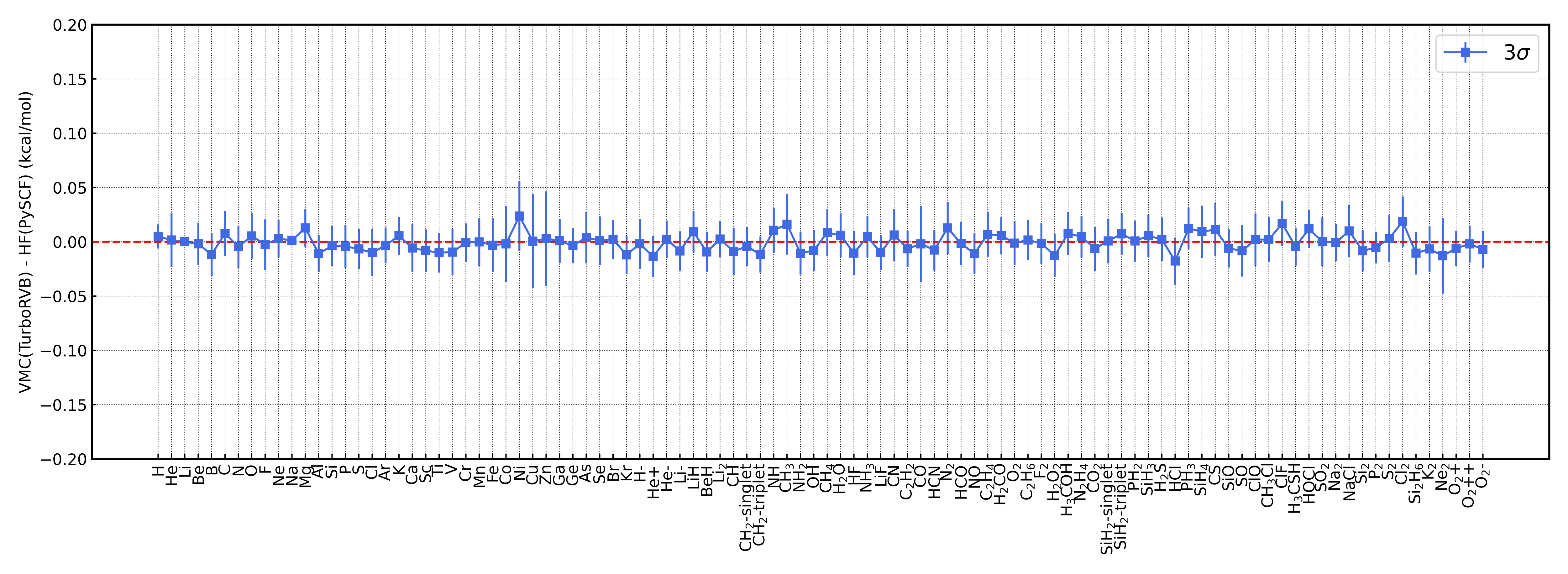}
  \caption{The comparison of the RHF(ROHF) energies obtained by \pyscf\ and the VMC energies obtained by \tvb\ with the WFs converted from the corresponding \pyscf\ checkpoints files via \trexio\ for 100 molecules. The error bars refer to $3 \sigma$ of the VMC calculations.}
  \label{fig:sanity-check-molecules-RHF}
\end{figure}
%%%%%%%%%%%%%%%%%%%%%%%%%%%%%%%%%%%%%%%%%%%%%%%%%%%%
\begin{figure}[htbp]
  \centering
  \includegraphics[width=\textwidth]{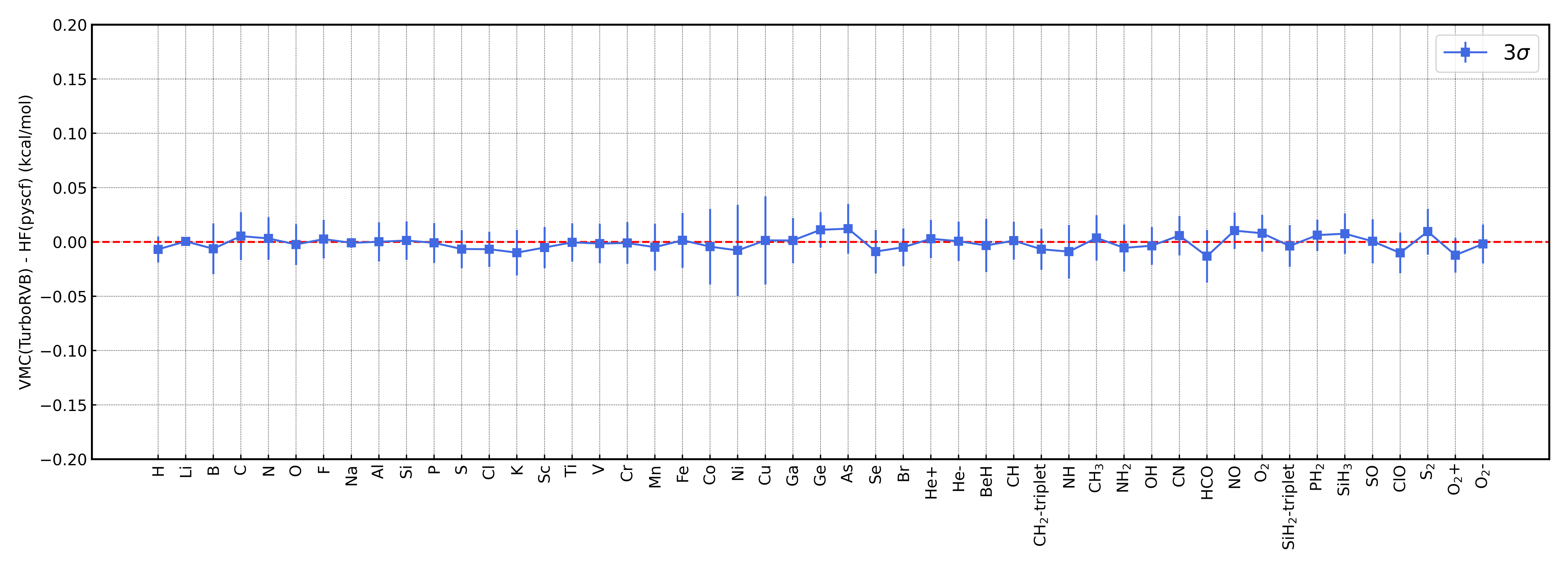}
  \caption{The comparison of the UHF energies obtained by \pyscf\ and the VMC energies obtained by \tvb\ with the WFs converted from the corresponding \pyscf\ checkpoints files via \trexio\ for 49 molecules. The error bars refer to $3 \sigma$ of the VMC calculations.}
  \label{fig:sanity-check-molecules-UHF}
\end{figure}
%%%%%%%%%%%%%%%%%%%%%%%%%%%%%%%%%%%%%%%%%%%%%%%%%%%%
\begin{figure}[htbp]
  \centering
  \includegraphics[width=0.5\columnwidth]{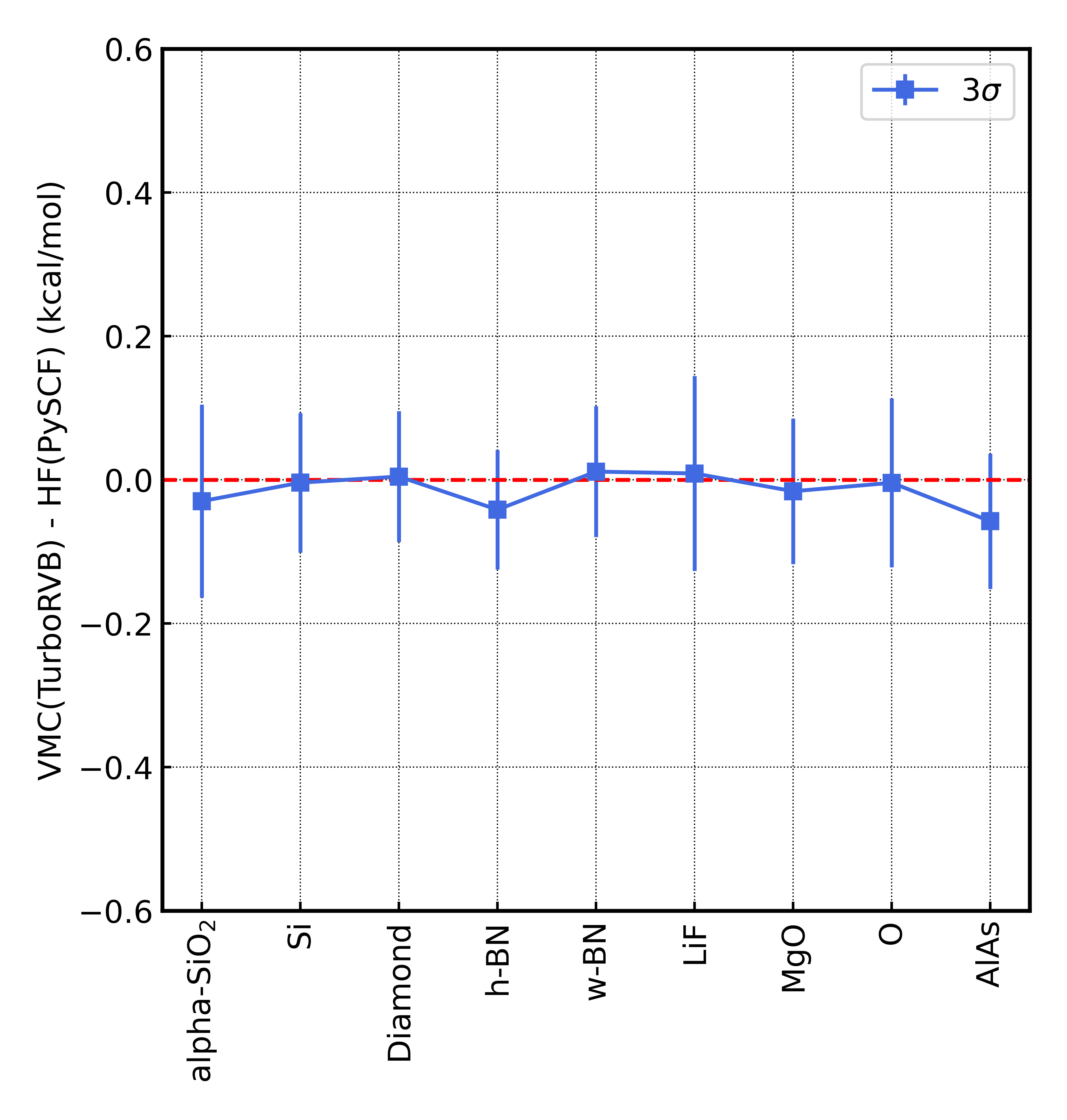}
  \caption{The comparison of the RHF(ROHF) energies obtained by \pyscf\ and the VMC energies obtained by \tvb\ with the WFs converted from the corresponding \pyscf\ checkpoints files via \trexio\ for 9 crystals at $k$ = $\Gamma$). The error bars refer to $3 \sigma$ of the VMC calculations.}
  \label{fig:sanity-check-crystals-k-gamma}
\end{figure}
%%%%%%%%%%%%%%%%%%%%%%%%%%%%%%%%%%%%%%%%%%%%%%%%%%%%
\begin{figure}[htbp]
  \centering
  \includegraphics[width=0.5\columnwidth]{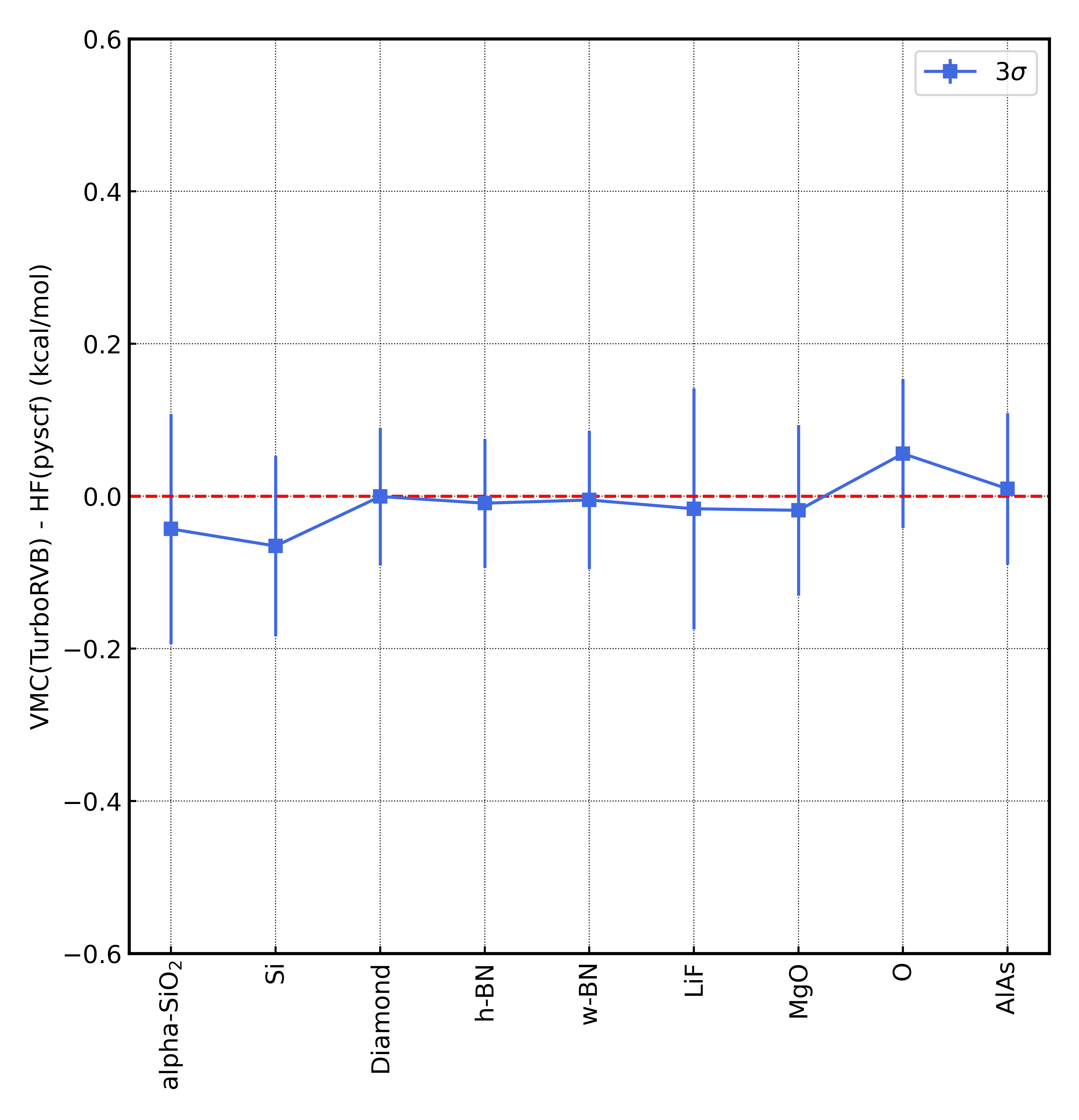}
  \caption{The comparison of the RHF(ROHF) energies obtained by \pyscf\ and the VMC energies obtained by \tvb\ with the WFs converted from the corresponding \pyscf\ checkpoints files via \trexio\ for 9 crystals at $k$ = (0.25, 0.25, 0.25)). The error bars refer to $3 \sigma$ of the VMC calculations.}
  \label{fig:sanity-check-crystals-k-complex}
\end{figure}
%%%%%%%%%%%%%%%%%%%%%%%%%%%%%%%%%%%%%%%%%%%%%%%%%%%%
\begin{figure}[htbp]
  \centering
  \includegraphics[width=0.5\columnwidth]{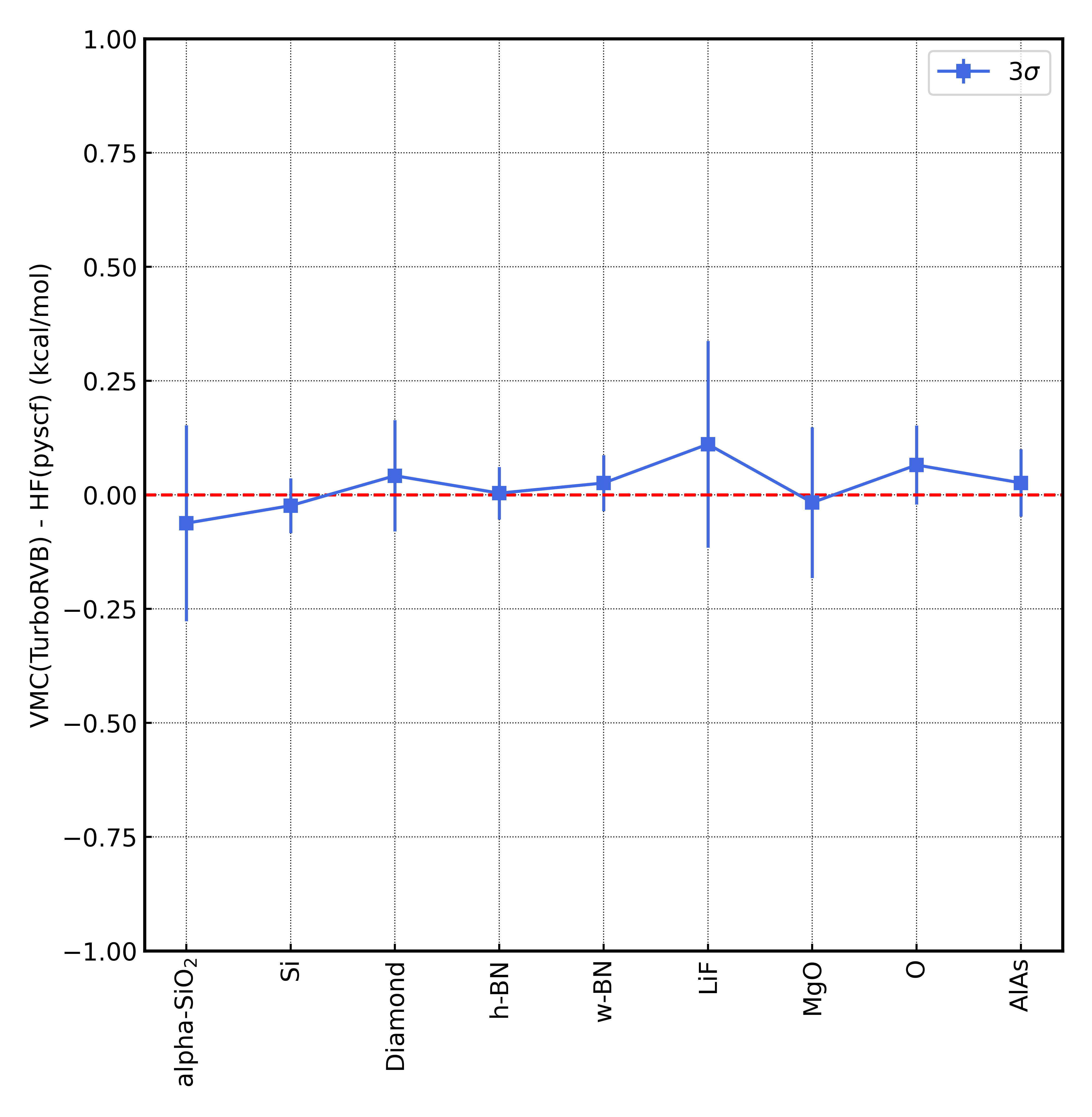}
  \caption{The comparison of the RHF(ROHF) energies obtained by \pyscf\ and the VMC energies obtained by \tvb\ with the WFs converted from the corresponding \pyscf\ checkpoints files via \trexio\ for 9 crystals with twisted average ($k$ = 4 $\times$ 4 $\times$ 4). The error bars refer to $3 \sigma$ of the VMC calculations.}
  \label{fig:sanity-check-crystals-k-twist}
\end{figure}
%%%%%%%%%%%%%%%%%%%%%%%%%%%%%%%%%%%%%%%%%%%%%%%%%%%%

\newpage

%%%%%%%%%%%%%%%%%%%%%%%%%%%%%%%%%%%%%%%%%%%%%%%%%%%%
\begin{table*}[h]
\caption{The atomization energies of 55 molecule contained in the G2-set. They were computed by LRDMC with the LDA-PZ nodal surfaces and the DLA. The experimental values and the differences between the LRDMC and experimental values are also shown.}
 \label{tab:G2-set}
 \centering
\begin{tabular}{llll}
\hline
    Molecule & LRDMC (kcal/mol) &  Experiment (kcal/mol) & Diff. (kcal/mol) \\
\hline
              LiH &                       57.85(17) &                                  58.00 &             -0.15(17) \\
              BeH &                       55.34(17) &                                  49.80 &              5.54(17) \\
         Li$_{2}$ &                       22.58(26) &                                  24.40 &             -1.82(26) \\
               CH &                       83.40(18) &                                  84.00 &             -0.60(18) \\
 CH$_{2}$-singlet &                      181.55(18) &                                 181.30 &              0.25(18) \\
 CH$_{2}$-triplet &                      196.26(17) &                                 190.80 &              5.46(17) \\
               NH &                       82.91(16) &                                  83.80 &             -0.89(16) \\
         CH$_{3}$ &                      313.79(17) &                                 308.00 &              5.79(17) \\
         NH$_{2}$ &                      183.57(18) &                                 182.20 &              1.37(18) \\
               OH &                      107.41(17) &                                 106.80 &              0.61(17) \\
         CH$_{4}$ &                      427.31(17) &                                 420.60 &              6.71(17) \\
         H$_{2}$O &                      233.64(16) &                                 233.35 &              0.29(16) \\
               HF &                      142.18(17) &                                 141.70 &              0.48(17) \\
         NH$_{3}$ &                      300.80(18) &                                 298.40 &              2.40(18) \\
              LiF &                      140.07(21) &                                 139.50 &              0.57(21) \\
               CN &                      179.35(21) &                                 181.20 &             -1.85(21) \\
   C$_{2}$H$_{2}$ &                      414.55(30) &                                 403.10 &             11.45(30) \\
               CO &                      260.04(21) &                                 259.80 &              0.24(21) \\
              HCN &                      316.75(21) &                                 311.80 &              4.95(21) \\
          N$_{2}$ &                      224.82(27) &                                 228.60 &             -3.78(27) \\
              HCO &                      281.62(22) &                                 279.00 &              2.62(22) \\
               NO &                      147.49(21) &                                 153.16 &             -5.67(21) \\
   C$_{2}$H$_{4}$ &                      573.69(28) &                                 564.10 &              9.59(28) \\
        H$_{2}$CO &                      378.84(22) &                                 374.50 &              4.34(22) \\
          O$_{2}$ &                      115.99(26) &                                 120.86 &             -4.87(26) \\
   C$_{2}$H$_{6}$ &                      726.17(28) &                                 713.40 &             12.77(28) \\
          F$_{2}$ &                       31.93(27) &                                  39.10 &             -7.17(27) \\
   H$_{2}$O$_{2}$ &                      266.00(26) &                                 269.60 &             -3.60(26) \\
       H$_{3}$COH &                      520.51(22) &                                 513.60 &              6.91(22) \\
   N$_{2}$H$_{4}$ &                      442.77(25) &                                 438.80 &              3.97(25) \\
         CO$_{2}$ &                      395.56(29) &                                 390.23 &              5.33(29) \\
SiH$_{2}$-singlet &                      153.01(16) &                                 152.30 &              0.71(16) \\
SiH$_{2}$-triplet &                      133.36(19) &                                 131.70 &              1.66(19) \\
         PH$_{2}$ &                      153.41(18) &                                 153.30 &              0.11(18) \\
        SiH$_{3}$ &                      228.81(17) &                                 227.80 &              1.01(17) \\
         H$_{2}$S &                      182.07(18) &                                 183.50 &             -1.43(18) \\
              HCl &                      107.43(18) &                                 107.54 &             -0.11(18) \\
         PH$_{3}$ &                      240.27(16) &                                 244.00 &             -3.73(16) \\
        SiH$_{4}$ &                      325.23(18) &                                 323.00 &              2.23(18) \\
               CS &                      168.81(22) &                                 171.90 &             -3.09(22) \\
              SiO &                      191.17(21) &                                 192.50 &             -1.33(21) \\
               SO &                      124.10(20) &                                 126.10 &             -2.00(20) \\
              ClO &                       62.05(21) &                                  65.92 &             -3.87(21) \\
       CH$_{3}$Cl &                      402.18(21) &                                 395.90 &              6.28(21) \\
              ClF &                       60.25(21) &                                  62.90 &             -2.65(21) \\
       H$_{3}$CSH &                      479.90(21) &                                 474.90 &              5.00(21) \\
             HOCl &                      164.20(21) &                                 166.10 &             -1.90(21) \\
         SO$_{2}$ &                      256.52(30) &                                 259.80 &             -3.28(30) \\
         Na$_{2}$ &                       15.64(28) &                                  17.00 &             -1.36(28) \\
             NaCl &                       99.12(22) &                                  98.90 &              0.22(22) \\
         Si$_{2}$ &                       74.33(27) &                                  75.70 &             -1.37(27) \\
          P$_{2}$ &                      110.89(25) &                                 117.00 &             -6.11(25) \\
          S$_{2}$ &                      100.93(28) &                                 103.06 &             -2.13(28) \\
         Cl$_{2}$ &                       57.88(28) &                                  58.98 &             -1.10(28) \\
  Si$_{2}$H$_{6}$ &                      536.20(27) &                                 532.60 &              3.60(27) \\
\hline
\end{tabular}
\end{table*}
%%%%%%%%%%%%%%%%%%%%%%%%%%%%%%%%%%%%%%%%%%%%%%%%%%%%
%
%%%%%%%%%%%%%%%%%%%%%%%%%%%%%%%%%%%%%%%%%%%%%%%%%%%%
\begin{table*}[h]
\caption{The binding energies of 22 molecule contained in the S22-set. They were computed by LRDMC with the LDA-PZ nodal surfaces and the DLA. Those computed by CCSD(T) and the differences between the LRDMC and CCSD(T) values are also shown.}
 \label{tab:S22-set}
 \centering
\begin{tabular}{llll}
\hline
    Molecule & LRDMC (kcal/mol) &  CCSD(T) (kcal/mol) & Diff. (kcal/mol) \\
\hline
                    81-Ammoniadimer &        -3.68(21) &               -3.17 &             -0.51(21) \\
                      82-Waterdimer &        -5.26(21) &               -5.02 &             -0.24(21) \\
           83-BenzeneMethanecomplex &        -1.77(18) &               -1.50 &             -0.27(18) \\
                 84-Formicaciddimer &       -20.45(23) &              -18.61 &             -1.84(23) \\
                  85-Formamidedimer &       -17.53(21) &              -15.96 &             -1.57(21) \\
              86-Uracildimerhbonded &       -21.47(34) &              -20.47 &             -1.00(34) \\
87-2pyridoxine2aminopyridinecomplex &       -17.80(36) &              -16.71 &             -1.09(36) \\
88-AdeninethymineWatsonCrickcomplex &       -18.03(40) &              -16.37 &             -1.66(40) \\
                    89-Methanedimer &        -0.35(20) &               -0.53 &              0.18(20) \\
                     90-Ethenedimer &        -1.75(23) &               -1.51 &             -0.24(23) \\
   91-Benzenedimerparalleldisplaced &        -2.22(23) &               -2.73 &              0.51(23) \\
                   92-Pyrazinedimer &        -4.37(25) &               -4.42 &              0.05(25) \\
                93-Uracildimerstack &       -10.04(34) &               -9.88 &             -0.16(34) \\
       94-Indolebenzenecomplexstack &        -4.07(28) &               -5.22 &              1.15(28) \\
      95-Adeninethyminecomplexstack &       -11.31(40) &              -12.23 &              0.92(40) \\
             96-Etheneethynecomplex &        -1.47(20) &               -1.53 &              0.06(20) \\
             97-Benzenewatercomplex &        -3.73(21) &               -3.28 &             -0.45(21) \\
           98-Benzeneammoniacomplex &        -3.09(21) &               -2.35 &             -0.74(21) \\
               99-BenzeneHCNcomplex &        -4.51(21) &               -4.46 &             -0.05(21) \\
            100-BenzenedimerTshaped &        -2.83(22) &               -2.74 &             -0.09(22) \\
     101-IndolebenzeneTshapecomplex &        -5.57(26) &               -5.73 &              0.16(26) \\
                    102-Phenoldimer &        -7.54(28) &               -7.05 &             -0.49(28) \\
\hline
\end{tabular}
\end{table*}
%%%%%%%%%%%%%%%%%%%%%%%%%%%%%%%%%%%%%%%%%%%%%%%%%%%%
%
%%%%%%%%%%%%%%%%%%%%%%%%%%%%%%%%%%%%%%%%%%%%%%%%%%%%
\begin{table*}[h]
\caption{The binding energies of 24 molecule contained in the A24-set. They were computed by LRDMC with the LDA-PZ nodal surfaces and the DLA. Those computed by CCSD(T) and the differences between the LRDMC and CCSD(T) values are also shown.}
 \label{tab:A24-set}
 \centering
\begin{tabular}{llll}
\hline
    Molecule & LRDMC (kcal/mol) &  CCSD(T) (kcal/mol) & Diff. (kcal/mol) \\
\hline
      3957-01waterammonia &        -7.25(20) &              -6.493 &             -0.76(20) \\
        3958-02waterdimer &        -5.34(22) &              -5.006 &             -0.33(22) \\
          3959-03HCNdimer &        -5.34(20) &              -4.745 &             -0.60(20) \\
           3960-04HFdimer &        -4.94(21) &              -4.581 &             -0.36(21) \\
      3961-05ammoniadimer &        -3.91(22) &              -3.137 &             -0.77(22) \\
         3962-06HFmethane &        -1.92(17) &              -1.654 &             -0.26(17) \\
    3963-07ammoniamethane &        -1.09(19) &              -0.765 &             -0.32(19) \\
      3964-08watermethane &        -0.85(19) &              -0.663 &             -0.18(19) \\
 3965-09formaldehydedimer &        -4.81(21) &              -4.554 &             -0.26(21) \\
       3966-10waterethene &        -2.73(18) &              -2.557 &             -0.17(18) \\
3967-11formaldehydeethene &        -2.27(21) &              -1.621 &             -0.65(21) \\
       3968-12ethynedimer &        -1.15(22) &              -1.524 &              0.38(22) \\
     3969-13ammoniaethene &        -1.01(22) &              -1.374 &              0.36(22) \\
       3970-14ethenedimer &        -1.14(23) &              -1.090 &             -0.05(23) \\
     3971-15methaneethene &        -0.61(18) &              -0.502 &             -0.11(18) \\
     3972-16boranemethane &        -1.03(19) &              -1.485 &              0.46(19) \\
     3973-17methaneethane &        -0.47(19) &              -0.827 &              0.36(19) \\
     3974-18methaneethane &        -0.61(20) &              -0.607 &             -0.00(20) \\
      3975-19methanedimer &        -0.63(20) &              -0.533 &             -0.09(20) \\
         3976-20Armethane &        -0.46(18) &              -0.405 &             -0.06(18) \\
          3977-21Arethene &         0.23(18) &              -0.364 &              0.59(18) \\
      3978-22etheneethyne &         1.02(22) &               0.821 &              0.20(22) \\
       3979-23ethenedimer &         0.91(24) &               0.934 &             -0.02(24) \\
       3980-24ethynedimer &         1.31(22) &               1.115 &              0.20(22) \\
\hline
\end{tabular}
\end{table*}
%%%%%%%%%%%%%%%%%%%%%%%%%%%%%%%%%%%%%%%%%%%%%%%%%%%%
%
%%%%%%%%%%%%%%%%%%%%%%%%%%%%%%%%%%%%%%%%%%%%%%%%%%%%
\begin{table*}[h]
\caption{The binding energies of 24 molecule contained in the SCAI-set. They were computed by LRDMC with the LDA-PZ nodal surfaces and the DLA. Those computed by CCSD(T) and the differences between the LRDMC and CCSD(T) values are also shown.}
 \label{tab:SCAI-set}
 \centering
\begin{tabular}{llll}
\hline
    Molecule & LRDMC (kcal/mol) &  CCSD(T) (kcal/mol) & Diff. (kcal/mol) \\
\hline
 702-RD1 &      -111.64(29) &             -110.80 &             -0.84(29) \\
 703-KE1 &      -109.00(27) &             -108.40 &             -0.60(27) \\
 704-DH1 &       -31.97(26) &              -30.64 &             -1.33(26) \\
705-DHN1 &       -20.12(26) &              -17.97 &             -2.15(26) \\
706-RDN1 &       -16.60(30) &              -16.32 &             -0.28(30) \\
707-KEN1 &       -11.48(28) &              -10.76 &             -0.72(28) \\
 708-QN1 &        -7.27(25) &               -7.37 &              0.10(25) \\
 709-TT1 &        -6.91(22) &               -6.50 &             -0.41(22) \\
 710-YY1 &        -4.55(39) &               -4.66 &              0.11(39) \\
 711-TS1 &        -4.79(22) &               -4.50 &             -0.29(22) \\
 712-LW1 &        -4.37(36) &               -4.04 &             -0.33(36) \\
 713-YP1 &        -3.70(33) &               -3.79 &              0.09(33) \\
 714-FF1 &        -2.26(30) &               -2.33 &              0.07(30) \\
 715-MM1 &        -1.78(24) &               -2.03 &              0.25(24) \\
 716-LY1 &        -1.73(32) &               -1.72 &             -0.01(32) \\
 717-LL1 &        -1.76(26) &               -1.62 &             -0.14(26) \\
 718-MC1 &        -1.24(21) &               -1.46 &              0.22(21) \\
 719-VV1 &        -1.28(24) &               -1.39 &              0.11(24) \\
 720-IL1 &        -0.96(25) &               -1.39 &              0.43(25) \\
 721-II1 &        -1.09(24) &               -1.24 &              0.15(24) \\
 722-LT1 &        -0.92(23) &               -1.09 &              0.17(23) \\
 723-VL1 &        -0.78(22) &               -1.08 &              0.30(22) \\
 724-AL1 &        -0.92(23) &               -1.07 &              0.15(23) \\
 725-LG1 &        -0.39(28) &               -0.77 &              0.38(28) \\
\hline
\end{tabular}
\end{table*}
%%%%%%%%%%%%%%%%%%%%%%%%%%%%%%%%%%%%%%%%%%%%%%%%%%%%
\newpage
%
%%%%%%%%%%%%%%%%%%%%%%%%%%%%%%%%%%%%%%%%%%%%%%%%%%%%
% EOS, basis check, s=1x1x1
%%%%%%%%%%%%%%%%%%%%%%%%%%%%%%%%%%%%%%%%%%%%%%%%%%%%
\begin{figure}[htbp]
  \centering
  \includegraphics[width=\textwidth]{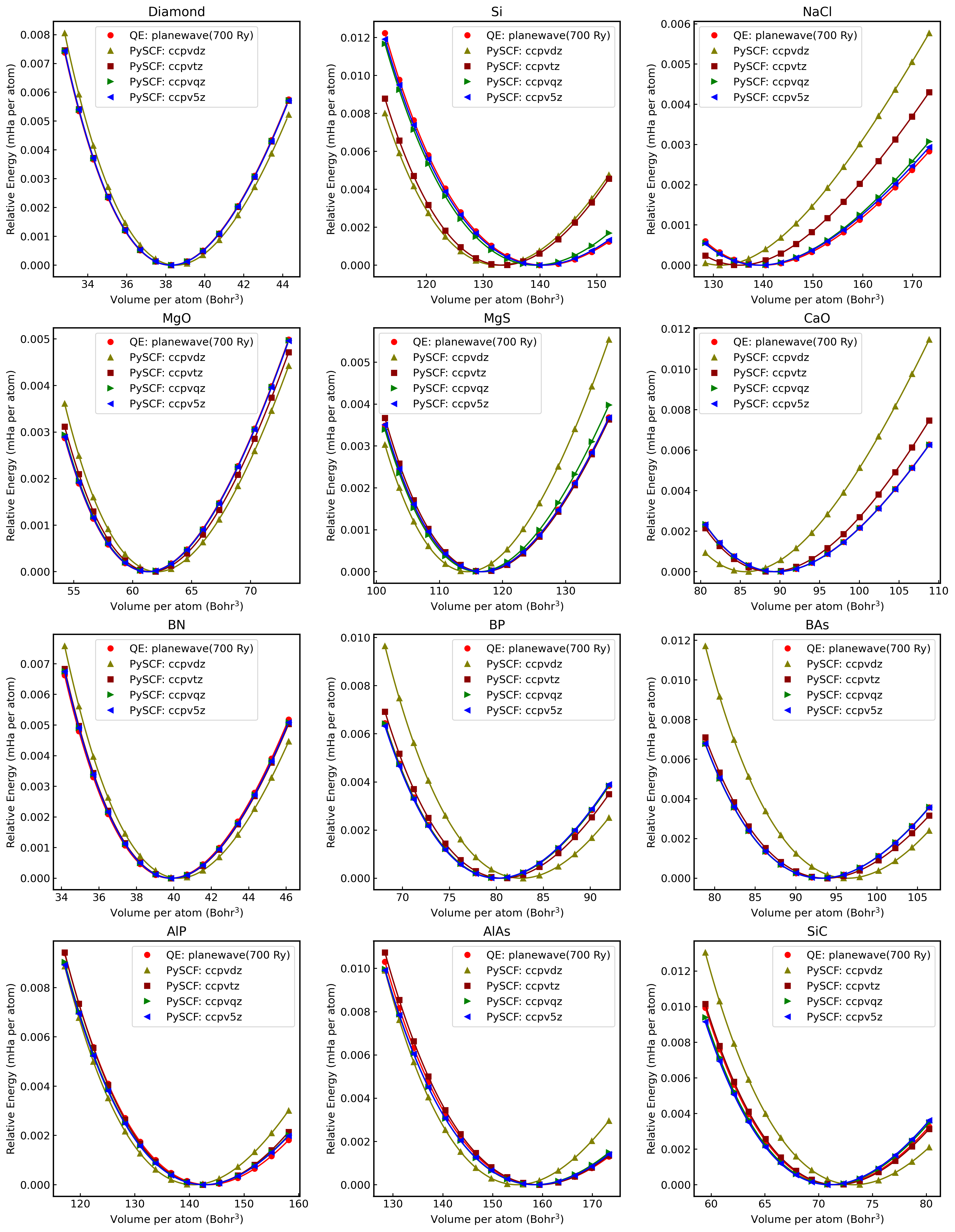}
  \caption{Comparisons of the EOSs computed by \qe\ and \pyscf\ with 1$\times$1$\times$1 conventional cells and $k$ = 2$\times$2$\times$2. The ccECP pseudo potentials were employed for the calculations. The cutoff energy 800~Ry was chosen for the plane-wave \qe\ calculations as the ccECP PPs are hard. For the \pyscf\ calculations, various basis sets from ccp-pVDZ to ccp-pV5Z were tested. The LDA-PZ exchange-correlation functional was used.}
  \label{fig:eos-basis-check-s-1-1-1}
\end{figure}
%%%%%%%%%%%%%%%%%%%%%%%%%%%%%%%%%%%%%%%%%%%%%%%%%%%%
%
%%%%%%%%%%%%%%%%%%%%%%%%%%%%%%%%%%%%%%%%%%%%%%%%%%%%
% EOS, basis check, s=2x2x2
%%%%%%%%%%%%%%%%%%%%%%%%%%%%%%%%%%%%%%%%%%%%%%%%%%%%
\begin{figure}[htbp]
  \centering
  \includegraphics[width=\textwidth]{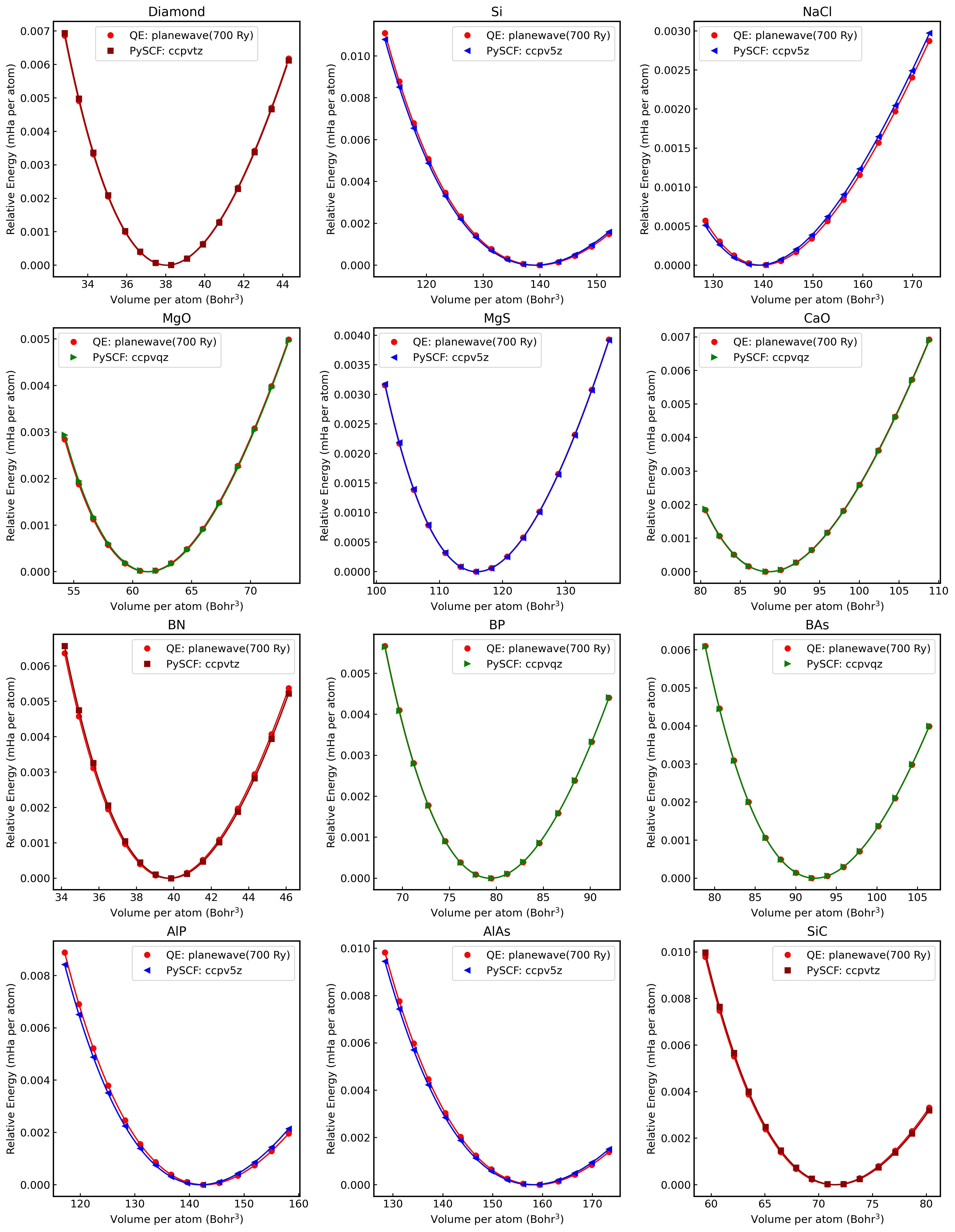}
  \caption{Comparisons of the EOSs computed by \qe\ and \pyscf\ with 2$\times$2$\times$2 conventional cells and $k$ = 2$\times$2$\times$2. The ccECP pseudo potentials were employed for the calculations. The cutoff energy 800~Ry was chosen for the plane-wave \qe\ calculations as the ccECP PPs are hard. For the \pyscf\ calculations, the basis sets converged at with 1$\times$1$\times$1 conventional cells were chosen. The LDA-PZ exchange-correlation functional was used.}
  \label{fig:eos-basis-check-s-2-2-2}
\end{figure}
%%%%%%%%%%%%%%%%%%%%%%%%%%%%%%%%%%%%%%%%%%%%%%%%%%%%
%
%%%%%%%%%%%%%%%%%%%%%%%%%%%%%%%%%%%%%%%%%%%%%%%%%%%%
% EOS, finite-size effect
%%%%%%%%%%%%%%%%%%%%%%%%%%%%%%%%%%%%%%%%%%%%%%%%%%%%
\begin{figure}[htbp]
  \centering
  \includegraphics[width=0.5\columnwidth]{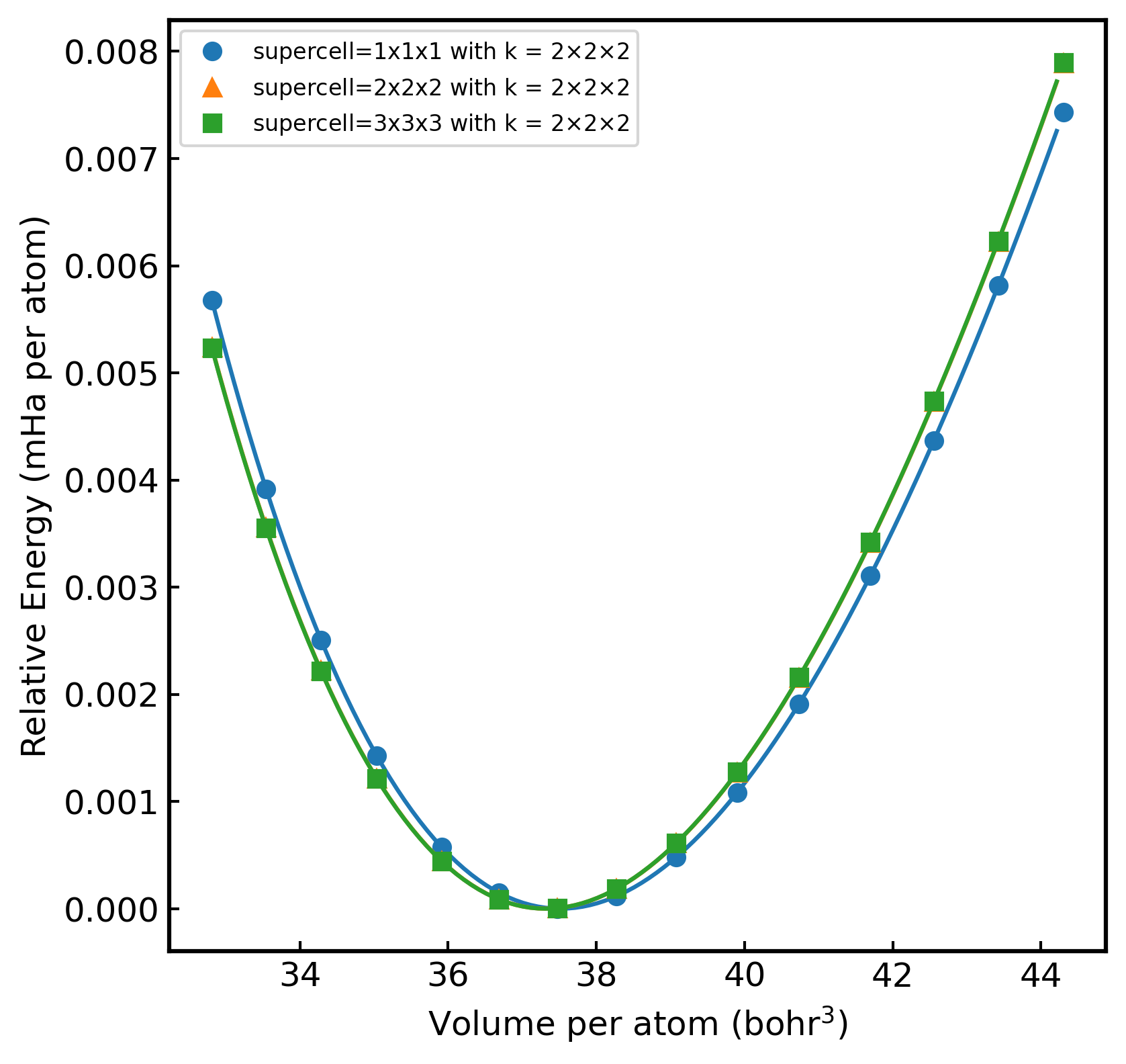}
  \caption{Comparisons of the EOSs computed by \qe\ with several supercells and $k$ = 2$\times$2$\times$2. The LDA-PZ exchange-correlation functional was used. The EOSs obtained by 2$\times$2$\times$2 and 3$\times$3$\times$3 conventional cells are overlapped, indicating that the 2$\times$2$\times$2 cell is enough large to mitigate the one-body finite-size effect.}
  \label{fig:eos-onebody-check-diamond}
\end{figure}
%%%%%%%%%%%%%%%%%%%%%%%%%%%%%%%%%%%%%%%%%%%%%%%%%%%%

%%%%%%%%%%%%%%%%%%%%%%%%%%%%%%%%%%%%%%%%%%%%%%%%%%%%%%%%%%%%%%%%%%
\bibliography{references}
%%%%%%%%%%%%%%%%%%%%%%%%%%%%%%%%%%%%%%%%%%%%%%%%%%%%%%%%%%%%%%%%%%